\documentclass[
 reprint,
 amsmath,amssymb,
 aps,
 prb,
floatfix,
]{revtex4-2}

\usepackage{graphicx}
\usepackage{dcolumn}
\usepackage{bm}%
\usepackage[bookmarks=true,
colorlinks=true,linkcolor=blue,citecolor=blue,urlcolor=blue]{hyperref}
\usepackage{comment}
\usepackage{xcolor}

\begin{document}

\preprint{APS/123-QED}

\title{
Ising superconductors: Interplay of magnetic field, triplet channels and disorder
}

\author{David M\"{o}ckli}
\email{d.mockli@gmail.com}
\author{Maxim Khodas}%
 
\affiliation{%
 The Racah Institute of Physics, The Hebrew University of Jerusalem, Jerusalem 9190401, Israel
}%

\date{\today}

\begin{abstract}
We study the superconducting instability in disordered non-centrosymmetric monolayers with intrinsic Ising spin-orbit coupling (SOC) subjected to an in-plane Zeeman magnetic field. 
The pairing interaction contains the channels allowed by crystal symmetry, such that in general, the pairing state is a mixture of singlet and triplet Cooper pairs. 
The joint action of the SOC and Zeeman field selects a specific in-plane $\mathbf{d}$-vector triplet component to couple with the singlets, which gains robustness against disorder through the coupling. The out-of-plane $\mathbf{d}$-vector component, that in the clean case is immune to both the Zeeman field and SOC, is obliterated by a very small impurity scattering rate.
We formulate the quasi-classical theory of Ising superconductors and solve the linearized Eilenberger equations to obtain the pair-breaking equations that determine the Zeeman field -- temperature dependence of the continuous superconducting transition. Our discussion emphasizes how the Zeeman field, SOC and disorder affect the different superconducting order parameters, and we show how the spin-fields inevitably induce odd-frequency pairing correlations.
\end{abstract}

\maketitle

\section{Introduction}

During the last decade, two-dimensional (2D) superconductivity became an active field of research.
The renewed interest in the field is a result of technological advances in the fabrication of quasi-2D devices fabricated from the Van der Waals materials \cite{Geim2013}.
Such systems are comprised of one-to-several atomically thin monolayers exfoliated on substrates in a nearly perfect atomic registry \cite{Ugeda2015,Xi2015,Saito2016,Dvir2017,Liu2018,Sohn2018,Nakata2018,DelaBarrera2018}.
Many of the properties of the bulk persist down to the monolayer limit. Both bulk and monolayer NbSe$_2$ are charge density wave metallic superconductors \cite{Ugeda2015,Dvir2017}.
Yet, due to the different effective dimensionality, monolayers react differently to the applied fields \cite{Fulde1964,DelaBarrera2018,Shimozawa2016}.
Besides, monolayers containing atoms of more than one kind often lack the inversion center although the bulk may have such a center \cite{Smidman2017}.

In this paper, we consider non-centrosymmetric superconducting monolayers having in-plane mirror $\sigma_h$ symmetry, referred to as \textit{Ising superconductors}. 
The Bloch states, $|\mathbf{k}\!\uparrow\rangle$ ($|\mathbf{k}\!\downarrow\rangle$) are labeled by the in-plane momentum $\mathbf{k}$ and up (down) out-of-plane spin polarization. 
Thanks to the time reversal symmetry, $\mathcal{T}$ the state $|\bf{k}\!\uparrow\rangle$ is degenerate with $|-\bf{k}\!\downarrow\rangle$. 
The states $|\bf{k}\!\downarrow\rangle$ and $|-\bf{k}\!\uparrow\rangle$ are similarly degenerate.
When the lattice breaks parity, the spin-orbit coupling (SOC) causes the spin splitting of Bloch states with the typical energy difference of $\Delta_\mathrm{so}$.
Therefore, the probability amplitude of the Cooper pair to be in a state  $|\bf{k}\!\uparrow;-\bf{k}\!\downarrow\rangle$ differs from the corresponding amplitude for the state 
$|\bf{k}\!\downarrow;-\bf{k}\!\uparrow\rangle$.
Here we denote the anti-symmetrized two-electron states as $|\alpha;\beta\rangle = |\alpha\rangle|\beta\rangle - |\beta\rangle|\alpha\rangle$.
Alternatively, the  parity-even singlets and parity-odd triplets $|\Psi_\mathrm{s,t}\rangle\propto |\mathbf{k}\uparrow;-\mathbf{k}\downarrow\rangle \mp |\mathbf{k}\downarrow;-\mathbf{k}\uparrow\rangle$ coexist  \cite{Gorkov2001,Frigeri2004d}.

Apart from inducing singlet-triplet mixing, the SOC makes the superconducting state robust against the in-plane Zeeman field $\mathbf{B}$.
Because of the negligible thickness of the monolayer, orbital limiting effects do not contribute, and the only way a magnetic field can affect the electronic states is via the paramagnetic effect \cite{Fulde1964}. 
In many instances, the SOC induced splitting greatly exceeds the superconducting gap and may be tuned \cite{Shimozawa2016,DelaBarrera2018}.
The large SOC enhances the critical in-plane field $B_\mathrm{c}$ beyond the Pauli limit. This has been studied theoretically \cite{Bulaevskii976,Frigeri2004d,Samokhin2008,Ilic2017,Mockli2018,Mockli2019} and demonstrated experimentally \cite{Ugeda2015,Xi2015,Saito2016,Dvir2017,Liu2018,Sohn2018,Nakata2018,DelaBarrera2018}.

The pair-breaking equation that determines the dependence of $B_\mathrm{c}$ on the temperature $T$ and the disorder in superconductors with finite SOC was first obtained in Ref.~\cite{Bulaevskii976}.
The critical field of the superconductor with two spin-polarized valleys with SOC which is comparable to the Fermi energy, $E_\mathrm{F}$, $\Delta_\mathrm{so} \gtrsim E_\mathrm{F}$ has been studied in Ref.~\cite{Sosenko2017}.
Subsequently, the effect of the inter-valley scattering on $B_\mathrm{c}$ in the opposite limit $\Delta_\mathrm{so} \ll E_\mathrm{F}$ has been discussed \cite{Ilic2017}.
In this case, the pair breaking equation is identical to the one in Ref.~\cite{Bulaevskii976}.

Here, we extend the results of Ref. \cite{Bulaevskii976} to include the interaction in the triplet channel.
We find that weak disorder obliterates the $|\Psi_{\mathrm{t}}\rangle$ triplets that in the clean case are immune to both SOC and the Zeeman field. 
We show, however, that the Zeeman field induces a different triplet component
$|\Psi_\mathrm{tB}\rangle \propto |\mathbf{k},\uparrow; -\mathbf{k},\uparrow\rangle + |\mathbf{k},\downarrow;-\mathbf{k},\downarrow\rangle$ that couples to the singlets $|\Psi_{\mathrm{s}}\rangle$.
The properties of $|\Psi_\mathrm{tB}\rangle$ and $|\Psi_{\mathrm{t}}\rangle$ triplets are drastically different.
The latter depends on the difference in the density of states of spin-split bands and decouple from the singlets and Zeeman field in the $\Delta_\mathrm{so}\ll E_\mathrm{F}$ limit \cite{Frigeri2004d}. 
In contrast, the field-induced triplets $|\Psi_{\mathrm{tB}}\rangle$ are present regardless of the band structure details, and survive the moderate disorder scattering.
We show that even weak interaction of electrons comprising $|\Psi_{\mathrm{tB}}\rangle$ triplets 
affect decisively and yet differently the phase boundary $B_\mathrm{c}(T)$ of clean and dirty Ising superconductors. 

We study the combined effect of the triplet correlations, non-magnetic disorder and Fermi surface topology on $B_\mathrm{c}$. The $|\Psi_\mathrm{tB}\rangle$ triplets play a much more prominent role than the $|\Psi_\mathrm{t}\rangle$ in the response of the clean or dirty Ising superconductor to the in-plane field.  
Furthermore, $B_\mathrm{c}$ is significantly lower in materials with simply connected Fermi surface hosting symmetry protected zeros of SOC compared with materials with multi-pocket Fermi surfaces without such zeros.  
Indeed, close to the zeros of SOC the superconductivity is not protected against the Zeeman field.
We argue that Fermi surface connectivity qualitatively modifies the effect of the disorder on $B_\mathrm{c}$.

{\it Basic definitions.} 
The lack of inversion symmetry causes
a spin splitting of the bands at a Bloch wave-vector $\mathbf{k}$ that can be described by an effective $\mathbf{k}$-dependent SOC vector $\boldsymbol{\gamma}(\mathbf{k})$. The normal state Hamiltonian acquires the form \cite{Bauer2012}
\begin{align}\label{H}
H_0=\sum_{\mathbf{k},\sigma}\xi(\mathbf{k})c^\dag_{\mathbf{k}\sigma} c_{\mathbf{k}\sigma}+
\sum_{\mathbf{k},\sigma,\sigma'}\boldsymbol{\gamma}(\mathbf{k})\cdot\boldsymbol{\sigma}_\mathbf{\sigma\sigma'}\,c^\dag_{\mathbf{k}\sigma} c_{\mathbf{k}\sigma'}.
\end{align}
Here $\boldsymbol{\sigma}=(\sigma_x,\sigma_y,\sigma_z)$ is the vector of Pauli matrices.
The states are spin polarized along $\boldsymbol{\gamma}(\mathbf{k})$ and the magnitude of the splitting is $2|\boldsymbol{\gamma}(\mathbf{k})|$. As the spin polarization flips under the time-reversal operation, the SOC vector is axial $\boldsymbol{\gamma}(\mathbf{k})=-\boldsymbol{\gamma}(-\mathbf{k})$.
Also, since the spins remain unaffected by the inversion operation, the SOC splitting requires the breaking of parity.

In the superconducting state, the presence of $\boldsymbol{\gamma}(\mathbf{k})$ inevitably leads to parity-mixed pairing correlations \cite{Gorkov2001}.  
Traditionally, the resulting superconducting order parameters are organized in matrix form in spin space as \cite{Balian963,Sigrist1991,Yip2014a}
\begin{align}
\Delta(\mathbf{k})=\left[\psi(\mathbf{k})\sigma_0+\mathbf{d}(\mathbf{k})\cdot\boldsymbol{\sigma} \right ]i\sigma_y.
\label{eq:gapMatrix}
\end{align}
Here, $\psi(\mathbf{k})=\psi(-\mathbf{k})$ parametrizes singlets, and $\mathbf{d}(\mathbf{k})=-\mathbf{d}(-\mathbf{k})$ parametrizes triplets. The singlet (triplet) order parameter $\psi(\mathbf{k})$ ($\mathbf{d}(\mathbf{k})$) is even (odd) in momentum $\mathbf{k}$ to comply with the Pauli principle. The triplet order parameter $\mathbf{d}(\mathbf{k})$ has three components and is usually referred to as the $\mathbf{d}$-vector.
According to Eq. \eqref{eq:gapMatrix},
the most general superconducting state-vector can be written as
 \begin{align}
|\Psi(\mathbf{k})\rangle & = \left[-d_x(\mathbf{k})+id_y(\mathbf{k}) \right ]|\mathbf{k\uparrow;-\mathbf{k\uparrow}}\rangle \notag \\
& + \left[d_x(\mathbf{k})+id_y(\mathbf{k}) \right ]|\mathbf{k\downarrow;-\mathbf{k\downarrow}}\rangle \notag \\
& + \left[\psi(\mathbf{k})+d_z(\mathbf{k}) \right ]|\mathbf{k\uparrow;-\mathbf{k\downarrow}}\rangle \notag \\
&+ \left[-\psi(\mathbf{k})+d_z(\mathbf{k}) \right ]|\mathbf{k\downarrow;-\mathbf{k\uparrow}}\rangle.
\label{eq:state_vector}
\end{align}
In the Ising superconductor, parity-even singlets $|\Psi_\mathrm{s}\rangle$ described by finite $\psi$ and zero $\mathbf{d}$-vector coexists with the parity-odd triplets $|\Psi_\mathrm{t}\rangle$ characterized by $\psi(\mathbf{k})=0$ and finite $\mathbf{d}(\mathbf{k}) \propto \boldsymbol{\gamma}(\mathbf{k})$ \cite{Gorkov2001,Frigeri2004d,Yip2014a}.

Let us assume for simplicity that $\Delta_\mathrm{so} \ll E_\mathrm{F}$ so that we can consider each of the order parameters  separately \cite{Frigeri2004d,Smidman2017}. 
A Zeeman field limits singlet order parameters paramagnetically. 
In contrast, triplets with $\mathbf{d}(\mathbf{k})\perp \mathbf{B}$ remain immune to the Zeeman field \cite{Yip2014a,Ramires2016,aline_fitness2}. 
The $s$-wave singlets are robust against the disorder \cite{Anderson1959}. In purely triplet superconductors, the $\mathbf{d}$-vector averages over the Fermi surface to zero, $\langle \mathbf{d}(\mathbf{k})\rangle=0$. This causes the disorder to suppress triplet order parameters \cite{Mackenzie1998}. Table \ref{tab:intro} summarizes how SOC, Zeeman fields and the disorder affects singlet and triplet superconductors. 

In this work we uncover the prominent role of a different kind of triplet, namely $|\Psi_{\mathrm{tB}}\rangle$, discussed in the introduction. It is characterized by $\mathbf{d}(\mathbf{k}) \propto i \boldsymbol{\gamma}(\mathbf{k}) \times \mathbf{B}$.  
In what follows, we analyze the dramatic modification of the table \ref{tab:intro} brought about by field induced triplets $|\Psi_{\mathrm{tB}}\rangle$ as summarized in table \ref{tab:table}.

The paper is outlined as follows.
In section \ref{sec:model} we present the Hamiltonian and derive the Gor'kov equations; in section \ref{sec:quasi} we introduce the quasi-classical theory and the Eilenberger equations; in section \ref{sec:clean} we solve the linearized Eilenberger equations for the clean case and discuss several technical details that serve as basis to discuss the disordered case; in section \ref{sec:disorder} we solve the disordered case and analyze the main results of this paper. The main result is followed by a discussion \ref{sec:discussion} and concluding remarks \ref{sec:conclusion}. The appendices provide further technical details.

\begin{table}
\caption{\label{tab:intro}%
Effect of SOC, Zeeman field and disorder on the singlet and triplet superconducting order parameters (OPs) considered separately. $\mathbf{d}_\mathrm{im}(\mathbf{k})$ denotes the immune triplet component.}
\begin{ruledtabular}
\begin{tabular}{cccc}
OP & SOC & Zeeman field & Disorder  \\
\hline
$s$-wave $\psi_0$ & Immune & Pauli-limited & Immune  \\ 
$\mathbf{d}(\mathbf{k})$ &$\mathbf{d}_\mathrm{im}(\mathbf{k})\parallel \boldsymbol{\gamma}(\mathbf{k})$& $\mathbf{d}_\mathrm{im}(\mathbf{k})\perp\mathbf{B}$ &  Suppressed  \\
\end{tabular}
\end{ruledtabular}
\end{table}

\section{The model \label{sec:model}}

\subsection{The Hamiltonian}

Our model Hamiltonian has two parts: $H=H_0+H_\mathrm{int}$, where $H_0$ describes the normal state and $H_\mathrm{int}$ contains the interaction channels giving rise to superconductivity. 
We treat the scalar impurities via a self-energy approach within the self-consistent Born approximation. Then, the discussion can be carried out in momentum space, since the role of the disorder is to essentially broaden the spectral function around the Fermi level. However, for completeness, we introduce the Hamiltonian in real-space, and in the next section we Fourier transform to momentum space. 
The real-space normal state Hamiltonian is
\begin{align}
H_0 = \sum_{\sigma,\sigma'} \int\mathrm{d}\mathbf{r}\int\mathrm{d}\mathbf{r}'\,\psi^\dag_\sigma(\mathbf{r})
h_{\sigma\sigma'}(\mathbf{r}-\mathbf{r}')
\psi_{\sigma'}(\mathbf{r}'),
\label{eq:hamiltonian}
\end{align}
where $h_{\sigma\sigma'}(\mathbf{r}-\mathbf{r}')$ contains the single-particle processes
\begin{align}
&h_{\sigma\sigma'}(\mathbf{r}-\mathbf{r}')  = \hat{K}\delta(\mathbf{r}-\mathbf{r}')\delta_{\sigma\sigma'}+\boldsymbol{\gamma}(\mathbf{r}-\mathbf{r}')\cdot\boldsymbol{\sigma}_{\sigma\sigma'} \notag \\
&-\mathbf{B}\cdot\boldsymbol{\sigma}_{\sigma\sigma'}\,\delta(\mathbf{r}-\mathbf{r}')+\sum_j u(\mathbf{r}-\mathbf{R}_j)\delta(\mathbf{r}-\mathbf{r}')\delta_{\sigma\sigma'}.  
\label{eq:single_particle}
\end{align}
Here, $\psi^\dag_\sigma(\mathbf{r})(\psi_\sigma(\mathbf{r}))$ is the field-operator creating (annihilating) a particle with spin-projection $\sigma$ at position $\mathbf{r}$.
The spin indices $\{\sigma,\sigma'\}$ run over the values $\{\uparrow,\downarrow\}$.
The kinetic term $\hat{K}=\left(-(2m)^{-1}\nabla^2-\mu \right )$, where $m$ is the mass of the electron and $\mu$ is the chemical potential.
We use units where the magnetic Zeeman field $\mathbf{B}$ absorbs the usual prefactor with the $g$-factor and the Bohr magneton $g\mu_\mathrm{B}/2$.
Because we are interested in the case of in-plane Zeeman fields applied to monolayers, orbital couplings to the charge are absent. 
The SOC term $\boldsymbol{\gamma}(\mathbf{r}-\mathbf{r}')$ arises due to the lack of an inversion center in the unit cell and its Fourier transform
\begin{align}
    \boldsymbol{\gamma}(\mathbf{k}) = \int\mathrm{d}(\mathbf{r}-\mathbf{r}')\,e^{-i\mathbf{k}\cdot(\mathbf{r}-\mathbf{r}')}\boldsymbol{\gamma}(\mathbf{r}-\mathbf{r}')
    \label{eq:ftg}
\end{align}
was introduced in Eq. \eqref{H}.
Without $\mathbf{B}$ and $u$, the Fourier transform to momentum space of Eq. \eqref{eq:hamiltonian} yields Eq. \eqref{H}. 
We include the effect of disorder by a scalar impurity potential $u(\mathbf{r}-\mathbf{R}_j)$, where the impurity positions $\mathbf{R}_j$ are randomly distributed. Later, we treat the impurities in the self-consistent Born approximation \cite{kopnin2001,bruus2004many,Kita2015}.

The superconducting interaction Hamiltonian in real-space can be written as
\begin{align}
 H_\mathrm{int}&  =
\frac{1}{2}\sum_{\sigma_i,\sigma_i^\prime}\int\mathrm{d}\mathbf{r}\int\mathrm{d}\mathbf{r}'\times 
 \notag \\
& \times V_{\sigma_1'\sigma_2'}^{\sigma_1\sigma_2}(|\mathbf{r}-\mathbf{r}'|) \psi_{\sigma_1}^\dag(\mathbf{r})\psi_{\sigma_2}^\dag(\mathbf{r}')\psi_{\sigma_2'}(\mathbf{r}')\psi_{\sigma_1'}(\mathbf{r}),
\label{eq:Vssss}
\end{align}
where $V_{\sigma_1'\sigma_2'}^{\sigma_1\sigma_2}(|\mathbf{r}-\mathbf{r}'|)$ is a pairing interaction that includes the singlet and triplet pairing channels allowed by symmetry. It has the properties
\begin{align}
V^{\sigma_1\sigma_2}_{\sigma_1'\sigma_2'}\left(|\mathbf{r}-\mathbf{r}'| \right )=V^{\sigma_2\sigma_1}_{\sigma_2'\sigma_1'}\left(|\mathbf{r}'-\mathbf{r}| \right )
=\left[V^{\sigma_1'\sigma_2'}_{\sigma_1\sigma_2}\left(|\mathbf{r}-\mathbf{r}'| \right ) \right ]^*.
\label{eq:intProps}
\end{align}
The first equality follows from the Pauli principle, and the second from hermiticity.

\subsection{Gor'kov equations \label{sec:gorkov}}

We now present the Heisenberg equations of motion for the Matsubara Green's functions, which are called \textit{the Gor'kov equations}. For a detailed derivation, see appendix \ref{app:eqsMoton}.
We wish to determine the normal and superconducting Matsubara Gor'kov Green's functions defined as
\begin{align}
& G_{\sigma\sigma'}(\mathbf{r},\mathbf{r}';\tau,\tau')=-\langle \mathcal{T}\psi_\sigma(\mathbf{r},\tau)\psi_{\sigma'}^\dag(\mathbf{r}',\tau')\rangle; \label{eq:GorkovG} \\
& F_{\sigma\sigma'}(\mathbf{r},\mathbf{r}';\tau,\tau')=-\langle \mathcal{T}\psi_\sigma(\mathbf{r},\tau)\psi_{\sigma'}(\mathbf{r}',\tau')\rangle;\label{eq:GorkovF} \\
&F^*_{\sigma\sigma'}(\mathbf{r},\mathbf{r}';\tau,\tau')=\langle \mathcal{T}\psi^\dag_\sigma(\mathbf{r},\tau')\psi^\dag_{\sigma'}(\mathbf{r}',\tau)\rangle.
\end{align}
Here $\psi_\sigma(\mathbf{r},\tau) = e^{H\tau}\psi_\sigma(\mathbf{r})e^{-H\tau}$ are the field-operators in the Heisenberg representation, where the real number $\tau = it$ ($\hbar=1$) is imaginary time. $\mathcal{T}$ is the time-ordering operator and $\langle \ldots \rangle$ indicate thermal averages. 
We denote $2\times 2$ matrices in spin-space by omitting the spin indices, such that Eq. \eqref{eq:GorkovG} can be expressed as
\begin{align}
G=\begin{bmatrix}
G_{\uparrow\uparrow} & G_{\uparrow\downarrow}\\ 
G_{\downarrow\uparrow} & G_{\downarrow\downarrow}
\end{bmatrix},
\end{align}
and similarly for $F$ and $F^*$.

We study the clean case first. The effects of the disorder can then be easily added via a self-energy that is introduced in Sec. \ref{sec:quasi}.
We Fourier transform the Green's functions to momentum $\mathbf{k}$ and Matsubara frequencies $\omega_n=(2n+1)\pi/\beta$, where $\beta=1/T$ ($k_\mathrm{B}=1$) as
\begin{align}
G(\mathbf{k};\omega_n) & = \int_V\mathrm{d}\mathbf{R}\int_0^\beta\mathrm{d}\mathcal{\tau}\, e^{-\mathbf{k}\cdot\mathbf{R}+i\omega_n\tau }G(\mathbf{R};\tau),
\end{align}
and similarly for $F^*(\mathbf{k};\omega_n)$ and $F(\mathbf{k};\omega_n)$. These Green's function have the properties (see appendix \ref{app:eqsMoton} and Ref. \cite{Kita2015})
\begin{align}
G(\mathbf{k};\omega_n)=G^\dag(\mathbf{k};-\omega_n),\quad 
F(\mathbf{k};\omega_n)=-F^\mathrm{T}(-\mathbf{k};-\omega_n).
\label{eq:propertiesGk}
\end{align}

We perform a mean-field decoupling for the superconducting correlations and using $V(\mathbf{k},\mathbf{k}')=\int\mathrm{d}\mathbf{R}\,e^{-i(\mathbf{k}-\mathbf{k}')\cdot\mathbf{R}}\,V(|\mathbf{R}|)$, we obtain the self-consistent order parameters given by
\begin{align}
\Delta_{\sigma_1\sigma_2}(\mathbf{k})=\frac{T}{ V}\sum_{n=-\infty}^\infty \sum_{\mathbf{k}'}\sum_{\sigma_1'\sigma_2'}V_{\sigma_1'\sigma_2'}^{\sigma_1\sigma_2}(\mathbf{k},\mathbf{k}')F_{\sigma_1'\sigma_2'}(\mathbf{k}';\omega_n).
\label{eq:selfconsistent}
\end{align}
The order parameters \eqref{eq:selfconsistent} can be organized in matrix form as in Eq. \eqref{eq:gapMatrix}. 
From the equations of motion, we obtain the
(left) Gor'kov equation 
\begin{align}
\hat{G}_\mathrm{n}^{-1}(\mathbf{k};\omega_n)\hat{G}(\mathbf{k};\omega_n) -\hat{U}_\mathrm{BdG}(\mathbf{k})\hat{G}(\mathbf{k};\omega_n) =\hat{\sigma}_0,
\label{eq:leftG}
\end{align}
where the hats indicate $4\times 4$ matrices. Each of the $4\times 4$ matrices can be expressed in terms of $2\times 2$ matrices as
\begin{align}
\hat{G}_\mathrm{n}^{-1}(\mathbf{k};\omega_n)=\begin{bmatrix}
G_\mathrm{n}^{-1}(\mathbf{k};\omega_n) & 0\\ 
0 & -G_\mathrm{n}^{\mathrm{T},-1}(-\mathbf{k};-\omega_n)
\end{bmatrix},
\end{align}
where $G_\mathrm{n}^{-1}(\mathbf{k};\omega_n) = \left[i\omega_n-\xi(\mathbf{k}) \right ]\sigma_0-\left[ \boldsymbol{\gamma}(\mathbf{k})-\mathbf{B}\right ]\cdot\boldsymbol{\sigma}$. Here, $\xi(\mathbf{k})$ is the dispersion measured form the chemical potential.
The other matrices are
\begin{align}
&\hat{G}(\mathbf{k};\omega_n)=\begin{bmatrix}
G(\mathbf{k};\omega_n) & F(\mathbf{k};\omega_n)\\ 
-F^*(-\mathbf{k};\omega_n) & -G^*(-\mathbf{k};\omega_n),
\end{bmatrix};\\
& \hat{U}_\mathrm{BdG}(\mathbf{k})
=\begin{bmatrix}
0 & \Delta(\mathbf{k}) \\ 
\Delta^\dag(\mathbf{k}) & 0
\end{bmatrix}.
\end{align}
Note that $\hat{U}_\mathrm{BdG}(\mathbf{k})=\hat{U}^\dag_\mathrm{BdG}(\mathbf{k})$, $\hat{G}(\mathbf{k};\omega_n)=\hat{G}^\dag(\mathbf{k};-\omega_n) $,
and 
$[G_\mathrm{n}^{-1}(\mathbf{k};\omega_n)]^{\dagger} =G_\mathrm{n}^{-1}(\mathbf{k};-\omega_n)$. 
The latter follows from the definition given above as well as $\boldsymbol{\gamma}^*(\mathbf{k}) = \boldsymbol{\gamma}(\mathbf{k})$ ensured by the hermiticity of the Hamiltonian \eqref{H}.
These properties allows us to write the (right) Gor'kov equation
\begin{align}
\hat{G}(\mathbf{k};\omega_n)\hat{G}_\mathrm{n}^{-1}(\mathbf{k};\omega_n) -\hat{G}(\mathbf{k};\omega_n)\hat{U}_\mathrm{BdG}(\mathbf{k}) =\hat{\sigma}_0.
\label{eq:rightG}
\end{align}
The left \eqref{eq:leftG} and right \eqref{eq:rightG} Gor'kov equations provide the starting point to develop the quasi-classical formalism.

\section{Quasi-classical theory \label{sec:quasi}}


We investigate the interplay of energy scales related to superconductivity $\{\psi(\mathbf{k}),|\mathbf{d}(\mathbf{k})|\}$, SOC $\Delta_\mathrm{so}$, Zeeman field $B$ and elastic spin-conserving impurity scattering rate $\Gamma$. 
We consider the regime  
\begin{align}
\psi(\mathbf{k}),|\mathbf{d}(\mathbf{k})|,\sqrt{\Delta_\mathrm{so}^2+B^2},\Gamma \ll E_\mathrm{F},
\label{eq:quasiRegime}
\end{align}
where $E_\mathrm{F}$ is the Fermi energy.
Based on this regime, we develop the quasi-classical theory that concentrates on phenomena close to the Fermi surface and eliminates the variables responsible for physics far away from the Fermi surface such as $\xi(\mathbf{k})$ \cite{Eilenberger1968,Larkin,kopnin2001}. 

To obtain the quasi-classical Green's functions that \textit{operate} at the Fermi surface at the Fermi momentum $\mathbf{k}=\mathbf{k}_\mathrm{F}$, we follow the standard procedure, see Refs. \cite{Eilenberger1968,Kita2015}. We manipulate the left \eqref{eq:leftG} and right \eqref{eq:rightG} Gor'kov equations in the following way: (i) multiply the left-Gor'kov equation by $\hat{\sigma}_z=\mathrm{diag}(\sigma_0,-\sigma_0)$ from the left; (ii) multiply the right-Gor'kov equation by $\hat{\sigma}_z$ from the right; (iii) subtract the right-Gor'kov equation from the left-Gor'kov equation; (iv) multiply the result by $\hat{\sigma}_z$ from the left, and identify the commutators. The procedure yields
\begin{align}
\left[\left(i\omega_n\hat{\sigma}_0-\hat{S}(\mathbf{k})-\hat{U}_\mathrm{BdG}(\mathbf{k}) \right )\hat{\sigma}_z,\hat{\sigma}_z\hat{G}(\mathbf{k};\omega_n) \right ]=0,
\label{eq:eliminated}
\end{align}
which eliminated the variable $\xi(\mathbf{k})$ and the \textit{spin-fields} are contained in
\begin{align}
\hat{S}(\mathbf{k})=\begin{bmatrix}
(\boldsymbol{\gamma}(\mathbf{k})-\mathbf{B})\cdot\boldsymbol{\sigma} & 0\\ 
0 & (\boldsymbol{\gamma}(\mathbf{k})+\mathbf{B})\cdot\boldsymbol{\sigma}^\mathrm{T}
\end{bmatrix}.
\end{align}

We now introduce the dimensionless quasi-classical Green's functions
\begin{align}
\hat{g}(\mathbf{k}_\mathrm{F},\omega_n)& =\oint \frac{\mathrm{d}\xi_\mathbf{k}}{\pi}\,i\hat{\sigma}_z \,\hat{G}(\mathbf{k},\omega_n) \notag \\
& =\begin{bmatrix}
g(\mathbf{k}_\mathrm{F},\omega_n) & -if(\mathbf{k}_\mathrm{F},\omega_n)\\ 
-if^*(-\mathbf{k}_\mathrm{F},\omega_n) & -g^*(-\mathbf{k}_\mathrm{F},\omega_n)
\end{bmatrix},
\label{eq:quasiclassicalG}
\end{align}
Here, $\oint$ only takes contributions from the poles close to the Fermi surface \cite{kopnin2001}. Henceforth, we simply write $\mathbf{k}_{\mathrm{F}}=\mathbf{k}$, such that it is implicitly understood that $\hat{g}(\mathbf{k}_\mathrm{F},\omega_n)$ only has an angular dependence. 
The new $2\times 2$ quasi-classical Green's functions have the properties $g(\mathbf{k},\omega_n) = -g^\dag(\mathbf{k},-\omega_n)$ and $f(\mathbf{k},\omega_n) = -f^\mathrm{T}(-\mathbf{k},-\omega_n)$, which are inherited from the Gor'kov Green's functions. 
Using Eqs. \eqref{eq:quasiclassicalG} and \eqref{eq:eliminated}, we can write the clean Eilenberger equation
\begin{align}
\left[\left(i\omega_n\hat{\sigma}_0-\hat{S}(\mathbf{k})-\hat{U}_\mathrm{BdG}(\mathbf{k}) \right )\hat{\sigma}_z,\hat{g}(\mathbf{k};\omega_n) \right ]=0.
\label{eq:EilenbergerEq_clean}
\end{align}

The effect of scalar disorder in the self-consistent Born approximation can be easily incorporated via the self-energy \cite{kopnin2001,Kita2015}
\begin{align}
\left[\left(i\omega_n\hat{\sigma}_0-\hat{\Sigma}(\omega_n)-\hat{S}(\mathbf{k})-\hat{U}_\mathrm{BdG}(\mathbf{k}) \right )\hat{\sigma}_z,\hat{g}(\mathbf{k};\omega_n) \right ]=0,
\label{eq:EilenbergerEq}
\end{align}
where
\begin{align}
\hat{\Sigma}(\omega_n)=-i\Gamma\left\langle \hat{g}(\mathbf{k};\omega_n) \right\rangle_\mathrm{FS}\hat{\sigma}_z,\quad \Gamma =\frac{1}{2\tau}. 
\end{align}
Here, $\Gamma$ is the scattering rate, $\tau$ is the scattering time, and $\langle \ldots \rangle_\mathrm{FS}\rightarrow \int\frac{\mathrm{d}\varphi_\mathbf{k}}{2\pi}$ indicates the average over the Fermi surface, where $\varphi_\mathbf{k}$ is the polar angle. We henceforth omit the subscript "FS" for the angular Fermi surface averages. 
The Eilenberger equation needs to be supplied with the normalization condition $\hat{g}^2(\mathbf{k};\omega_n)=\hat{\sigma}_0$, which then allows the determination of $\hat{g}(\mathbf{k};\omega_n)$. 
We describe the properties and normalization condition of the quasi-classical Green's function in appendix \ref{app:properties}.

\section{The clean case: limiting effects, odd-frequency pairing and singlet-triplet conversion \label{sec:clean}}

Our objective is to find the critical field $B_\mathrm{c}(T,\Delta_\mathrm{so},\Gamma)$ that marks the continuous normal state -- superconducting transition.
The equation that determines the transition line is referred to as a \textit{pair-breaking equation}.
In this section, we consider the clean case $\Gamma=0$.
For this, we linearize the Eilenberer equation \eqref{eq:EilenbergerEq_clean} and solve for $\{f_0(\mathbf{k};\omega_n),\mathbf{f}(\mathbf{k};\omega_n)\}$. 
We discuss the limiting effects of the spin-fields on superconductivity, the difference between pairing correlations and order parameters, and the self-consistency conditions coming from the interaction channels. 
The reader familiar with these concepts and the results of Ref. \cite{Mockli2019}, and interested in the effect of the disorder may skip section \ref{sec:clean}.

\subsection{The linearized Eilenberger equations}

We parametrize the Green's functions in terms of Pauli matrices as $g(\mathbf{k};\omega_n)=g_0(\mathbf{k};\omega_n)\sigma_0+\mathbf{g}(\mathbf{k};\omega_n)\cdot\boldsymbol{\sigma}$ and $f(\mathbf{k};\omega_n)=\left[f_0(\mathbf{k};\omega_n)\sigma_0+\mathbf{f}(\mathbf{k};\omega_n)\cdot\boldsymbol{\sigma})\right] i\sigma_y$.
The $(1,2)$ component of the Eilenberger equation \eqref{eq:EilenbergerEq} gives the two coupled equations
\begin{align}
& 2\omega_n f_0(\mathbf{k};\omega_n)  =  \psi(\mathbf{k})\left[g_0^*(-\mathbf{k};\omega_n)+g_0(\mathbf{k};\omega_n) \right] \notag \\
& +\mathbf{d}(\mathbf{k})\cdot\left[\mathbf{g}(\mathbf{k};\omega_n)-\mathbf{g}^*(-\mathbf{k};\omega_n) \right]
+2i\mathbf{f}(\mathbf{k};\omega_n)\cdot\mathbf{B};
\end{align}
\begin{align}
 2\omega_n\mathbf{f}(\mathbf{k};\omega_n) & = i\left[\mathbf{g}(\mathbf{k};\omega_n)+\mathbf{g}^*(-\mathbf{k};\omega_n) \right]\times \mathbf{d}(\mathbf{k}) \notag \\ 
& +\psi(\mathbf{k})\left[\mathbf{g}(\mathbf{k};\omega_n)-\mathbf{g}^*(-\mathbf{k};\omega_n) \right] \notag \\
& +\left[ g_0^*(-\mathbf{k};\omega_n)+g_0(\mathbf{k};\omega_n)\right]\mathbf{d}(\mathbf{k})  \notag \\
&+2if_0(\mathbf{k};\omega_n)\mathbf{B}+2\boldsymbol{\gamma}(\mathbf{k})\times\mathbf{f}(\mathbf{k};\omega_n).
\end{align}
More components of the Eilenberger matrix equations would be needed if we went beyond linearization. 
We now linearize the problem by retaining only the linear contribution ($\nu=1$) of the expansion series
\begin{align}
f_0(\mathbf{k};\omega_n)=\sum_{\nu=1}^\infty 
f_0^{(\nu)}(\mathbf{k};\omega_n).
\end{align}
To maintain a clean notation, we omit the $\nu^\mathrm{th}$-order superscript by rewriting
$f_0^{(1)}(\mathbf{k};\omega_n)\rightarrow f_0(\mathbf{k};\omega_n)$, $\mathbf{f}^{(1)}(\mathbf{k};\omega_n)\rightarrow \mathbf{f}(\mathbf{k};\omega_n)$, $\mathbf{g}^{0}(\mathbf{k};\omega_n)=\mathbf{g}^{*0}(-\mathbf{k};\omega_n)\rightarrow 0$, $g^{(0)}(\mathbf{k};\omega_n)=g^{*(0)}(-\mathbf{k};\omega_n)\rightarrow\mathrm{sgn}(\omega_n)$ (see appendix \ref{app:properties}),
which then yields
\begin{align}
& \omega_n f_0(\mathbf{k};\omega_n)=\mathrm{sgn}(\omega_n)\psi(\mathbf{k})+i\mathbf{f}(\mathbf{k};\omega_n)\cdot\mathbf{B}; \label{eil1} \\
& \omega_n\mathbf{f}(\mathbf{k};\omega_n) =  \mathrm{sgn}(\omega_n)\mathbf{d}(\mathbf{k})+if_0(\mathbf{k};\omega_n)\mathbf{B} \notag \\
& \qquad\qquad\qquad
+\boldsymbol{\gamma}(\mathbf{k})\times\mathbf{f}(\mathbf{k};\omega_n).\label{eil2}
\end{align}
These are the linearized Eilenberger equations that determine $\{f_0(\mathbf{k};\omega_n),\mathbf{f}(\mathbf{k};\omega_n)\}$ in the presence of the spin-fields $\{\boldsymbol{\gamma}(\mathbf{k}),\mathbf{B}\}$. In the next section, we discuss these equations because they highlight important differences between the \textit{pairing correlations} $\{f_0(\mathbf{k};\omega_n),\mathbf{f}(\mathbf{k};\omega_n)\}$ and the \textit{order parameters} $\{\psi(\mathbf{k}),\mathbf{d}(\mathbf{k})\}$, and how the spin-fields affect them.

\subsection{Limiting of order parameters by spin-fields}

\begin{figure*} 
\centering
\includegraphics[width=0.95\textwidth]{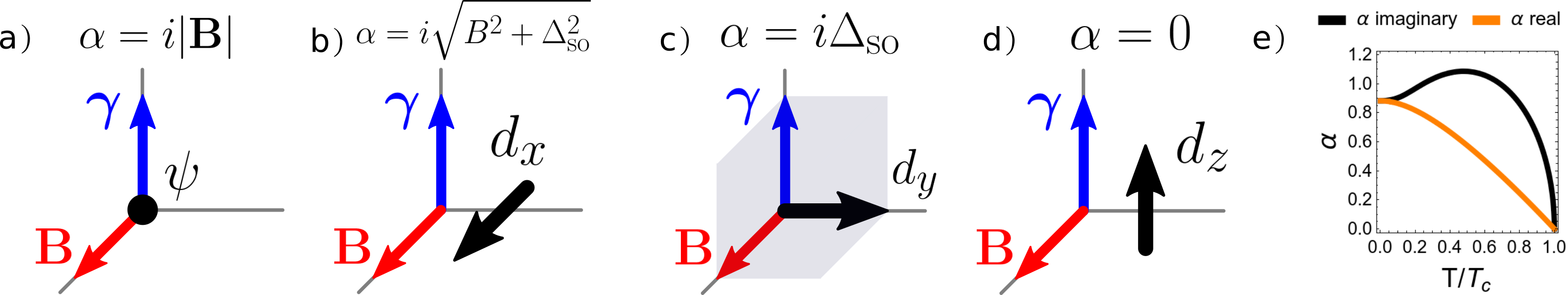}
\caption{\label{fig:limiting} 
Limiting effects quantified by the pair-breaking strength $\alpha$ of the joint action of SOC (blue arrow) and Zeeman field (red arrow) on the singlet $\psi$ (black circle) and triplet components $\{d_x,d_y,d_z\}$ (black arrows). a) The singlet component is limited by the Zeeman field. 
b) The $d_x$ component is limited by both the Zeeman field and SOC. c) The Zeeman field induces the $d_y$, but SOC limits it.
The singlet to triplet conversion rate is proportional to the parallelepiped volume $\mathbf{B}\times\boldsymbol{
\gamma}(\mathbf{k})\cdot\mathrm{Im}\mathbf{d}(\mathbf{k})$.
d) The $d_z$ component is \textit{protected} ($\alpha=0$) from the spin-fields. e) Plot of the pair-breaking equation \eqref{eq:fulde} with generic pair-breaking strength $\alpha$.
The orange curve with $\alpha$ real is sometimes called the Abrikosov-Gor'kov pair-breaking equation and appears in the discussion about disorder.}
\end{figure*}

To discuss the central pair of equations \eqref{eil1} and \eqref{eil2}, we analyze the following situations: (i) paramagnetic limiting of singlets; (ii) paramagnetic limiting of triplets; (iii) limiting of triplets via SOC, see Fig. \ref{fig:limiting}. 

In case (i) we set $\mathbf{d}(\mathbf{k})=\boldsymbol{\gamma}(\mathbf{k})=0$, which restores the inversion symmetry to the Hamiltonian. An important point to notice here is: although the triplet order parameter $\mathbf{d}(\mathbf{k})$ is absent, triplet pairing correlations $\mathbf{f}(\mathbf{k};\omega_n)$ are necessarily present at finite $\mathbf{B}$. The solutions of Eqs. \eqref{eil1} and \eqref{eil2} are
\begin{align}
f_0(\omega_n) = \frac{|\omega_n|}{\omega_n^2+B^2}\psi_0,
\,\, 
\mathbf{f}(\omega_n)=i\mathrm{sgn}(\omega_n)\frac{\psi_0}{\omega_n^2+B^2}\mathbf{B}.
\label{zeeman}
\end{align}
The triplet correlations are odd in frequency because they are induced by the Zeeman field and must comply with the Pauli principle. Although the odd-frequency pairing-correlations are present, there is no interaction in the odd-frequency channel, which means that there is no self-consistency condition for $\mathbf{f}$.
Therefore, we only feed $f_0(\omega_n)$ to the self-consistency condition \eqref{eq:selfconsistent} 
, evaluate the Matsubara sum, which leads to the pair-breaking equation describing paramagnetic limiting of a singlet order parameter \cite{Fulde1964}
\begin{align}
\ln\frac{T}{T_\mathrm{c}}+\mathrm{Re}\,\psi\left(\frac{1}{2}+\frac{\alpha }{2\pi T} \right )-\psi\left(\frac{1}{2} \right )=0,
\label{eq:fulde}
\end{align}
where $T_\mathrm{c}$ is the superconducting transition temperature, $\alpha = i|\mathbf{B}|$, $\psi(z)$ is the digamma function, $\psi\left(\frac{1}{2}\right)=-\ln 4e^\gamma$, and
$\gamma=0.577\ldots$ is the Euler--Mascheroni constant. The parameter $\alpha$ is the \textit{pair-breaking strength} \cite{tinkham}.
The transition line $B_\mathrm{c}(T)$ correspondent to Eq. \eqref{eq:fulde} is plotted in Fig. \ref{fig:limiting}e. 
From Eq. \eqref{eq:fulde} we can extract the zero temperature Pauli limit
\begin{align}
B_\mathrm{P} =\left(\frac{\pi}{2e^\gamma} \right )T_\mathrm{c}=\frac{\Delta_0}{2},
\label{eq:PauliLimit}
\end{align}
where $\Delta_0$ is the familiar zero temperature BCS gap. 
$B_\mathrm{P}$ is the continuous phase transition Pauli limit that is lower than the first-order Clogston limit by a factor of $1/\sqrt{2}$ \cite{Clogston1962,saint1969type}.

In case (ii), with $\psi(\mathbf{k})=\boldsymbol{\gamma}(\mathbf{k})=0$, Eq. \eqref{eil2} shows that only $\mathbf{d}$-vector components that share a parallel component to $\mathbf{B}$ suffer paramagnetic limiting.

In case (iii), we have $\psi(\mathbf{k})=\mathbf{B}=0$, which shows that $\mathbf{d}$-vector components perpendicular to $\boldsymbol{\gamma}(\mathbf{k})$ suffer limiting by SOC. The pair-breaking equation by SOC of such a perpendicular component is (see appendix \ref{app:limitingSOC}) given by Eq. \eqref{eq:fulde} with $\alpha =i\Delta_\mathrm{so}$,
where $\Delta_\mathrm{so}^2=\langle \boldsymbol{\gamma}^2(\mathbf{k})\rangle_\mathrm{FS}$. 
Therefore, SOC limits $\mathbf{d}$-vector components that are perpendicular to $\boldsymbol{\gamma}(\mathbf{k})$ in the same way a Zeeman field limits singlets.
One special case occurs when the $\mathbf{d}$-vector satisfies $\mathbf{d}(\mathbf{k})\parallel \boldsymbol{\gamma}(\mathbf{k})\perp \mathbf{B}$, which then escapes limiting by both $\mathbf{B}$ and $\boldsymbol{\gamma}(\mathbf{k})$ \cite{Frigeri2004,Ramires2016,Smidman2017}. 
Fig. \ref{fig:limiting} summarizes how the joint effect of SOC and the Zeeman field affect
each superconducting component when considered separately.

In this work, we consider Ising SOC $\boldsymbol{\gamma}(\mathbf{k})=\Delta_\mathrm{so}\hat{\gamma}(\mathbf{k})\hat{\boldsymbol{z}}$, where $\hat{\gamma}(\mathbf{k})$ is a basis function that has the symmetries of the crystal and normalized according to $\int_0^{2\pi}\frac{\mathrm{d}\varphi_\mathbf{k}}{2\pi}\,\hat{\gamma}^2(\mathbf{k})=1$. 
Also, without loss of generality, we fix the in-plane Zeeman field $\mathbf{B}=B\hat{\boldsymbol{x}}$. In this case, $d_z(\mathbf{k})$ remains immune against both $\boldsymbol{\gamma}(\mathbf{k})$ and $\mathbf{B}$. On the other hand, the Zeeman field induces an imaginary $d_y(\mathbf{k})$, which is limited by SOC. 
Despite the limiting by SOC, we will show that in the presence of the disorder, $d_y(\mathbf{k})$ is robust, whereas $d_x(\mathbf{k})$ and $d_z(\mathbf{k})$ are obliterated. 

\subsection{Solving the Eilenberger equations}

With SOC and Zeeman field specified, we cast the linearized Eilenberger equations \eqref{eil1} and \eqref{eil2} into matrix form
\small
\begin{align}
\begin{bmatrix}
\omega_n & -iB & 0 & 0\\ 
-iB & \omega_n & \gamma(\mathbf{k}) & 0\\ 
0 & -\gamma(\mathbf{k}) & \omega_n & 0\\ 
0 & 0 & 0 & \omega_n
\end{bmatrix}
\begin{bmatrix}
f_0(\mathbf{k};\omega_n)\\ 
f_x(\mathbf{k};\omega_n)
\\ 
f_y(\mathbf{k};\omega_n)
\\ 
f_z(\mathbf{k};\omega_n)
\end{bmatrix}
=
\mathrm{sgn}(\omega_n)
\begin{bmatrix}
\psi(\mathbf{k})\\ 
d_x(\mathbf{k})\\ 
d_y(\mathbf{k})
\\ 
d_z(\mathbf{k})
\end{bmatrix}.
\label{eq:cleanEil}
\end{align}
\normalsize
The structure of the linear system \eqref{eq:cleanEil} reveals interesting properties of the components $\{f_0,\mathbf{f}\}$. 
The $f_z$ component is decoupled from all the others and remains unaffected by $\gamma(\mathbf{k})$ and $B$. The decoupling of $f_z$ stems from the built-in condition $\Delta_\mathrm{so}\ll E_\mathrm{F}$ in the quasi-classical formalism. If this condition is relaxed, then $f_z$ couples to all other components and is in this way indirectly affected by $\gamma(\mathbf{k})$ and $B$.  
The $f_0$ component is directly affected by $B$. In contrast, $f_y$ is directly affected by $\gamma(\mathbf{k})$. The $f_x$ component can be thought of as a mediator between $f_0$ and $f_y$, since it couples them via $B$ and $\gamma(\mathbf{k})$. The limiting of $f_0$ by $B$ and $f_y$ by $\gamma(\mathbf{k})$ establishes an interesting interplay of the $\{f_0,f_y\}$ sub-system.

To find the order parameters $\{\psi(\mathbf{k}),\mathbf{d}(\mathbf{k})\}$, we first have to solve the Eilenberger matrix \eqref{eq:cleanEil} for the Green's functions $\{f_0,\mathbf{f}\}$, and then supply self-consistency \eqref{eq:selfconsistent}.
In appendix \ref{app:relaxingDx0}, we show that while the pairing correlations $\{f_0,\mathbf{f}\}$ are composed of two independent sub-systems, $\{f_0,f_x,f_y\}$ and $\{f_z\}$, the order parameters separate into three sub-systems, $\{d_x(\mathbf{k})\}$, $\{\psi(\mathbf{k}),d_y(\mathbf{k})\}$ and $\{d_z(\mathbf{k})\}$.  
Since the role of $f_x$ is to couple $f_0$  and $f_y$ and $d_x(\mathbf{k})$ remains uncoupled from the other order parameters, we can safely set $d_x(\mathbf{k})=0$ to solve for $\{f_0,f_y\}$. 
In fact, we will soon see that part of $f_x$ that mediates between $f_0$ and $d_y$ is an odd-frequency pairing correlation.
With this, we can eliminate $f_x$ in favor of $f_0$ and $f_y$ in Eq. \eqref{eq:cleanEil}, and obtain the sub-system

\begin{align}
\begin{bmatrix}
\omega_n^2+B^2 & iB\gamma(\mathbf{k}) \\ 
-iB\gamma(\mathbf{k}) & \omega_n^2+\gamma^2(\mathbf{k})
\end{bmatrix}
\begin{bmatrix}
f_0(\mathbf{k};\omega_n)\\ 
f_y(\mathbf{k};\omega_n)
\end{bmatrix}
=
|\omega_n|
\begin{bmatrix}
\psi(\mathbf{k})\\ 
d_y(\mathbf{k})
\end{bmatrix}.
\label{eq:conversion}
\end{align}
The form of Eq. \eqref{eq:conversion} shows that while $B$ suppresses $f_0$, $\gamma(\mathbf{k})$ suppresses $f_y$. The two components \textit{convert} between each other through the joint presence of $B$ and $\gamma(\mathbf{k})$. 
The system \eqref{eq:conversion} has solution
\begin{align}
& f_0(\mathbf{k};\omega_n) = \frac{1}{|\omega_n|}\frac{\psi(\mathbf{k})\left(\omega_n^2+\gamma^2(\mathbf{k}) \right )-iB\gamma(\mathbf{k})d_y(\mathbf{k})}{\omega_n^2+B^2+\gamma^2(\mathbf{k})}; \label{eq:f0} \\
& f_y(\mathbf{k};\omega_n) = \frac{1}{|\omega_n|}\frac{d_y(\mathbf{k})\left(\omega_n^2+B^2 \right )+iB\gamma(\mathbf{k})\psi(\mathbf{k})}{\omega_n^2+B^2+\gamma^2(\mathbf{k})}.\label{eq:fy}
\end{align}
The singlet component $f_0(\mathbf{k};\omega_n)$ is even in $\mathbf{k}$, while the triplet component $f_y(\mathbf{k};\omega_n)$ is odd. Both components depend on the singlet $\psi(\mathbf{k})$ and triplet $d_y(\mathbf{k})$ order parameters. 

Even if $d_x(\mathbf{k})=0$, the $f_x$ correlations are inevitably present, with solution
\begin{align}
f_x(\mathbf{k};\omega_n) = \mathrm{sgn}(\omega_n)\,\frac{\gamma(\mathbf{k})d_y(\mathbf{k})-iB\psi(\mathbf{k})}{\omega_n^2+B^2+\gamma^2(\mathbf{k})}.
\label{eq:fx}
\end{align}
Note that $f_x(\mathbf{k};\omega_n)$ is even in $\mathbf{k}$ and odd in $\omega_n$.

\subsection{Symmetry pairing channels and self-consistency}

Once we solved the linearized Eilenberger equations for $\{f_0,\mathbf{f}\}$, we use the self-consistent gap equation \eqref{eq:selfconsistent} to determine the order parameters for which there is a pairing channel. 
To do this, we specify the pairing channels and relate Eq. \eqref{eq:selfconsistent} to $\{f_0,\mathbf{f}\}$. 

The pairing interaction can be written in terms of crystal symmetry compatible singlet and triplet channels as
\begin{align}
V^{\sigma_1\sigma_2}_{\sigma_1'\sigma_2'}(\mathbf{k},\mathbf{k}')& =
\sum_{\Gamma,j}v_{s,\Gamma}\left[\hat{\tau}_{\Gamma_j}(\mathbf{k}) \right ]_{\sigma_1\sigma_2}\left[\hat{\tau}_{\Gamma_j}(\mathbf{k}') \right ]_{\sigma_1'\sigma_2'}^*  \notag \\
&+\sum_{\Gamma,j}v_{t,\Gamma}\left[\hat{\boldsymbol{\tau}}_{\Gamma_j}(\mathbf{k}) \right ]_{\sigma_1\sigma_2}\left[\hat{\boldsymbol{\tau}}_{\Gamma_j}(\mathbf{k}') \right ]_{\sigma_1'\sigma_2'}^*.
\label{eq:interaction}
\end{align}
Here, $\hat{\tau}_{\Gamma_j}(\mathbf{k})=\hat{\psi}_{\Gamma_j}(\mathbf{k})i\sigma_y $ and $\hat{\boldsymbol{\tau}}_{\Gamma_j}(\mathbf{k})=\hat{\mathbf{d}}_{\Gamma_j}(\mathbf{k})\cdot\boldsymbol{\sigma}\,i\sigma_y$, where $j$ labels the basis functions of an irreducible representation (irrep) $\Gamma$ of a point symmetry group, and $v_{s(t),\Gamma}<0$ are attractive interactions in each channel. 
We address the case of repulsion in the triplet channel in appendix \ref{app:repulsion}.
In principle, additional parity-mixed channels that convert between singlets and triplets are also allowed \cite{Frigeri2006}.
We do not include parity-mixed
interaction channels here to show that singlet-triplet conversion occurs due to the presence of spin-fields alone.

For concreteness, here we discuss the Cooper channels of the $D_{3h}$ point-group lacking the inversion element. Yet, the pair-breaking equations obtained in this paper are universal to all point-groups lacking inversion.
We assume a dominant singlet ($s$-wave) channel and write the singlet order parameter in terms of the basis function $\psi_{A_1'}(\mathbf{k})=\psi_0\,\hat{\psi}_{A_1'}(\mathbf{k})=\psi_0$. Generally, $\psi_0$ is a complex number, but here we choose $\psi_0$ to be real. We denote the superconducting transition temperature associated to $\psi_0$ as $T_\mathrm{cs}$. For the triplet part, we are interested in the order parameter that gives a finite contribution to the triple product $\mathbf{B}\times\boldsymbol{\gamma}(\mathbf{k})\cdot\mathrm{Im}\,\mathbf{d}(\mathbf{k})$ that only keeps the imaginary triplet component that is induced by $\mathbf{B}$ \cite{Mockli2019}. If $\mathbf{B}=B\hat{\boldsymbol{x}}$, then $\mathbf{d}_{E''}(\mathbf{k})=i\eta_y\hat{\gamma}(\mathbf{k})\hat{\boldsymbol{y}}=d_y(\mathbf{k})\hat{\boldsymbol{y}}$, where $\eta_y$ is real. Here, $\hat{\gamma}(\mathbf{k})$ is the same basis function used for SOC $\boldsymbol{\gamma}(\mathbf{k})=\Delta_\mathrm{so}\hat{\gamma}(\mathbf{k})\hat{\boldsymbol{z}}$. We denote the critical temperature associated to $\eta_y$ by $T_\mathrm{ct}<T_\mathrm{cs}$.

Usually, when the superconductivity does not lower the symmetry of the lattice, all the Cooper pairs transforming trivially are simultaneously present in the condensate. In this standard situation, the broken parity allows for singlets and certain triplets to belong to the trivial irrep ($A_1'$). Hence, the $A_1'$ singlets and triplets coexist \cite{Yip2014a,Smidman2017}. We emphasize that the in-plane Zeeman field lowers the symmetry, and selects a specific two-dimensional irrep ($E''$) to mix with the one-dimensional lattice symmetric irrep ($A_1'$). 
In a previous paper, we showed that while the singlet-triplet coupling within a same irrep depends on the difference of the density of states of the spin-split bands, the mixing of different irreps due to the magnetic field is always present \cite{Mockli2019}. In this regard, the inter-irrep parity-mixing discussed here is more generic than intra-irrep mixing.

With the singlet (triplet) interactions $v_\mathrm{s(t)}<0$ and the density of states per spin at the Fermi level $N_0$, we define the singlet (triplet) coupling constants $\lambda_\mathrm{s(t)}=2N_0 v_\mathrm{s(t)}/V$. 
We can express the dimensionless coupling constants in favor of the critical temperatures as (see appendix \ref{app:cpling} for details)
\begin{align}
-\frac{1}{\lambda_\mathrm{s(t)}}=\ln\left(\frac{T}{T_\mathrm{cs(t)}} \right )+\pi T\sum_{n=-n_\mathrm{c}-1}^{n_\mathrm{c}} \frac{1}{|\omega_n|} ,
\label{eq:manipulated}
\end{align}
where the cutoff $n_\mathrm{c}$ is determined by the characteristic energy scale of the pairing interaction $\epsilon_\mathrm{c}$ via $(2n_\mathrm{c}+1)\pi=\epsilon_\mathrm{c}/T$.

Next, we express the self-consistency condition \eqref{eq:selfconsistent} in terms of the critical temperatures and the quasi-classical Green's functions. To do this we use the definition of the quasi-classical Green's functions \eqref{eq:quasiclassicalG} and parametrize it according to \eqref{eq:paramf}. 
Also, given the quasi-classical regime, we write the momentum sum as
\begin{align}
\sum_\mathbf{k}\rightarrow N_0\int\frac{\mathrm{d}\varphi_\mathbf{k}}{2\pi}\int_{-\infty}^\infty\mathrm{d}\xi_\mathbf{k}.
\label{eq:dos}
\end{align}
This allows us to obtain the self-consistency conditions for the singlet and triplet order parameters, which are
\begin{align}
\psi_0\ln\frac{T}{T_\mathrm{cs}}+\pi T\sum_{n=-\infty}^\infty\left(\frac{\psi_0}{|\omega_n|}-\langle f_0(\mathbf{k};\omega_n)\rangle \right )=0,
\label{eq:selfcSinglet}
\end{align}
\begin{align}
& d_y(\mathbf{k})\ln\frac{T}{T_\mathrm{ct}} \label{eq:selcCtriplet} \\
& +\pi T\sum_{n=-\infty}^\infty
\biggr(\frac{d_y(\mathbf{k})}{|\omega_n|}  -\hat{\gamma}(\mathbf{k})\left\langle\hat{\gamma}(\mathbf{k}')f_y(\mathbf{k}';\omega_n) \right\rangle \notag
\biggr)=0.
\end{align}
Here, the averages $\langle\dots\rangle\equiv \int\frac{\mathrm{d}\varphi_\mathbf{k}}{2\pi}(\ldots)$ are taken over the Fermi surface.
The argument of the Matsubara sum is now convergent so that we can make $n_\mathrm{c}\rightarrow\infty$. 
Together with the Eilenberger equations, Eqs. \eqref{eq:selfcSinglet} and \eqref{eq:selcCtriplet} yield a coupled system of equations for $\{\psi_0,\eta_y\}$.

\subsection{The pair-breaking equation}

\begin{figure*} 
\centering
\includegraphics[width=0.9\textwidth]{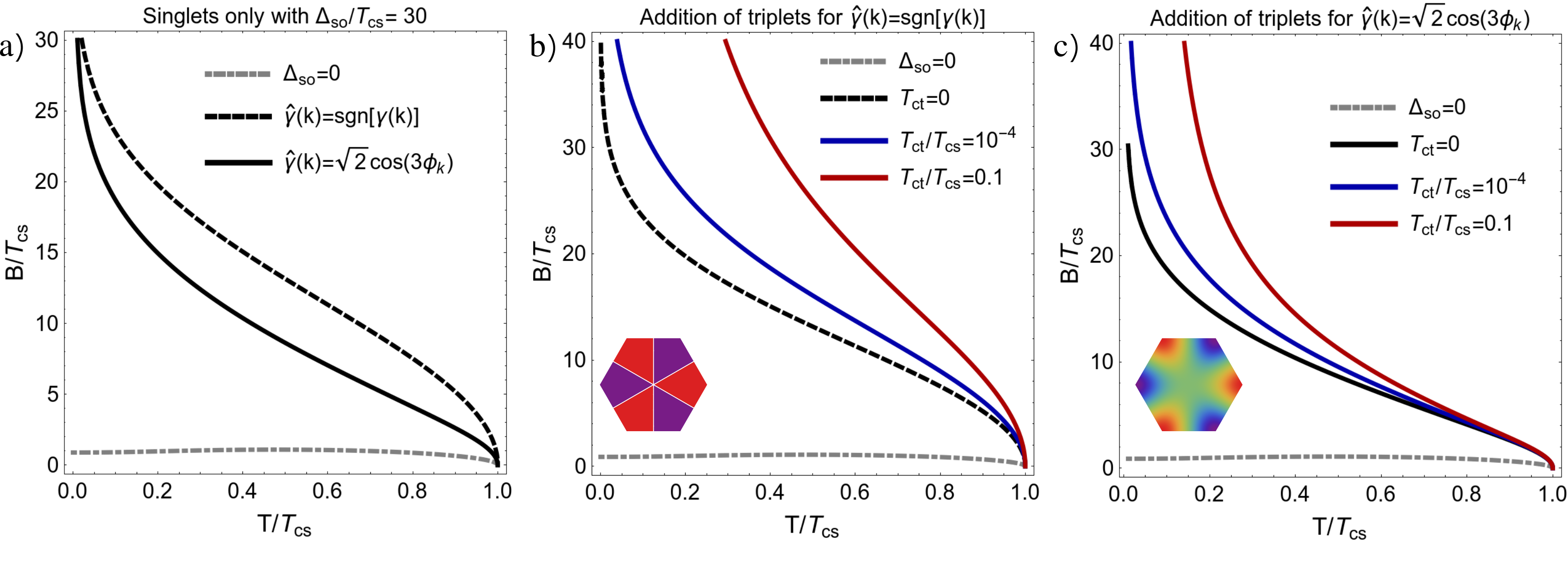}
\caption{\label{fig:1} 
Transition lines $B_\mathrm{c}(T)$ for clean Ising superconductors comparing (b) $\hat{\gamma}(\mathbf{k})=\mathrm{sgn}[\gamma(\mathbf{k})]$ and (c) $\hat{\gamma}(\mathbf{k})=\sqrt{2}\cos(3\varphi_\mathbf{k})$.
The clean transition lines are very sensitive to $T_\mathrm{ct}$. The hexagonal insets show the schematic structure of the basis functions $\hat{\gamma}(\mathbf{k})$ in the first Brillouin zone. 
}
\end{figure*}

We now feed the solutions in Eqs. \eqref{eq:f0} and \eqref{eq:fy} to the self-consistency conditions \eqref{eq:selfcSinglet} and \eqref{eq:selcCtriplet}.
Writing $d_y(\mathbf{k})=i\eta_y\hat{\gamma}(\mathbf{k})$ and $\gamma(\mathbf{k})=\Delta_\mathrm{so}\hat{\gamma}(\mathbf{k})$, the required Fermi surface averages are 
\begin{align}
\langle f_0(\mathbf{k};\omega_n)\rangle  = \frac{\psi_0}{|\omega_n|}A_1+\frac{\eta_y}{|\omega_n|}\left(\frac{B}{\Delta_\mathrm{so}} \right )A_3 
 ;
 \label{cc1}
\end{align}
\begin{align}
& \langle \hat{\gamma}(\mathbf{k})f_y(\mathbf{k};\omega_n)\rangle  = \notag \\  
& \frac{i\psi_0}{|\omega_n|}\left(\frac{B}{\Delta_\mathrm{so}} \right )A_3+\frac{i\eta_y}{|\omega_n|}\left(\frac{\omega_n^2+B^2}{\Delta_\mathrm{so}^2} \right )A_3,
\label{cc2}
\end{align}
where we define the dimensionless angular averages
\begin{align}
A_1(\omega_n, B,\Delta_\mathrm{so}) & =\left\langle \frac{\omega_n^2+\gamma^2(\mathbf{k})}{\omega_n^2+B^2+\gamma^2(\mathbf{k})}\right\rangle; \label{a1} \\
A_2(\omega_n, B,\Delta_\mathrm{so}) &  = \left\langle \frac{\omega_n^2}{\omega_n^2+B^2+\gamma^2(\mathbf{k})}\right\rangle;\label{a2} \\
A_3(\omega_n, B,\Delta_\mathrm{so}) & =\left\langle \frac{\gamma^2(\mathbf{k})}{\omega_n^2+B^2+\gamma^2(\mathbf{k})}\right\rangle. \label{a3}
\end{align}
The first two averages are related by $(B^2/\omega_n^2)A_2=1-A_1$.
We frequently prefer expressing $A_1$ in favour of $A_2$.
With these definitions, the linearized coupled self-consistency conditions in Eqs. \eqref{eq:selfcSinglet} and \eqref{eq:selcCtriplet} give
\begin{align}
\begin{bmatrix}
\ln\frac{T}{T_\mathrm{cs}}+\mathcal{S}_\mathrm{s} & \mathcal{S}_\mathrm{st}\\ 
\mathcal{S}_\mathrm{st} & \ln\frac{T}{T_\mathrm{ct}}+\mathcal{S}_\mathrm{t}
\end{bmatrix}
\begin{bmatrix}
\psi_0\\ 
\eta_y
\end{bmatrix}
=0, 
\label{eq:coupled_self_consistency}
\end{align}
with the Matsubara sums $\mathcal{S}=\mathcal{S}(T,B,\Delta_\mathrm{so})$ defined as
\begin{align}
& \mathcal{S}_\mathrm{s} = \pi T B^2\sum_{n=-\infty}^\infty \frac{A_2}{|\omega_n|^3}; \\
& \mathcal{S}_\mathrm{st}=\pi T \frac{B}{\Delta_\mathrm{so}}\sum_{n=-\infty}^\infty \frac{A_3}{|\omega_n|}; \\
& \mathcal{S}_\mathrm{t}=\pi T \sum_{n=-\infty}^\infty \left(\frac{1}{|\omega_n|}-\frac{\omega_n^2+B^2}{\Delta_\mathrm{so}^2}\frac{A_3}{|\omega_n|} \right ).
\end{align}
It is useful to keep in mind that $A_3$ is only present with SOC.

\subsubsection{The structure of SOC}

We now show that the specific choice of the SOC basis function $\hat{\gamma}(\mathbf{k})$ affects the shape of the transition line $B_\mathrm{c}(T)$. To illustrate this, we work with two basis function examples: (i) $\hat{\gamma}(\mathbf{k})=\mathrm{sgn}[\gamma(\mathbf{k})]$, for which $\gamma^2(\mathbf{k})\rightarrow \Delta_\mathrm{so}^2$ and $\langle \hat{\gamma}^2(\mathbf{k})\rangle=1$. This toy example is extensively used throughout the literature and in some situations gives qualitatively correct results \cite{Xiao2012,Ilic2017,Mockli2019}; (ii) $\hat{\gamma}(\mathbf{k})=\sqrt{2}\cos(3\varphi_\mathbf{k})$, which implements a more realistic SOC for the point group $D_{3h}$. 
The case (i) is frequently used for multi-pocket Fermi surfaces, while case (ii) is suitable for simply connected Fermi surfaces. 
We now solve for both cases and contrast the solutions.

In case (i), the Matsubara sums can be carried out analytically. 
To simplify notations, we define the function involving the digamma function $C(y)=\mathrm{Re}\,\psi\left(\frac{1}{2}+i\frac{y}{2} \right )-\psi\left(\frac{1}{2}\right)\geq 0$.
With this, the pair-breaking equation is 
\begin{align}
\det
\begin{bmatrix}
\ln\frac{T}{T_\mathrm{cs}}+\frac{B^2}{B^2+\Delta_\mathrm{so}^2} \,C(y )& \frac{B\Delta_\mathrm{so}}{B^2+\Delta_\mathrm{so}^2}\, C(y )\\ 
\frac{B\Delta_\mathrm{so}}{B^2+\Delta_\mathrm{so}^2}\, C(y ) & \ln\frac{T}{T_\mathrm{ct}}+\frac{\Delta_\mathrm{so}^2}{B^2+\Delta_\mathrm{so}^2}\,C(y )
\end{bmatrix}
=0,
\label{eq:pairbreakingClean}
\end{align}
with $y=\sqrt{B^2+\Delta_\mathrm{so}^2}/(\pi T)$.
The $B_\mathrm{c}(T)$ transition lines obtained from Eq. \eqref{eq:pairbreakingClean} are plotted in Fig. \ref{fig:1}(b). 
At finite SOC, all curves diverge at low temperatures. 
Note that the singlet and triplet components $\{\psi_0,\eta_y\}$ only couple through the joint action of SOC and the Zeeman field. 

In case (ii), the averages yield
\begin{align}
& A_1 = 1-\frac{B^2}{\sqrt{(\omega_n^2+B^2)(\omega_n^2+B^2+2\Delta_\mathrm{so}^2)}}; \\
& A_2 = \frac{\omega_n^2}{\sqrt{(\omega_n^2+B^2)(\omega_n^2+B^2+2\Delta_\mathrm{so}^2)}}; \\
& A_3 = 1-\sqrt{\frac{\omega_n^2+B^2}{\omega_n^2+B^2+2\Delta_\mathrm{so}^2}}.
\end{align}
The sums are convergent and can be performed numerically.

In Fig. \ref{fig:1} we compare the transition lines $B_\mathrm{c}(T)$ using $\Delta_\mathrm{so}/T_\mathrm{cs}=30$ with $\hat{\gamma}(\mathbf{k})=\mathrm{sgn}[\gamma(\mathbf{k})]$ (dashed-black lines in a and b) and $\hat{\gamma}(\mathbf{k})=\sqrt{2}\cos(3\varphi_\mathbf{k})$ (solid-black lines in a and c). 
The gray dashed-dotted line indicates the pure Pauli-limit, also see black line in Fig. \ref{fig:limiting}e.
Fig. \ref{fig:1}a shows that the critical field is lower with $\hat{\gamma}(\mathbf{k})=\sqrt{2}\cos(3\varphi_\mathbf{k})$. 
In Fig. \ref{fig:1}b, we illustrate the effect of an attractive triplet channel with $\hat{\gamma}(\mathbf{k})=\mathrm{sgn}[\gamma(\mathbf{k})]$. Similarly, in Fig. \ref{fig:1}c we show the case with $\hat{\gamma}(\mathbf{k})=\sqrt{2}\cos(3\varphi_\mathbf{k})$.
In section \ref{sec:disorder} we study how the disorder affects the transition lines and the triplet channels.

The off-diagonal terms in Eq. \eqref{eq:pairbreakingClean} show the interplay of SOC, the Zeeman field and its role to induce equal-spin triplets.
According to Eq. \eqref{eq:state_vector},
the Zeeman field converts singlet Cooper pairs with a state-vector
\begin{align}
|\Psi_\mathrm{s}\rangle = \psi_0\left(| \mathbf{k}\uparrow; -\mathbf{k}\downarrow\rangle - |\mathbf{k}\downarrow;-\mathbf{k}\uparrow\rangle\right ) 
\label{s1}
\end{align}
to equal-spin triplet Cooper pairs
\begin{align}
|\Psi_\mathrm{tB}(\mathbf{k})\rangle = i\eta_y\hat{\gamma}(\mathbf{k})\left(|\mathbf{k}\uparrow; -\mathbf{k}\uparrow\rangle \!+\! |\mathbf{k}\downarrow;-\mathbf{k}\downarrow\rangle  \right ).
\label{s2}
\end{align}
This conversion can be understood by following the spin realignment caused by $\mathbf{B}$. The imaginary $i$ is the total Berry phase accumulated by the spins in the course of realignment, see Ref. \cite{Mockli2019}.

\section{The effect of disorder \label{sec:disorder}}

In this section, we address the effect of scalar impurity scattering $\Gamma$ on the components $\{\psi_0,\mathbf{d}(\mathbf{k})\}$. We show that while the parity-mixed $\{\psi_0,d_y(\mathbf{k})\}$ sub-system displays robustness, the independent triplet $d_x(\mathbf{k})$ and $d_z(\mathbf{k})$ components are obliterated.

\subsection{The Eilenberger equations }

To solve the disordered case, we linearize the Eilenberger equation \eqref{eq:EilenbergerEq} using the same procedure used to obtain Eqs. \eqref{eil1} and \eqref{eil2}. This gives
\begin{align}
& \tilde{\omega}_n f_0(\mathbf{k};\omega_n) = \mathrm{sgn}(\omega_n)\tilde{\psi}(\omega_n)+i\mathbf{B}\cdot \mathbf{f}(\mathbf{k};\omega_n) \label{eq:f0_tilde} \\
&\tilde{\omega}_n \mathbf{f}(\mathbf{k};\omega_n)=\mathrm{sgn}(\omega_n)\tilde{\mathbf{d}}(\mathbf{k};\omega_n)+i\mathbf{B}f_0(\mathbf{k};\omega_n) \notag \\
&\qquad \qquad \quad 
+\boldsymbol{\gamma}(\mathbf{k})\times \mathbf{f}(\mathbf{k};\omega_n),
\label{eq:f_tilde}
\end{align}
with the effective frequencies and order parameters defined as
\begin{align}
& \tilde{\omega}_n=\omega_n+\mathrm{sgn}(\omega_n)\Gamma; \\
& \tilde{\psi}(\omega_n) = \psi_0+\Gamma\langle f_0(\mathbf{k};\omega_n)\rangle; \\
& \tilde{\mathbf{d}}(\mathbf{k};\omega_n) = \mathbf{d}(\mathbf{k})+\Gamma\langle \mathbf{f}(\mathbf{k};\omega_n)\rangle.
\end{align}
These equations now also involve the angular Fermi surface averages of the pairing correlations $\{\langle f_0(\mathbf{k};\omega_n)\rangle,\langle \mathbf{f}(\mathbf{k};\omega_n)\rangle\}$. 
These averages determine how the disorder affects the superconducting state. Larger averages imply more robustness. 
As in the clean case, the $d_z(\mathbf{k})$ component decouples from all the others. 
In the next sections, we obtain the pair-breaking equation by the disorder for $d_z(\mathbf{k})$, and then study the coupled sub-system $\psi(\mathbf{k})+i d_y(\mathbf{k})$. 

\begin{figure*} 
\centering
\includegraphics[width=0.7\textwidth]{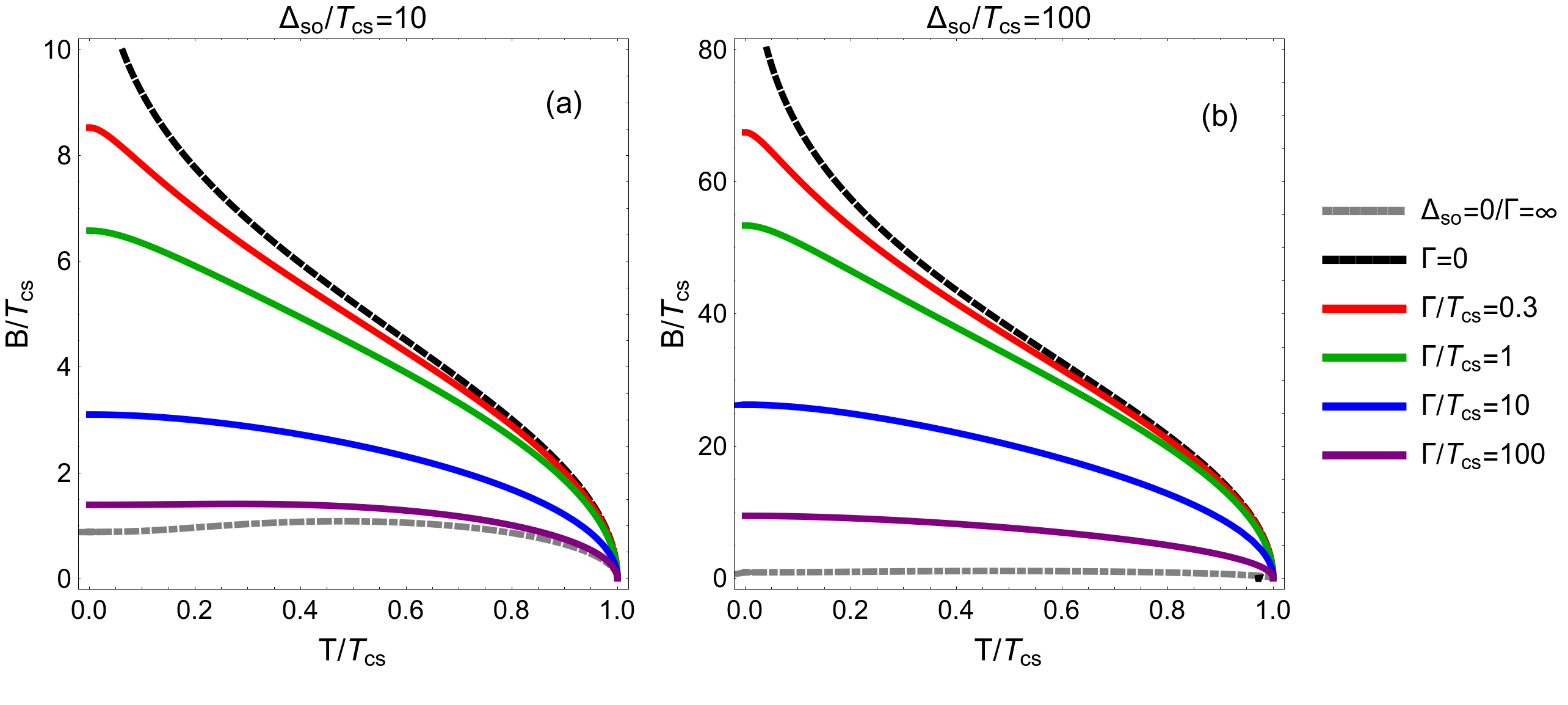}
\caption{\label{fig:2} 
Effect of the disorder on the singlet transition lines $B_\mathrm{c}(T)$ with $\hat{\gamma}(\mathbf{k})=\mathrm{sgn}[\gamma(\mathbf{k})]$.
The scattering rate undoes the enhancement caused by SOC. In Fig. \ref{fig:4} we show the case
for $\hat{\gamma}(\mathbf{k})=\sqrt{2}\cos(3\varphi_\mathbf{k})$ .
}
\end{figure*}

\subsection{Solution for \texorpdfstring{$d_z(\mathbf{k})$}{} and \texorpdfstring{$d_x(\mathbf{k})$}{}}

By taking the average of Eq. \eqref{eq:f_tilde} we see that $\langle f_z(\mathbf{k};\omega_n) \rangle =0$. 
The solution for the $f_z$ component is then $f_z(\mathbf{k};\omega_n)=d_z(\mathbf{k})(|\omega_n|+\Gamma)^{-1}$. Substituting this in the self-consistency equation for $d_z(\mathbf{k})$ analogous to \eqref{eq:selcCtriplet}, we obtain the pair-breaking equation by the disorder, which is Eq. \eqref{eq:fulde} with $\alpha=\Gamma$ and $T_\mathrm{c}=T_\mathrm{ct}$. This result is universal to all superconductors where the Fermi surface average of the order parameter vanishes \cite{Radtke1993,kopnin2001}.
An estimate for the critical scattering rate $\Gamma_\mathrm{c}$ at which the disorder obliterates $d_z(\mathbf{k})$ can be obtained from the asymptotic behavior of the digamma function $\psi(z)\approx \ln|z|$ ($z\gg 1$), for which
\begin{align}
     \Gamma_\mathrm{c} = \left(\frac{\pi}{2e^\gamma}\right)T_\mathrm{ct}.
     \label{eq:gammaCt}
\end{align}
Therefore, although $d_z(\mathbf{k})$ remains immune to both SOC and Zeeman field (see Fig. \ref{fig:limiting}d),
if $T_\mathrm{ct}\ll T_\mathrm{cs}$ and the quasi-classical regime $\Delta_\mathrm{so}/E_\mathrm{F}\ll 1$ is satisfied, $d_z(\mathbf{k})$ is obliterated by a very small scattering rate $\Gamma\sim T_\mathrm{ct}$. 
Beyond the quasi-classical regime ($\Delta_\mathrm{so}\sim E_\mathrm{F}$), $d_z(\mathbf{k})$ couples to the other order parameters and is expected to be less sensitive to the disorder through the coupling.

In appendix \ref{app:relaxingDx0} we show that the $d_x(\mathbf{k})$ component is even more sensitive and has a pair-breaking strength of $\alpha = \Gamma +i\sqrt{B^2+\Delta_\mathrm{so}^2}$. 
This is very different for the $\psi(\mathbf{k})+i d_y(\mathbf{k})$ state, which is the subject of the next section.

\subsection{Solution for \texorpdfstring{$\psi(\mathbf{k})+i d_y(\mathbf{k})$}{}}

\subsubsection{Solving for the averages}

The remaining three components $\{f_0,f_x,f_y\}$ are coupled. 
We rewrite the Eilenberger equations in Eqs. \eqref{eq:f0_tilde} and \eqref{eq:f_tilde} in matrix form as
\small
\begin{align}
\begin{bmatrix}
\tilde{\omega}_n & -iB & 0\\ 
-iB & \tilde{\omega}_n & \gamma(\mathbf{k})\\ 
0 & -\gamma(\mathbf{k}) & \tilde{\omega}_n
\end{bmatrix}
\begin{bmatrix}
f_0(\mathbf{k};\omega_n)\\ 
f_x(\mathbf{k};\omega_n)\\ 
f_y(\mathbf{k};\omega_n)
\end{bmatrix}
=
\mathrm{sgn}(\omega_n)\begin{bmatrix}
\tilde{\psi}(\omega_n)\\ 
\tilde{d_x}(\omega_n)\\ 
\tilde{d_y}(\mathbf{k};\omega_n)
\end{bmatrix}.
\label{eq:EilMatDis}
\end{align}
\normalsize 

This equation has the same structure as in the clean case in Eq. \eqref{eq:cleanEil}. 
In appendix \ref{app:relaxingDx0} we solved for the $d_x(\mathbf{k})$ component, and since only the odd-frequency part of $f_x(\mathbf{k})$ mediates between $f_0$ and $f_y$, it is safe to set $d_x(\mathbf{k})=0$.
However, because of the disorder,
we have now a finite $\tilde{d_x}(\omega_n) = \Gamma\langle f_x(\mathbf{k};\omega_n)\rangle$. 
Solving Eq. \eqref{eq:EilMatDis} in terms of the averages, we obtain

\begin{widetext}
\begin{align}
f_0(\mathbf{k};\omega_n) & = 
\frac{\tilde{\omega}_n^2+\gamma^2(\mathbf{k})}{|\tilde{\omega}_n|\left(\tilde{\omega}_n^2+B^2+\gamma^2(\mathbf{k}) \right )}\tilde{\psi}(\omega_n)
+\frac{i\mathrm{sgn}(\omega_n)B}{\tilde{\omega}_n^2+B^2+\gamma^2(\mathbf{k}) }\tilde{d_x}(\omega_n)
-\frac{iB\gamma(\mathbf{k})}{|\tilde{\omega}_n|\left(\tilde{\omega}_n^2+B^2+\gamma^2(\mathbf{k}) \right )}\tilde{d_y}(\mathbf{k};\omega_n); 
\label{eq:f0d} \\
f_x(\mathbf{k};\omega_n) & = 
\frac{i\mathrm{sgn}(\omega_n)B}{\tilde{\omega}_n^2+B^2+\gamma^2(\mathbf{k}) }\tilde{\psi}(\omega_n)
+\frac{|\tilde{\omega}_n|}{\tilde{\omega}_n^2+B^2+\gamma^2(\mathbf{k}) }\tilde{d_x}(\omega_n)
-\frac{\mathrm{sgn}(\omega_n)\gamma(\mathbf{k})}{\tilde{\omega}_n^2+B^2+\gamma^2(\mathbf{k})}\tilde{d_y}(\mathbf{k};\omega_n);
\label{eq:fxd} \\
f_y(\mathbf{k};\omega_n) & = 
\frac{iB\gamma(\mathbf{k})}{|\tilde{\omega}_n|\left(\tilde{\omega}_n^2+B^2+\gamma^2(\mathbf{k}) \right )}\tilde{\psi}(\omega_n)
+\frac{\mathrm{sgn}(\omega_n)\gamma(\mathbf{k})}{\tilde{\omega}_n^2+B^2+\gamma^2(\mathbf{k}) }\tilde{d_x}(\omega_n)
+\frac{\left(\tilde{\omega}_n^2+B^2 \right )}{|\tilde{\omega}_n|\left(\tilde{\omega}_n^2+B^2+\gamma^2(\mathbf{k}) \right )}\tilde{d_y}(\mathbf{k};\omega_n).
\label{eq:fyd}
\end{align}
\end{widetext}
By taking the average of Eq. \eqref{eq:fyd}, we obtain $\langle f_y(\mathbf{k};\omega_n)\rangle =0$. This means that if the $f_y$ component were uncoupled from $\{f_0,f_x\}$, it would be affected by the disorder the same way $f_z$ is. 
One can already get a hint which components suffer from the disorder. The $\{f_y,f_z\}$ averages vanish, which shows the tendency of disorder to obliterate them. However, unlike $f_z$, $f_y$ couples to $\psi_0$ (via $f_x$), which provides robustness. By the same token, $f_0$ is expected to loose some of its original robustness due to its coupling to $f_y$. We emphasize that all pairing correlations $\{f_0,\mathbf{f}\}$ are inevitably present, even in the absence of a pairing interaction in the triplet channels. 
Eq. \eqref{eq:fxd} shows us that the triplet correlations $ f_x(\mathbf{k};\omega_n)$ are odd in frequency.

The finite average that enters the self-consistency for $d_y(\mathbf{k})$ is
\begin{align}
&\langle \hat{\gamma}(\mathbf{k})f_y(\mathbf{k};\omega_n)\rangle = i\frac{B}{\Delta_\mathrm{so}}\frac{\psi_0+\Gamma \langle f_0\rangle}{|\tilde{\omega}_n|}\tilde{A}_3 \notag \\
& +\mathrm{sgn}(\omega_n)\frac{\Gamma}{\Delta_\mathrm{so}}\langle f_x\rangle \tilde{A_3}
+i\eta_y\frac{\tilde{\omega}_n^2+B^2}{|\tilde{\omega}_n|\Delta_\mathrm{so}^2}\tilde{A_3}.
\end{align}
Here, all the averages $\tilde{A}=A(\tilde{\omega}_n,B,\Delta_\mathrm{so})$ are taken at the disorder affected frequencies $\tilde{\omega}_n$. The average $\langle \hat{\gamma}(\mathbf{k})f_y(\mathbf{k};\omega_n)\rangle$ is determined once we know $\{\langle f_0\rangle, \langle f_x\rangle \}$.
Using the averages defined in Eqs. \eqref{a1}, \eqref{a2} and \eqref{a3}, and taking the averages of Eqs. \eqref{eq:f0d}, \eqref{eq:fxd} and \eqref{eq:fyd}, we obtain the system of equations for the averages of $\{\langle f_0\rangle, \langle f_x\rangle \}$, which reads

\begin{widetext}

\begin{align}
\begin{bmatrix}
\tilde{\omega}_n^2-\Gamma|\tilde{\omega}_n|\tilde{A}_1 & -i\mathrm{sgn}(\omega_n)\Gamma B \tilde{A}_2\\ 
-i\mathrm{sgn}(\omega_n)\Gamma B \tilde{A}_2 & \tilde{\omega}_n^2-\Gamma|\tilde{\omega}_n|\tilde{A}_2
\end{bmatrix}
\begin{bmatrix}
\langle f_0(\mathbf{k};\omega_n)\rangle\\ 
\langle f_x(\mathbf{k};\omega_n)\rangle
\end{bmatrix}
= 
\begin{bmatrix}
|\tilde{\omega}_n|\tilde{A}_1\psi_0+|\tilde{\omega}_n|\tilde{A}_3\frac{B}{\Delta_\mathrm{so}}\eta_y\\ 
i\mathrm{sgn(\omega_n)}\left(\tilde{A}_2 B\psi_0-\tilde{A}_3\frac{\tilde{\omega}_n^2}{\Delta_\mathrm{so}}\eta_y \right )
\end{bmatrix}.
\end{align}
We define the recurrent occurring quantity $C_l'=\tilde{A}_l\Gamma\left(B^2-|\omega_n||\tilde{\omega}_n| \right )$. 
For conciseness, we now eliminate $\tilde{A}_1$ in favor of $\tilde{A}_2$, such that the solutions are

\begin{align}
& \langle f_0(\mathbf{k};\omega_n)\rangle  = \frac{\tilde{\omega}_n^2-\tilde{A}_2\left( \Gamma|\tilde{\omega}_n|+B^2\right)}{|\omega_n|\tilde{\omega}_n^2+C_2'}\psi_0  +\frac{\tilde{A}_3(B/\Delta_\mathrm{so})\tilde{\omega}_n^2}{|\omega_n|\tilde{\omega}_n^2+C_2'}\eta_y;
\label{eq:f0kw} \\
& \langle f_x(\mathbf{k};\omega_n)\rangle  = \frac{i\mathrm{sgn}(\omega_n)|\tilde{\omega}_n|B \tilde{A}_2}{|\omega_n|\tilde{\omega}_n^2+C_2'}\psi_0 -\frac{i\mathrm{sgn}(\omega_n)\tilde{\omega}_n^2\left(|\omega_n|/\Delta_\mathrm{so} \right )\tilde{A}_3}{|\omega_n|\tilde{\omega}_n^2+C_2'}\eta_y; \\
& \langle \hat{\gamma}(\mathbf{k})f_y(\mathbf{k};\omega_n)\rangle = \frac{i\tilde{A}_3(B/\Delta_\mathrm{so})\tilde{\omega}_n^2}{|\omega_n|\tilde{\omega}_n^2+C_2'}\psi_0+\frac{\tilde{A}_3}{\Delta_\mathrm{so}^2|\tilde{\omega}_n|}\frac{C_2'\left(\tilde{\omega}_n^2+B^2 \right )+\tilde{\omega}_n^2\left[C_3'+|\omega_n|\left(\tilde{\omega}_n^2+B^2 \right ) \right ] }{|\omega_n|\tilde{\omega}_n^2+C_2'}i\eta_y. 
\label{eq:fykw}
\end{align}

\end{widetext}

\subsubsection{Self-consistency}

\begin{figure*} 
\centering
\includegraphics[width=0.7\textwidth]{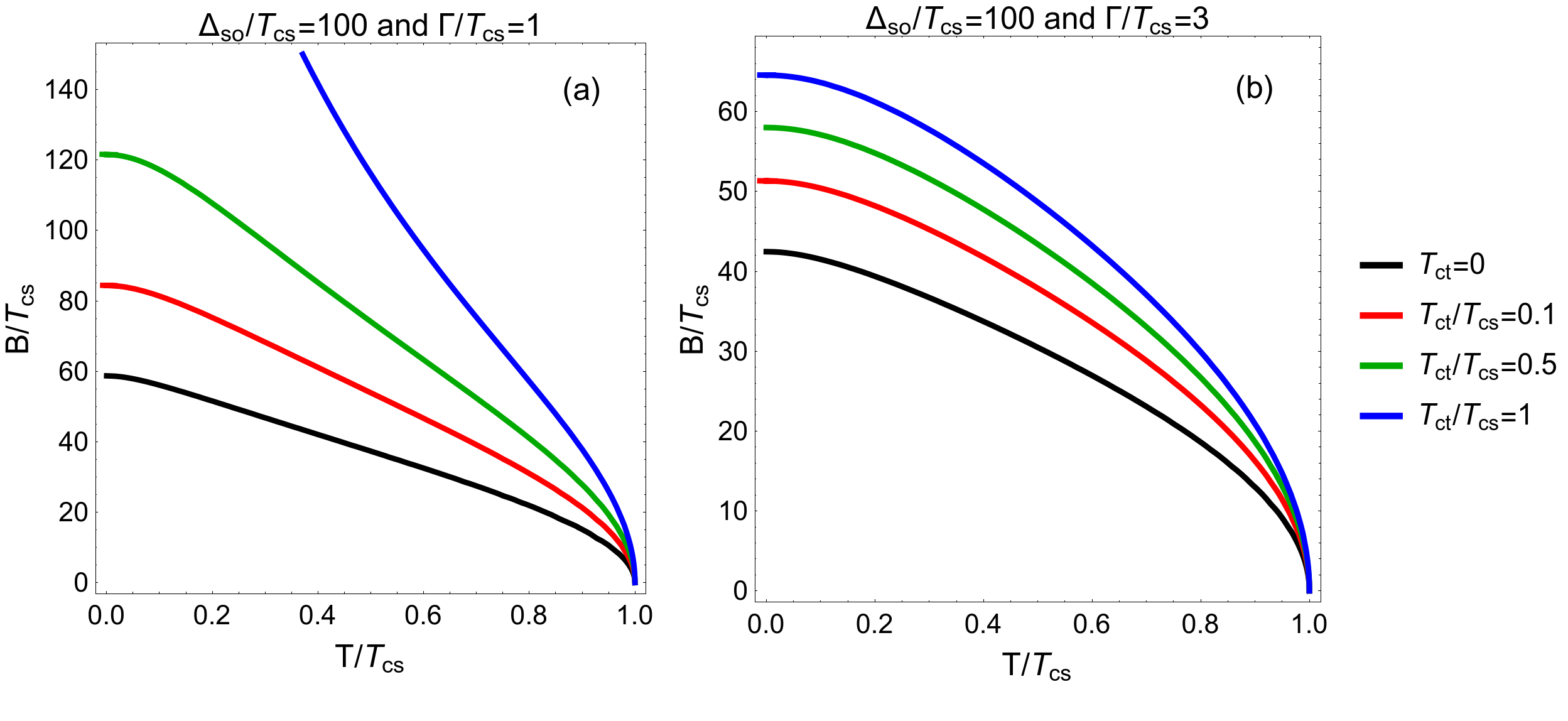}
\caption{\label{fig:3} 
Effect of the triplet channel on the disordered transition lines $B_\mathrm{c}(T)$ with $\hat{\gamma}(\mathbf{k})=\mathrm{sgn}[\gamma(\mathbf{k})]$. The cleaner the system, the greater the enhancement of the critical field caused by the triplets.
In Fig. \ref{fig:4} we show the case
for $\hat{\gamma}(\mathbf{k})=\sqrt{2}\cos(3\varphi_\mathbf{k})$. See the supplemental material for an animated version. 
}
\end{figure*}

We now use Eqs. \eqref{eq:f0kw} and \eqref{eq:fykw} for the self-consistency equations \eqref{eq:selfcSinglet} and \eqref{eq:selcCtriplet}.
To maintain the same form as in Eq. \eqref{eq:coupled_self_consistency}, we define the Matsubara sums $\mathcal{S}=\mathcal{S}(T,B,\Delta_\mathrm{so},\Gamma)$
\begin{align}
\mathcal{S}_\mathrm{s} & = \pi T  \sum_{n=-\infty}^\infty\frac{|\tilde{\omega}_n|}{|\omega_n|}\frac{\tilde{A}_2 B^2}{|\omega_n|\tilde{\omega}_n^2+C_2'} ;
\label{ss} \\
\mathcal{S}_\mathrm{st} & =\pi T  \sum_{n=-\infty}^\infty\frac{\tilde{A}_3(B/\Delta_\mathrm{so})\tilde{\omega}_n^2}{|\omega_n|\tilde{\omega}_n^2+C_2'};
\label{sst}
\end{align}
\begin{align}
& \mathcal{S}_\mathrm{t} = \pi T  \sum_{n=-\infty}^\infty\biggr[\frac{1}{|\omega_n|}\label{st} \\
&-\frac{\tilde{A}_3}{\Delta_\mathrm{so}^2|\tilde{\omega}_n|}\frac{C_2'\left(\tilde{\omega}_n^2+B^2 \right )+\tilde{\omega}_n^2\left[C_3'+|\omega_n|\left(\tilde{\omega}_n^2+B^2 \right ) \right ] }{|\omega_n|\tilde{\omega}_n^2+C_2'} \biggr ]. \notag
\end{align}
These are the most general Matsubara sums in this paper. 
In the suitable limit, they allow us to obtain all pair-breaking equations studied here. 
The values of the averages $\{\tilde{A}_2,\tilde{A}_3\}$ change depending on the choice $\hat{\gamma}(\mathbf{k})=\mathrm{sgn}[\gamma(\mathbf{k})]$ or $\hat{\gamma}(\mathbf{k})=\sqrt{2}\cos(3\varphi_\mathbf{k})$. Both cases contain the relevant interplay of the different energy scales. However, the precise value of $B_\mathrm{c}(T,\Delta_\mathrm{so},\Gamma)$ and the way it is affected by the disorder depends on the specific choice of $\hat{\gamma}(\mathbf{k})$. 
In the following, we choose the simpler $\hat{\gamma}(\mathbf{k})=\mathrm{sgn}[\gamma(\mathbf{k})]$ case for the sake of discussion, and present the plots for the $\hat{\gamma}(\mathbf{k})=\sqrt{2}\cos(3\varphi_\mathbf{k})$ case in Fig. \ref{fig:4}.

\subsection{Main results for the case \texorpdfstring{$\hat{\gamma}(\mathbf{k})=\mathrm{sgn}[\gamma(\mathbf{k})]$}{}}

Using $\hat{\gamma}^2(\mathbf{k})=1$ in the averages \eqref{a2} and \eqref{a3}, we rewrite the Matsubara sums in Eqs. \eqref{ss}, \eqref{sst} and \eqref{st} as
\begin{align}
\mathcal{S}_\mathrm{s} & = \pi T  \sum_{n=-\infty}^\infty\left[\frac{1}{|\omega_n|}-\frac{|\omega_n||\tilde{\omega}_n|+\Delta_\mathrm{so}^2}{|\tilde{\omega}_n|(\omega_n^2+B^2+\Delta_\mathrm{so}^2)-\Gamma\Delta_\mathrm{so}^2} \right ];
\label{eq:matsS} \\
\mathcal{S}_\mathrm{st} & = \pi T  \sum_{n=-\infty}^\infty \frac{B\Delta_\mathrm{so}}{|\tilde{\omega}_n|(\omega_n^2+B^2+\Delta_\mathrm{so}^2)-\Gamma\Delta_\mathrm{so}^2}; \\
\mathcal{S}_\mathrm{t} & = \pi T  \sum_{n=-\infty}^\infty\left[\frac{1}{|\omega_n|}-\frac{\omega_n^2+B^2}{|\tilde{\omega}_n|(\omega_n^2+B^2+\Delta_\mathrm{so}^2)-\Gamma\Delta_\mathrm{so}^2} \right ].
\label{eq:matsT}
\end{align}
Eq. \eqref{eq:matsS} is identical to the main result of Ref. \cite{Ilic2017}.
The three sums converge, but cannot be expressed in terms of the digamma functions like in the clean case. Nonetheless, one can easily implement these sums using \textit{Wolfram Mathematica} that can express these sums as a sum of roots of digamma functions.
Together with the self-consistency conditions \eqref{eq:selfcSinglet} and \eqref{eq:selcCtriplet}, the pair-breaking equation including the effect of disorder is
\begin{align}
\det
\begin{bmatrix}
\ln\frac{T}{T_\mathrm{cs}}+\mathcal{S}_\mathrm{s} & \mathcal{S}_\mathrm{st}\\ 
\mathcal{S}_\mathrm{st} & \ln\frac{T}{T_\mathrm{ct}}+\mathcal{S}_\mathrm{t}
\end{bmatrix}
=0,
\label{eq:pairbreaking_disordered}
\end{align}
This equation generalizes Eq. \eqref{eq:pairbreakingClean} to the disordered case. 
The special case of $\gamma(\mathbf{k})=0$ is a good sanity test for which
$f_0$ decouples from $f_y$ and the resulting pair-breaking equation for $\psi_0$ reduces to Eq. \eqref{eq:fulde}, which is independent of the scattering rate $\Gamma$. 

Note that in the clean situation $\Gamma=0$, Eqs. \eqref{eq:matsS} and \eqref{eq:matsT} tell us that $B$ limits singlet superconductivity in the same functional way $\Delta_\mathrm{so}$ limits the equal-spin triplet component. The presence of a finite $\Gamma$ changes this, since the triplets suffer more from the disorder than the singlets. Nonetheless, the triplets gain robustness against the disorder through the coupling with the singlets, which are favoured by the SOC. 

In the opposite limit with $\Gamma\rightarrow\infty$, the conversion term vanishes $\mathcal{S}_\mathrm{st}=0$, so that the singlets decouple from the triplets. Then, the disorder obliterates the triplets and the pair-breaking equation for the singlets reduces to Eq. \eqref{eq:fulde}. This shows that the role of spin-conserving impurity scattering $\Gamma$ is to undo the enhancement caused by SOC.

In Fig. \ref{fig:2} we show the case considering singlets only with the pair-breaking equation given by $\ln (T/T_\mathrm{cs})+\mathcal{S}_\mathrm{s}=0$. The plots are given for $\Delta_\mathrm{so}/T_\mathrm{cs}=10$ and $\Delta_\mathrm{so}/T_\mathrm{cs}=100$. The same scattering rate color legend applies to both plots. These plots illustrate that the disorder works against the enhancement caused by SOC. For $\Gamma\gg \Delta_\mathrm{so}$, the transition line saturates at the Pauli-limit, at which the enhancement effect due to SOC has been undone by the disorder. The latter strongly affects the transition line when the scattering rate becomes comparable to SOC.

The effect of an increasing attraction in the triplet channel for $\Delta_\mathrm{so}/T_\mathrm{cs}=100$ is shown in Fig. \ref{fig:3}. The left panel shows the case when the scattering rate compares to the singlet critical temperature. Usually, the signature of any triplets would be obliterated in this regime, see Eq. \eqref{eq:gammaCt}. However, the magnetic field induced triplet channel still yields a stark enhancement of the critical field. Still, the more disordered the system, the less relevant the triplet channel becomes. This is illustrated in the right panel with $\Gamma/T_\mathrm{cs}=3$. Also, see the supplemental material to see an animated version showing a wider range of scattering rates \footnote{See Supplemental Material at [URL will be inserted by publisher] for an animated version \label{foot}}.

\subsection{Expansion close to \texorpdfstring{$T_\mathrm{cs}$}{}}

We can estimate the behaviour of $B_\mathrm{c}(T)$ close to $T_\mathrm{cs}$ by considering $T_\mathrm{cs}\ll\Delta_\mathrm{so}$ and $T_\mathrm{cs}\Gamma\ll\Delta_\mathrm{so}^2$. The expansion can be written as
\begin{align}
\frac{B^2_\mathrm{c}(T)}{\Delta^2_\mathrm{so}}=
\mathcal{C}_{1(2)}
\left(1-\frac{T}{T_\mathrm{cs}} \right ),
\label{eq:expansionTc}
\end{align}
where $\mathcal{C}_1$ is the coefficient for the case with $\hat{\gamma}(\mathbf{k})=\mathrm{sgn}[\gamma(\mathbf{k})]$, and $\mathcal{C}_2$ corresponds to $\hat{\gamma}(\mathbf{k})=\sqrt{2}\cos(3\varphi_\mathbf{k})$. Up to logarithmic accuracy, the coefficients are  
\begin{align}
\mathcal{C}_1 &=
\left[\ln\frac{T_\mathrm{cs}}{T_\mathrm{ct}}\,
\frac{\ln\frac{\Delta_\mathrm{so}}{ T_\mathrm{cs}}}{\ln\frac{\Delta_\mathrm{so}}{T_\mathrm{ct}}}
+\frac{\pi\Gamma}{4T_\mathrm{cs}} \right ]^{-1};\label{c1} \\
\mathcal{C}_2 & =
\left(\frac{\pi\Delta^2_\mathrm{so}}{4T_\mathrm{cs}\left(\sqrt{\Gamma^2+2\Delta^2_\mathrm{so}}-\Gamma \right )}
-\frac{2\ln\frac{\Delta_\mathrm{so}}{ T_\mathrm{cs}} }{\ln\frac{\Delta_\mathrm{so}}{ T_\mathrm{ct}} }
\right )^{-1}.
\label{c2}
\end{align}
In both $\mathcal{C}_1$ and $\mathcal{C}_2$, the triplet critical temperature $T_\mathrm{ct}$ only occurs in the argument of logarithms, whereas the scattering rate $\Gamma$ does not. This implies that the larger $\Gamma$, the more insensitive $B_\mathrm{c}(T)$ becomes to $T_\mathrm{ct}$. This is illustrated in Fig. \ref{fig:3}. 
In the purely singlet case we can take the limit $T_\mathrm{ct}\rightarrow 0$ to obtain
\begin{align}
\mathcal{C}_1 & = \left(\ln\frac{\Delta_\mathrm{so}}{T_\mathrm{cs}}+\frac{\pi\Gamma}{4T_\mathrm{cs}} \right )^{-1};\label{eq:nodeless} \\
\mathcal{C}_2 & = \frac{4}{\pi} \frac{T_\mathrm{cs}}{\Delta_\mathrm{so}^2}\left(\sqrt{\Gamma^2+2\Delta_\mathrm{so}^2}-\Gamma \right ). \label{eq:bulaevski}
\end{align}

For the nodeless SOC, $\hat{\gamma}(\mathbf{k})=\mathrm{sgn}[\gamma(\mathbf{k})]$, Eq.~\eqref{c1} shows that the characteristic scattering rate affecting the critical field is $\Gamma \sim T_\mathrm{cs}\ln(\Delta_\mathrm{so}/T_\mathrm{cs})$. 
This model of SOC is appropriate to the multi-pocket Fermi surfaces not crossing the high-symmetry lines where SOC vanishes, such as considered in Ref. \cite{Ilic2017}. 
In the systems with Fermi surfaces crossing the high-symmetry lines, the SOC has nodes and can be modelled in general as a series of odd Fourier harmonics consistent with a particular lattice symmetry.
The essential point is that the typical scattering rates affecting the critical field in this case $\Gamma \sim \Delta_\mathrm{so}$ is much larger than the corresponding scale for the nodeless case. 
This scaling is evident from Eq.~\eqref{eq:bulaevski}, which was first obtained in Ref.~\cite{Bulaevskii976} for the model $\hat{\gamma}(\mathbf{k})=\sqrt{2}\cos(3\varphi_\mathbf{k})$.
Indeed, at the nodes of SOC, the disorder has no effect on the Cooper pairs. For a nodal SOC, the critical field is lower, but more robust to the disorder as compared to the nodeless SOC.

\begin{table*}
\caption{\label{tab:table}%
Summary of the effect of Ising SOC, Zeeman field and the disorder on the superconducting order parameters. The first line shows the parity-mixed sub-system $\{\psi_0,\eta_y\}$ displaying robustness against the disorder. See the table's footnotes for a detailed description. The last two lines refer to the two independent triplet sub-systems $\{\{\eta_x\},\{\eta_y\}\}$, which are obliterated by the disorder. The last column shows the case for $\Gamma\rightarrow \infty$ for which $\psi_0$ decouples from $\eta_y$. }
\begin{ruledtabular}
\begin{tabular}{ccccc}
Order parameter & Ising SOC &In-plane $B$ &Disorder $\Gamma$ & $\Gamma\rightarrow \infty$ ($B,\Delta_\mathrm{so}\neq 0$) \\
\hline
$\{\psi_0,\eta_y\}$ & \{Immune, limited\} & \{Limited, induced\} & Suppressed\footnote{The disorder energy scale to substantially suppress the critical field is $\Gamma\sim \Delta_\mathrm{so}$ for $\hat{\gamma}(\mathbf{k})=\sqrt{2}\cos(3\varphi_\mathbf{k})$, and $\Gamma \sim T_\mathrm{cs}\ln(\Delta_\mathrm{so}/T_\mathrm{cs})$ for $\hat{\gamma}(\mathbf{k})=\mathrm{sgn}[\gamma(\mathbf{k})]$. In both cases, $\Gamma \gg \Delta_\mathrm{so}$ is needed to suppress the critical field down to the Pauli limit $B_\mathrm{P}$, below which $\psi_0$ is immune to the disorder. The $\eta_y$ triples are coupled to the $\psi_0$ singlets, such that they vanish at the same critical field as $\psi_0$. An infinite scattering rate is necessary to decouple $\eta_y$ from $\psi_0$, which then obliterates $\eta_y$.} & \{Immune, obliterated\} \\ 
$\eta_x$ & Limited & Limited & Obliterated\footnote{The disorder energy scale to obliterate $\Gamma\sim \left[(\pi/(2e^\gamma))^2T_\mathrm{ct}^2-(B^2+\Delta_\mathrm{so}^2) \right ]^\frac{1}{2}$. This is the most sensitive of all components, because it suffers from all spin-fields and the disorder. } & Obliterated \\ 
$\eta_z$
  & Immune
  & Immune & Obliterated\footnote{The disorder energy scale to obliterate $\eta_z$ is $\Gamma\sim T_\mathrm{ct}< T_\mathrm{cs}\ll \Delta_\mathrm{so}$.} & Obliterated  \\
\end{tabular}
\end{ruledtabular}
\end{table*}

\section{Discussion \label{sec:discussion}}

We now relate our results to the wider context of the field. We discuss the applicability of our results to monolayer transition metal dichalcogenides (TMDs) and epitaxial heterostructures,
comment on the nature of the phase transition at low temperatures, and point out the ubiquitous presence of odd-frequency pairing correlations. 

\begin{figure} 
\centering
\includegraphics[width=0.4\textwidth]{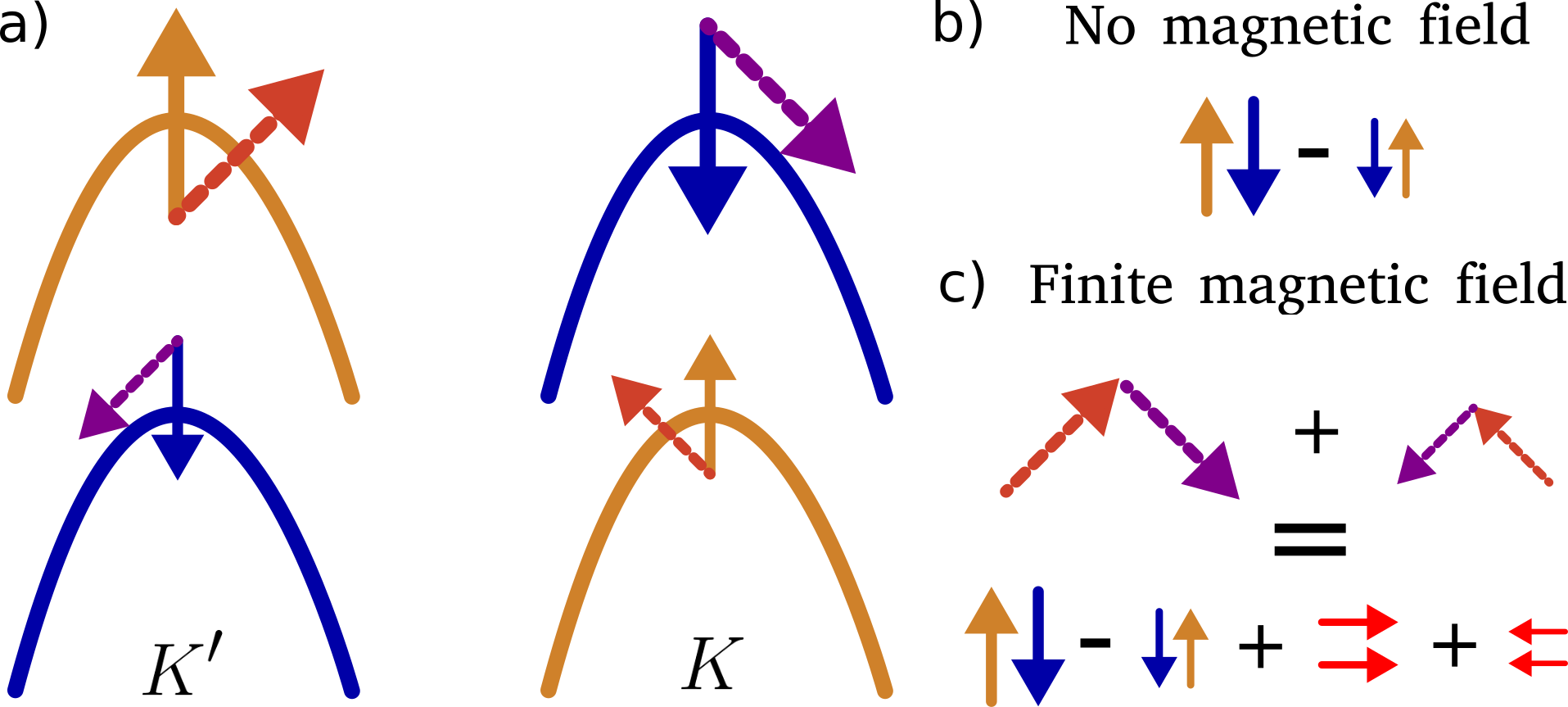}
\caption{\label{fig:tilt} 
Schematic illustration of the reorientation (conversion) of the Cooper pairs by the in-plane Zeeman field. The size of the arrows reflect the amplitude of the order parameters taking into account the difference in densities of states of the bands. The Cooper spin partners $\nearrow$ and $\searrow$ are related by the symmetry $\sigma_h\mathcal{T}$, which ensures their degeneracy at opposite momenta in 2D. Schematically, $\sigma_h\mathcal{T}\nearrow=\sigma_h\swarrow=\searrow$. a) Cooper partners are paired at the same energy located at $K$ and $K^\prime=\mathcal{T}K$. b) At $B=0$, Cooper pairs respect time-reversal symmetry. c) A finite $B$ induces equal-spin triplets.
}
\end{figure}

\subsection{The large SOC limit}

The theory presented so far applies in the regime specified by \eqref{eq:quasiRegime}.
It holds for any ratio $\Delta_{\mathrm{so}}/T_{\mathrm{cs(t)}}$.
We now discuss the applicability of our results to the case $\Delta_{\mathrm{so}} \gtrsim E_\mathrm{F}$.
In this case, the band basis formulation is more appropriate \cite{Samokhin2004,Samokhin2008,Samokhin2009,Bauer2012,Samokhin2015}.
The necessity for a band basis formulation arises when the densities of states of the spin-split bands differ significantly \cite{Frigeri2006}. 
In the extreme situation of strong SOC, one of the spin split electron (hole) bands may be pushed above (below) $E_\mathrm{F}$ resulting in valley polarized bands. 
The difference between the regime of weak and strong SOC is the amount of admixture of the  opposite-spin, $\eta_z$ triplets to singlets in the ground state  at $\mathbf{B}=0$.
While at $\Delta_{\mathrm{so}} \ll E_\mathrm{F}$ the ground state Cooper pairs are singlets, in the limiting case of valley polarized bands  the ground state correlations are approximately half singlets and half $\eta_z$, i.e. opposite-spin triplets. 

We argue that despite this difference the condition $\Delta_{\mathrm{so}} \ll E_\mathrm{F}$ may be relaxed without affecting our results qualitatively.
The reason for this is the independence of the singlet to triplet conversion on any of the details of the band structure. 
This point is illustrated in Fig. \ref{fig:tilt} \footnote{The illustration is a corrected version of a similar figure in Ref. \cite{Mockli2018}}. 
At finite in-plane $\mathbf{B}$, the 
electron spins tilt yet the states with opposite momenta that are paired remain degenerate thanks to the combined mirror and time-reversal symmetry $\mathcal{T} \sigma_h$ \cite{Fischer2018}.
As a result, the paired states at finite field are combinations of singlets and $\eta_y$ triplets.
In contrast, the $\eta_z$ triplets do not participate in the conversion process.
Therefore, the $\psi_0$ singlets  to $\eta_y$ triplets conversion and the gained robustness of the $B_c$ associated with it do not depend on the amount of $\eta_z$ triplets admixture in the ground state.
Alternatively, the reorientation of the spins in the upper hole band shown in Fig. \ref{fig:tilt} occurs regardless of the occupation of the lower bands.
This in turn means that our results apply qualitatively to the regime of valley polarized bands.
And even more so to the intermediate regime, $\Delta_{\mathrm{so}} \lesssim E_\mathrm{F}$.

\subsection{
The role of the orbital content in monolayer TMDs
}

Perhaps the most well known Ising superconductors are the monolayer TMDs such as NbSe$_2$, gated MoS$_2$ and all their cousins \cite{Xi2015,Lu2015,Bawden2016,Dvir2017}. 
In these materials, the critical in-plane magnetic field exceeds the Pauli-limit in Eq. \eqref{eq:PauliLimit} by several times, which is associated with the enhancement caused by Ising SOC. 

An additional ingredient in these systems is that
the effect of the disorder on thermodynamic properties depends on the orbital content of the Bloch bands.
Specifically, the orthogonality of the orbital wave functions of the bands reduces the amount of the inter-band scattering.
This is argued to be the cause of insensitivity of the zero-field transition temperature $T_\mathrm{c}$ in MgB$_2$ to the disorder \cite{Mazin2002}. In the two band superconductor such as MgB$_2$, the inter-band scattering is expected to suppress the critical temperature \cite{Golubov1997}. Yet, only a slight decrease of $T_\mathrm{c}$ for dirtier systems has been reported. A similar phenomenon has been recently reported in for NbSe$_2$ \cite{Cho2018}.

Although we disregard the orbital structure of the electronic bands in this work, we now show how our results give insight to the TMDs.
For illustration purposes, we focus on the situation in monolayer NbSe$_2$.
The conduction bands are derived from the $d_{xy}$, $d_{x^2-y^2}$ and $d_{z^2}$ Nb orbitals.
The hole pocket at $\Gamma$ has the $d_{z^2}$ orbital character while the hole pockets at $K$ and  $K'$ are approximately made of orbital states $d_{x^2-y^2} \pm i d_{xy}$ respectively with some  admixture of $d_{z^2}$ states.
The amount of orbital admixture at the Fermi level scales with the ratio of the band splitting at $K$($K'$) to $E_\mathrm{F}$ \cite{Xiao2012}.
The situation with the large orbital admixture and $\Delta_{\mathrm{so}} \ll E_F$ pertinent to MoS$_2$ has been analyzed in \cite{Ilic2017}. 
In this case, the admixture of $d_{z^2}$ orbitals to the conduction band makes the scattering between the $K$ and $K'$ an allowed process.
Naturally, such an inter-valley scattering results in the pair-breaking equation that was obtained previously for the systems with trivial orbital content \cite{Bulaevskii976}.

For the case of valence bands in MoS$_2$, or correspondingly the conduction bands in NbSe$_2$, the spin splitting is substantial $\Delta_{\mathrm{so}} \lesssim E_F$.
The inter-valley scattering still leads to the suppression of $B_\mathrm{c}$ \cite{Sosenko2017}.
Realistically, however, the short-range disorder needed for the large momentum inter-valley scattering is provided by scatterers normally found at high symmetry lattice positions.
As the admixture of the symmetric $d_{z^2}$ orbital at valence bands is negligible \cite{Xiao2012}, the $C_3$ symmetric scattering potential blocks the inter-valley scattering \cite{Mockli2018}.
In result, in multi-orbital systems, the actual $B_\mathrm{c}$ is higher than in the systems with the trivial orbital content.
Moreover, the only effect on $B_\mathrm{c}$ comes from the intra-valley scattering within the $\Gamma$ band.
In this case the amount of disorder needed to suppress the $B_\mathrm{c}$ down to the Pauli limit is quite large, $\Gamma \approx \Delta_{\mathrm{so}} \sim E_\mathrm{F}$, see \eqref{eq:bulaevski} in agreement with numerical results of Ref. \cite{Mockli2018}.

\subsection{Epitaxial heterostructures}
Beside the TMDs, another class of Ising superconductors are the epitaxial heterostructures, such as the interface between a Pb film and a Si substrate \cite{Eom2006,Qin2009,Zhang2010,Yamada2013,Brun2014}.
In these systems, the interface generates an Ising SOC component that satisfies $\Delta_\mathrm{so}\ll E_\mathrm{F}$ \cite{Liu2018}. 
Depending on the interface, a Rashba component can also occur that will tend to counteract the critical field enhancement caused by the Ising component. The Eilenberger equations \eqref{eq:f0_tilde} and \eqref{eq:f_tilde} provide the suitable starting point to study a general structure of SOC $\boldsymbol{\gamma}(\mathbf{k})$.

\subsection{Nature of the phase transition at low temperatures}

In the clean limit, it is known that in Pauli-limited superconductors the continuous phase transition changes to a first-order phase transition for temperatures $T^\dag \lesssim 0.56 T_\mathrm{c}$ \cite{Maki1964,Matsuda2007}. Below $T^\dag$ at high Zeeman fields, the superconducting phase enters the Fulde-Ferrell-Larkin-Ovchinnikov (FFLO) state, which is characterized by Cooper pairs with finite total momentum. 
At finite SOC, the spin-susceptibility of the superconducting state remains close to that of the normal state, such that the normal to superconducting state transition remains continuous for all temperatures and no FFLO phase stabilizes \cite{Frigeri2004,Samokhin2005,Samokhin2007,Sohn2018}. Moreover, the residual spin-susceptibility is even more enhanced by scalar impurities \cite{Samokhin2007}, which also suppresses the FFLO state \cite{Song2019}. 
One might ask if the first-order phase transition with the FFLO state reappears at sufficiently low SOC and impurity scattering rates. Such an analysis was carried out in Ref. \cite{Ilic2017} which found a small window of reappearance for the SOC energy scale smaller than the superconducting energy scales. 
This shows that it is generally reasonable to assume continuous superconducting phase transitions in non-centrosymmetric Ising superconductors.

\subsection{Ubiquitous odd-frequency pairing correlations}

It is important to differentiate between pairing correlations $\{f_0(\mathbf{k};\omega_n),\mathbf{f}(\mathbf{k};\omega_n)\}$ and order parameters $\{\psi(\mathbf{k}),\mathbf{d}(\mathbf{k})\}$. Order parameters are part of the Hamiltonian, while pairing correlations are not. Nonetheless, triplet paring correlations $\mathbf{f}(\mathbf{k};\omega_n)$ are in general finite in the presence of spin-fields and affect response functions \cite{Yip2014a,Zhou2016}. Even in BCS theory, a Zeeman field populates $\mathbf{f}(\mathbf{k};\omega_n)$, see Eq. \eqref{zeeman}. In non-centrosymmetric superconductors, $\mathbf{f}(\mathbf{k};\omega_n)$ is inevitably populated by SOC. 

Moreover, odd-frequency pairing correlations that are historically viewed as exotic pairing states are in fact ubiquitous \cite{Linder2017}. Any spin-field, either SOC (Eqs. \eqref{eq:soc1} and \eqref{eq:soc2}), or magnetic field (Eq. \eqref{zeeman}), generates them \cite{Gentile2011}. While odd-frequency pairing interactions possibly exist \cite{Linder2017}, they are excluded from the pairing interaction \eqref{eq:interaction} considered here. This allowed us to set $d_x(\mathbf{k})=0$, because this would have to be an odd-frequency order parameter. 

While the conditions to realize odd-frequency pairing correlations are usually related to multi-band systems, layered heterostructures, double quantum dots, double nanowires, Josephson junctions, etc. \cite{Black-Schaffer2013,Linder2017,Moghaddam,cayao2017,cayao2018}, here we showed that the spin-fields also generate odd-frequency pairing correlations.

\section{Conclusion \label{sec:conclusion}}

In this paper, we first described the magnetic field induced singlet Cooper pair to triplet conversion phenomena in the clean case and then studied the effect of scalar impurities on the superconducting transition at a finite magnetic field. 

For the clean situation, we showed that any spin-field generates odd-frequency pairing correlations. These correlations mediate the coupling between pairing-correlation components and strongly affect the response of the superconducting state to external perturbations. This does not depend on whether the odd-frequency correlations have corresponding superconducting order parameters. Furthermore, the magnetic field couples the singlets from the lattice symmetric irrep to triplets transforming as a two-dimensional or, more generally, vector irreps. It is this field-induced coupling that leads to the singlet-triplet conversion.

As for the effects of the scalar impurities when $\Delta_\mathrm{so}\ll E_\mathrm{F}$, for the zero-field (intra-irrep) parity-mixing, the triplets ($\eta_z$) are obliterated once the disorder scattering rate reaches the energy scale of the triplet order parameters.
In contrast, the field-induced triplets ($\eta_y$) cannot be obliterated due to their coupling with singlets at finite magnetic fields. The superconducting state is substantially suppressed by the disorder once the disorder scattering rate becomes comparable to SOC. The field-induced triplets have a strong effect on the critical magnetic field for disorder scattering rate comparable to the singlet (and not the triplet) critical temperature. Therefore, the properties and potentially topology of converted triplets are expected to be experimentally accessible even in moderately disordered systems. 

In the large SOC limit, the $\eta_z$ triplets are also expected to couple to the $\psi_0$ singlets and in this way gain some robustness against impurities. This coupling originates from the possible difference of the densities of states of the spin-split bands, and is not directly related to the Zeeman field. 

We finally argue that the field-induced singlet-triplet conversion occurs whenever the intrinsic anti-symmetric SOC has a component perpendicular to the applied field. This phenomenon is therefore generic to a large class of non-centrosymmetric materials including but not limited to Ising superconductors.

\begin{acknowledgments}
We thank G. Blumberg, T. Dvir, and H. Steinberg for enlightening discussions. We are indebted to M. Haim for a critical reading of the manuscript. D.M. acknowledges the support from the Swiss National Science Foundation, Project No. 184050, and authors
acknowledge the support from the Israel Science Foundation, Grant No. 1287/15.
\end{acknowledgments}

\newpage 

\onecolumngrid

\appendix

\section{Deriving the Gor'kov equations \label{app:eqsMoton}}

In this section, we derive the real-space Gor'kov equations. For a didactic introduction to the procedure followed here in similar notations, see Ref. \cite{Kita2015}.

\subsection{Matsubara Green's functions}

We use a compact notation for the Matsubara Green's functions in particle-particle ($i=j$) and particle-hole ($i\neq j$) space
\begin{align}
G_{ij}\left(\mathbf{r}_\sigma,\tau;\mathbf{r}'_{\sigma'},\tau' \right ) =  -\left\langle \mathcal{T}\psi^i_\sigma(\mathbf{r},\tau)\psi_{\sigma'}^{3-j}(\mathbf{r}',\tau')\right\rangle,
\end{align}
where $\psi^1\equiv \psi$ are annihilation operators, and $\psi^2\equiv \psi^\dag$ are creation operators. 
The Green's functions only depend on the time difference $\tau-\tau'\rightarrow \tau$.
One can verify that the Green's function has the general properties
\begin{align}
G_{ij}(\mathbf{r}_\sigma,\mathbf{r}'_{\sigma'};\tau)=G_{ji}^*(\mathbf{r}'_{\sigma'},\mathbf{r}_{\sigma};\tau)=-G_{3-j,3-i}(\mathbf{r}'_{\sigma'},\mathbf{r}_{\sigma};-\tau).
\label{eq:propsG}
\end{align}
We can write a Fourier transform to Matsubara frequencies $\omega_n=(2n+1)\pi/\beta$ as
\begin{align}
& G_{ij}\left(\mathbf{r}_\sigma,\mathbf{r}'_{\sigma'};\tau \right ) = \frac{1}{\beta}\sum_{n=-\infty}^\infty G_{ij}\left(\mathbf{r}_\sigma,\mathbf{r}'_{\sigma'};\omega_n \right )e^{-i\omega_n\tau};\quad  G_{ij}\left(\mathbf{r}_\sigma,\mathbf{r}'_{\sigma'};\omega_n \right ) =\int_0^\beta \mathrm{d}\tau\,G_{ij}\left(\mathbf{r}_\sigma,\mathbf{r}'_{\sigma'};\tau \right )e^{i\omega_n\tau}.
\label{eq:ftMatsubara}
\end{align}
Note that Note that $G_{ij}\left(\mathbf{r}_\sigma,\mathbf{r}'_{\sigma'};0^+ \right )=-\langle \psi_\sigma^i(\mathbf{r})\psi_{\sigma'}^{3-j}(\mathbf{r}')\rangle $ and $G_{ij}\left(\mathbf{r}_\sigma,\mathbf{r}'_{\sigma'};0^- \right )=\langle \psi_{\sigma'}^{3-j}(\mathbf{r}')\psi_\sigma^i(\mathbf{r})\rangle $. 
In frequency-space, the symmetries in Eq. \eqref{eq:propsG} translate to
\begin{align}
G_{ij}(\mathbf{r}_\sigma,\mathbf{r}'_{\sigma'};\omega_n)  =G_{ji}^*(\mathbf{r}'_{\sigma'},\mathbf{r}_{\sigma};-\omega_n)  =-G_{3-j,3-i}(\mathbf{r}'_{\sigma'},\mathbf{r}_{\sigma};-\omega_n).
\label{eq:symmetryOfG}
\end{align}
These general properties are extensively used throughout the paper.

\subsection{Equations of motion of the field operators}

With the normal state Hamiltonian given by Eq. \eqref{eq:single_particle} and the superconducting interaction in Eq. \eqref{eq:Vssss}, the full clean Hamiltonian can be written as
\begin{align}
H & = \int\mathrm{d}\mathbf{r}\,\psi_\sigma^\dag (\mathbf{r})\hat{K}\psi_\sigma (\mathbf{r})+ \sum_{\sigma,\sigma'} \int\mathrm{d}\mathbf{r}\int\mathrm{d}\mathbf{r}'\,\psi^\dag_\sigma(\mathbf{r})
\mathbf{g}(\mathbf{r}-\mathbf{r}')\cdot\boldsymbol{\sigma}_{\sigma\sigma '}\,
\psi_{\sigma'}(\mathbf{r}') \notag \\
&+\frac{1}{2}\sum_{\sigma_i,\sigma_i^\prime}\int\mathrm{d}\mathbf{r}\int\mathrm{d}\mathbf{r}'\,
V_{\sigma_1'\sigma_2'}^{\sigma_1\sigma_2}(|\mathbf{r}-\mathbf{r}'|) \psi_{\sigma_1}^\dag(\mathbf{r})\psi_{\sigma_2}^\dag(\mathbf{r}')\psi_{\sigma_2'}(\mathbf{r}')\psi_{\sigma_1'}(\mathbf{r}),
\end{align}
where $\mathbf{g}(\mathbf{r}-\mathbf{r}')=\boldsymbol{\gamma}(\mathbf{r}-\mathbf{r}')-\mathbf{B}\delta(\mathbf{r}-\mathbf{r}')$ is the effective spin-field. 
The Heisenberg equation of motion for the field operators is
\begin{align}
\frac{\partial \psi_\sigma^{(\dag)}(\mathbf{r},\tau)}{\partial \tau}=\frac{\partial}{\partial \tau}\left(e^{\tau H} \psi_\sigma^{(\dag)}(\mathbf{r})e^{-\tau H} \right )=e^{\tau H}\left[H,\psi_\sigma^{(\dag)}(\mathbf{r}) \right ]e^{-\tau H},
\end{align}
for which the commutators with $H$ must be evaluated. We obtain
\small
\begin{align}
& \frac{\partial \psi_\sigma(\mathbf{r},\tau)}{\partial \tau}=
-\hat{K}\psi_\sigma(\mathbf{r},\tau)
-\int\mathrm{d}\mathbf{r}'\,\sum_{\sigma'}\mathbf{g}(\mathbf{r}-\mathbf{r}')\cdot\boldsymbol{\sigma}_{\sigma\sigma'}\psi_{\sigma'}(\mathbf{r}',\tau)
-\int\mathrm{d}\mathbf{r}'\sum_{\sigma_1,\sigma_1',\sigma_2'}V^{\sigma_1\sigma}_{\sigma_1'\sigma_2'}\left(|\mathbf{r}-\mathbf{r}'| \right )\psi^\dag_{\sigma_1}(\mathbf{r}',\tau)\psi_{\sigma_1'}(\mathbf{r}',\tau)\psi_{\sigma_2'}(\mathbf{r},\tau); \label{eq:f1} \\
& \frac{\partial \psi_\sigma^\dag(\mathbf{r},\tau)}{\partial \tau} =\hat{K}^* \psi^\dag_\sigma(\mathbf{r},\tau) + \int\mathrm{d}\mathbf{r}'\sum_{\sigma'}\mathbf{g}(\mathbf{r}'-\mathbf{r})\cdot\boldsymbol{\sigma}_{\sigma'\sigma}\,\psi^\dag_{\sigma'}(\mathbf{r}',\tau)
+\sum_{\sigma_1\sigma_2,\sigma_1'}\int\mathrm{d}\mathbf{r}'\,V^{\sigma_1\sigma_2}_{\sigma_1'\sigma}\left(|\mathbf{r}-\mathbf{r}'| \right ) \psi^\dag_{\sigma_2}(\mathbf{r},\tau)\psi^\dag_{\sigma_1}(\mathbf{r}',\tau)\psi_{\sigma_1'}(\mathbf{r}',\tau),
\label{eq:f2}
\end{align}
\normalsize
where $\hat{K}$ is defined below Eq. \eqref{eq:single_particle}. 
To arrive at Eqs. \eqref{eq:f1} and \eqref{eq:f2}, we used the symmetries of the pairing interaction in Eq. \eqref{eq:intProps},  $[\hat{K},H]=0$ and inserted $1\rightarrow e^{-\tau H}e^{\tau H}$ where suitable.

\subsection{Equations of motion of the Green's functions: the Gor'kov equations}

We now compute the equations of motion for the Matsubara Green's function:
\begin{align}
 \frac{G_{ij}
\left(\mathbf{r}_\sigma,\tau;\mathbf{r}'_{\sigma'},\tau' \right )}{\partial \tau} = 
-\delta(\tau-\tau')
\biggr(\left\langle \psi^i_\sigma(\mathbf{r},\tau)\psi_{\sigma'}^{3-j}(\mathbf{r}',\tau')\right\rangle +\left\langle \psi_{\sigma'}^{3-j}(\mathbf{r}',\tau')\psi^i_\sigma(\mathbf{r},\tau)\right\rangle \biggr )
-\left\langle 
\mathcal{T}\frac{\partial \psi_\sigma^i(\mathbf{r},\tau)}{\partial \tau}\psi_{\sigma'}^{3-j}(\mathbf{r}',\tau')\right\rangle.
\label{eq:eqmot}
\end{align}
Using the equations of motion for the field operators in Eqs. \eqref{eq:f1} and \eqref{eq:f2}, we see that the last term in Eq. \eqref{eq:eqmot} is quartic in the field operators.
We use the Wick decomposition $\langle ABCD\rangle =\langle AD\rangle\langle BC\rangle-\langle AC\rangle\langle BD\rangle+\langle AB\rangle\langle CD\rangle$ for these terms and only retain the pairing correlations. We define the real-space mean-field order parameter as
\begin{align}
&\Delta_{\sigma_1\sigma_2}(\mathbf{r}',\mathbf{r})  =-\sum_{\sigma_1',\sigma_2'}V^{\sigma_1\sigma_2}_{\sigma_1'\sigma_2'}(|\mathbf{r}-\mathbf{r}'|)\left\langle \psi_{\sigma_1'}(\mathbf{r}')\psi_{\sigma_2'}(\mathbf{r}) \right\rangle  = \sum_{\sigma_1',\sigma_2'}V^{\sigma_1\sigma_2}_{\sigma_1'\sigma_2'}(|\mathbf{r}-\mathbf{r}'|)
\frac{1}{\beta}\sum_{n=-\infty}^\infty F_{\sigma_1'\sigma_2'}(\mathbf{r}',\mathbf{r};\omega_n). \label{eq:real_gap}
\end{align}
The Fourier transform of  Eq.\eqref{eq:real_gap} leads to Eq. \eqref{eq:selfconsistent}. The mean-field decoupled equations of motion then read
\small

\begin{align}
& \left(-\frac{\partial}{\partial\tau}-\hat{K} \right )G_{1j}\left(\mathbf{r}_\sigma,\tau;\mathbf{r}'_{\sigma'},\tau' \right ) 
-\sum_{s}\int\mathrm{d}\mathbf{r}''\,\mathbf{g}(\mathbf{r}-\mathbf{r}'')\cdot\boldsymbol{\sigma}_{\sigma s}
\,G_{1j}\left(\mathbf{r}_s'',\tau;\mathbf{r}'_{\sigma'},\tau' \right )  +\sum_{\sigma_1}\int\mathrm{d}\mathbf{r}''\Delta_{\sigma_1\sigma}(\mathbf{r}'',\mathbf{r})
G_{2j}\left(\mathbf{r}''_{\sigma_1},\tau ;\mathbf{r}'_{\sigma'},\tau' \right ) \notag \\
& = \delta(\tau-\tau')\delta(\mathbf{r}-\mathbf{r}')\delta_{\sigma\sigma'}\delta_{1j};
\label{eq:moT1}
\end{align}
\begin{align}
& \left(-\frac{\partial}{\partial\tau}+\hat{K}^* \right )G_{2j}\left(\mathbf{r}_\sigma,\tau;\mathbf{r}'_{\sigma'},\tau' \right )  
+\sum_{s}\int\mathrm{d}\mathbf{r}''\,\mathbf{g}(\mathbf{r}''-\mathbf{r})\cdot\boldsymbol{\sigma}_{s\sigma}
\,G_{2j}\left(\mathbf{r}_s'',\tau;\mathbf{r}'_{\sigma'},\tau' \right ) -\sum_{\sigma_1}\int\mathrm{d}\mathbf{r}''\Delta^*_{\sigma_1\sigma}(\mathbf{r}'',\mathbf{r})
G_{1j}\left(\mathbf{r}''_{\sigma_1},\tau ;\mathbf{r}'_{\sigma'},\tau' \right ) \notag \\
& = \delta(\tau-\tau')\delta(\mathbf{r}-\mathbf{r}')\delta_{\sigma\sigma'}\delta_{2j}.
\label{eq:moT2}
\end{align}
\normalsize
Using Eq. \eqref{eq:ftMatsubara},
we now Fourier transform from imaginary time to Matsubara frequencies, which gives
\small
\begin{align}
&\left(i\omega_n-\hat{K} \right )G_{1j}\left(\mathbf{r}_\sigma,\mathbf{r}'_{\sigma'};\omega_n\right )
-\sum_{s}\int\mathrm{d}\mathbf{r}''\mathbf{g}(\mathbf{r}-\mathbf{r}'')\cdot\boldsymbol{\sigma}_{\sigma s}
\,G_{1j}\left(\mathbf{r}_s'',\mathbf{r}'_{\sigma'};\omega_n \right )
+\sum_{\sigma_1}\int\mathrm{d}\mathbf{r}''\Delta_{\sigma_1\sigma}(\mathbf{r}'',\mathbf{r})
G_{2j}\left(\mathbf{r}''_{\sigma_1},\mathbf{r}'_{\sigma'};\omega_n \right )\notag \\
& =\delta(\mathbf{r}-\mathbf{r}')\delta_{\sigma\sigma'}\delta_{1j}; \\
& \left(i\omega_n+\hat{K}^* \right )G_{2j}\left(\mathbf{r}_\sigma,\mathbf{r}'_{\sigma'};\omega_n\right )
+\sum_{s}\int\mathrm{d}\mathbf{r}''\mathbf{g}(\mathbf{r}''-\mathbf{r})\cdot\boldsymbol{\sigma}_{s\sigma}
\,G_{2j}\left(\mathbf{r}_s'',\mathbf{r}'_{\sigma'};\omega_n \right )
-\sum_{\sigma_1}\int\mathrm{d}\mathbf{r}''\Delta^*_{\sigma_1\sigma}(\mathbf{r}'',\mathbf{r})
G_{1j}\left(\mathbf{r}''_{\sigma_1},\mathbf{r}'_{\sigma'};\omega_n \right ) \notag \\
& = \delta(\mathbf{r}-\mathbf{r}')\delta_{\sigma\sigma'}\delta_{2j}.
\end{align}
\normalsize
Next, we relabel the Green's functions to a more familiar form, and introduce matrix notations for conciseness.

\subsection{Matrix representation}

Because of symmetries \eqref{eq:symmetryOfG} of the four Green's functions $G_{ij}$, one can reduce the amount of different Green's function to two. We redefine the normal and anomalous functions explicitly as $G_{11}(\mathbf{r}_\sigma,\mathbf{r}'_{\sigma'};\omega_n)=G_{\sigma\sigma'}(\mathbf{r},\mathbf{r}';\omega_n)$, $G_{12}(\mathbf{r}_\sigma,\mathbf{r}'_{\sigma'};\omega_n)=F_{\sigma\sigma'}(\mathbf{r},\mathbf{r}';\omega_n)$, $G_{22}(\mathbf{r}_\sigma,\mathbf{r}'_{\sigma'};\omega_n)=-G^*_{\sigma\sigma'}(\mathbf{r},\mathbf{r}';\omega_n)$, and, as a detailed demonstration of the properties \eqref{eq:symmetryOfG}
\begin{equation}
G_{21}(\mathbf{r}_\sigma,\mathbf{r}_{\sigma'};\omega_n)=-G_{21}(\mathbf{r}'_{\sigma'},\mathbf{r}_\sigma;-\omega_n)=-G_{12}^*(\mathbf{r}_\sigma,\mathbf{r}'_{\sigma'};\omega_n)=-F^*_{\sigma\sigma'}(\mathbf{r},\mathbf{r}';\omega_n).
\end{equation}

From the re-labeling of the Green's functions, the properties in Eq. \eqref{eq:symmetryOfG}, and a Fourier transform to momentum space yields Eq. \eqref{eq:propertiesGk}. 
In the following notation $G_{\sigma\sigma'}$ is the matrix element of the $2\times 2$ matrix in spin space $G$. We can, therefore, construct a $4\times 4$ Nambu matrix
\begin{align}
\hat{G}(\mathbf{r},\mathbf{r}';\omega_n) = \begin{bmatrix}
G(\mathbf{r},\mathbf{r}';\omega_n) & F(\mathbf{r},\mathbf{r}';\omega_n) \\ 
-F^*(\mathbf{r},\mathbf{r}';\omega_n) & -G^*(\mathbf{r},\mathbf{r}';\omega_n)
\label{eq:44}
\end{bmatrix}. 
\end{align}
We perform the Fourier transformation of Eqs. \eqref{eq:moT1} and \eqref{eq:moT2} to the frequency domain and use the relation
$\Delta_{\alpha\beta}(\mathbf{r},\mathbf{r}') = -\Delta_{\beta\alpha}(\mathbf{r}',\mathbf{r})$ which gives
\small
\begin{align}
&\left(i\omega_n-\hat{K} \right )G_{\sigma\sigma'}\left(\mathbf{r},\mathbf{r}';\omega_n\right )
-\sum_{s}\int\mathrm{d}\mathbf{r}''\mathbf{g}(\mathbf{r}-\mathbf{r}'')\cdot\boldsymbol{\sigma}_{\sigma s}
\,G_{s\sigma'}\left(\mathbf{r}'',\mathbf{r}';\omega_n \right )
+\sum_{s}\int\mathrm{d}\mathbf{r}''\Delta_{ \sigma s}(\mathbf{r},\mathbf{r}'')
F^*_{s\sigma'}\left(\mathbf{r}'',\mathbf{r}';\omega_n \right )
=\delta(\mathbf{r}-\mathbf{r}')\delta_{\sigma\sigma'} ; \\
& \left(i\omega_n-\hat{K} \right )F_{\sigma\sigma'}\left(\mathbf{r},\mathbf{r}';\omega_n\right )
-\sum_{s}\int\mathrm{d}\mathbf{r}''\mathbf{g}(\mathbf{r}-\mathbf{r}'')\cdot\boldsymbol{\sigma}_{\sigma s}
\,F_{s\sigma'}\left(\mathbf{r}'',\mathbf{r}';\omega_n \right )
+\sum_{s}\int\mathrm{d}\mathbf{r}''\Delta_{ \sigma s}(\mathbf{r},\mathbf{r}'')
G^*_{s\sigma'}\left(\mathbf{r}'',\mathbf{r}';\omega_n \right )
=0; \\
& -\left(i\omega_n+\hat{K}^* \right )F^*_{\sigma\sigma'}\left(\mathbf{r},\mathbf{r}';\omega_n\right )
-\sum_{s}\int\mathrm{d}\mathbf{r}''\,\mathbf{g}(\mathbf{r}''-\mathbf{r})\cdot\boldsymbol{\sigma}_{s\sigma }
\,F_{s\sigma'}^*\left(\mathbf{r}'',\mathbf{r}';\omega_n \right )
+\sum_{s}\int\mathrm{d}\mathbf{r}''\Delta^*_{ \sigma s}(\mathbf{r},\mathbf{r}'')
G_{s\sigma'}\left(\mathbf{r}'',\mathbf{r}';\omega_n \right )
=0 \\
& -\left(i\omega_n+\hat{K}^* \right )G^*_{\sigma\sigma'}\left(\mathbf{r},\mathbf{r}';\omega_n\right )
-\sum_{s}\int\mathrm{d}\mathbf{r}''\,\mathbf{g}(\mathbf{r}''-\mathbf{r})\cdot\boldsymbol{\sigma}_{s\sigma }
\,G^*_{s\sigma'}\left(\mathbf{r}'',\mathbf{r}';\omega_n \right )
+\sum_{s}\int\mathrm{d}\mathbf{r}''\Delta^*_{ \sigma s}(\mathbf{r},\mathbf{r}'')
F_{s\sigma'}\left(\mathbf{r}'',\mathbf{r}';\omega_n \right ) \notag \\
& =\delta(\mathbf{r}-\mathbf{r}')\delta_{\sigma\sigma'}.
\end{align}
\normalsize
We Fourier now transform to momentum space using
\begin{align}
\hat{G}(\mathbf{r}-\mathbf{r}';\omega_n)  = \frac{1}{V}\sum_\mathbf{k}e^{i\mathbf{k}\cdot(\mathbf{r}-\mathbf{r}')} \hat{G}(\mathbf{k};\omega_n),\quad \delta_{\mathbf{k,k}'}=\frac{1}{V}\int\mathrm{d}\mathbf{R}\, e^{-i(\mathbf{k}-\mathbf{k}')\cdot \mathbf{R}},
\end{align}
and similarly for the other terms, which yields
\begin{align}
& \left(i\omega_n-\xi(\mathbf{k}) \right )G_{\sigma\sigma'}(\mathbf{k};\omega_n)-\sum_s\tilde{\mathbf{g}}(\mathbf{k})\cdot\boldsymbol{\sigma}_{\sigma s}\, G_{s\sigma'}(\mathbf{k};\omega_n)-\sum_s\Delta_{s\sigma}(-\mathbf{k})F^*_{s\sigma'}(-\mathbf{k};\omega_n)=\delta_{\sigma\sigma'}; \label{g1} \\
&  \left(i\omega_n-\xi(\mathbf{k}) \right )F_{\sigma\sigma'}(\mathbf{k};\omega_n)-\sum_s\tilde{\mathbf{g}}(\mathbf{k})\cdot\boldsymbol{\sigma}_{\sigma s}\, F_{s\sigma'}(\mathbf{k};\omega_n)-\sum_s\Delta_{s\sigma}(-\mathbf{k})G^*_{s\sigma'}(-\mathbf{k};\omega_n)=0;\label{g2} \\
& -\left(i\omega_n+\xi(\mathbf{k}) \right )F^*_{\sigma\sigma'}(-\mathbf{k};\omega_n)-\sum_s\tilde{\mathbf{g}}(-\mathbf{k})\cdot\boldsymbol{\sigma}^T_{\sigma s}\, F^*_{s\sigma'}(-\mathbf{k};\omega_n)-\sum_s\Delta^*_{s\sigma}(\mathbf{k})G_{s\sigma'}(\mathbf{k};\omega_n)=0;\label{g3} \\
& -\left(i\omega_n+\xi(\mathbf{k}) \right )G^*_{\sigma\sigma'}(-\mathbf{k};\omega_n)-\sum_s\tilde{\mathbf{g}}(-\mathbf{k})\cdot\boldsymbol{\sigma}^T_{\sigma s}\, G^*_{s\sigma'}(-\mathbf{k};\omega_n)-\sum_s\Delta^*_{s\sigma}(\mathbf{k})F_{s\sigma'}(\mathbf{k};\omega_n)=\delta_{\sigma\sigma'}.
\end{align}
The Pauli principle ensures $\Delta_{s\sigma}(\mathbf{k})=-\Delta_{\sigma s}(-\mathbf{k})$.
Here $\tilde{\mathbf{g}}(\mathbf{k})$ is the Fourier transform of $\mathbf{g}(\mathbf{r}-\mathbf{r}')$:
\begin{align}
\tilde{\mathbf{g}}(\mathbf{k})=\int\mathrm{d}(\mathbf{r}-\mathbf{r}')\,e^{-i\mathbf{k}\cdot(\mathbf{r}-\mathbf{r}')}\,\mathbf{g}(\mathbf{r}-\mathbf{r}')=\boldsymbol{\gamma}(\mathbf{k})-\mathbf{B},
\end{align}
where we used Eq. \eqref{eq:ftg}. Reorganizing the set of the four Gor'kov equations in $\mathbf{k}$-space given above in matrix form gives the left-Gor'kov equation \eqref{eq:leftG}.

\twocolumngrid

\section{
Properties and normalization condition of the quasi-classical Green's functions
\label{app:properties}
}

It is convenient to parametrize the quasi-classical Green's functions in terms of Pauli matrices as
\begin{align}
& g(\mathbf{k};\omega_n)=g_0(\mathbf{k};\omega_n)\sigma_0+\mathbf{g}(\mathbf{k};\omega_n)\cdot\boldsymbol{\sigma}; \label{eq:paramg} \\
& f(\mathbf{k};\omega_n)=\left[f_0(\mathbf{k};\omega_n)\sigma_0+\mathbf{f}(\mathbf{k};\omega_n)\cdot\boldsymbol{\sigma})\right] i\sigma_y.\label{eq:paramf}
\end{align}
The properties
$g(\mathbf{k};\omega_n) = -g^\dag(\mathbf{k},-\omega_n)$ and $f(\mathbf{k},\omega_n) = -f^\mathrm{T}(-\mathbf{k};-\omega_n)$ translate to
\begin{align}
& g_0(\mathbf{k};\omega_n)=-g_0^*(\mathbf{k};-\omega_n),\,\,\,
f_0(\mathbf{k};\omega_n)=f_0(-\mathbf{k};-\omega_n), \notag 
\\
& \mathbf{g}(\mathbf{k};\omega_n)=-\mathbf{g}^*(\mathbf{k};-\omega_n),\,\,\,\, \mathbf{f}(\mathbf{k};\omega_n)=-\mathbf{f}(-\mathbf{k};-\omega_n). 
\end{align}
Using Eq. \eqref{eq:quasiclassicalG}, the normalization condition $\hat{g}^2(\mathbf{k};\omega_n)=\hat{\sigma}_0$ reads
\begin{align}
\begin{bmatrix}
g^2-ff^* & -igf+ifg^*\\ 
-if^* g+ig^*f^* & -f^*f+g^{*2}
\end{bmatrix}
=
\begin{bmatrix}
\sigma_0 & 0\\ 
0 & \sigma_0
\end{bmatrix},
\label{eq:normMatrix}
\end{align}
where $g\equiv g(\mathbf{k};\omega_n)$ and $g^*\equiv g^*(-\mathbf{k};\omega_n)$ and similarly for $f$ ($f^*$). 
In general, the Eilenberger equation \eqref{eq:EilenbergerEq} together with the normalization \eqref{eq:normMatrix} yields a system of 32 equations to be solved. Here, we are interested in studying the superconducting instabilities at which the order parameters are small. 
Therefore, we can study the linearized version of Eqs. \eqref{eq:EilenbergerEq} and \eqref{eq:normMatrix}, which simplifies the problem considerably. Using the parametrizations \eqref{eq:paramg} and \eqref{eq:paramf}, the $(1,1)$ component of Eq. \eqref{eq:normMatrix} gives the two conditions
\begin{align}
g_0^2(\mathbf{k};\omega_n)+\mathbf{g}^2(\mathbf{k};\omega_n)  & =1-f_0(\mathbf{k};\omega_n)f_0^*(-\mathbf{k};\omega_n)
 \notag \\
& +\mathbf{f}(\mathbf{k};\omega_n)\cdot \mathbf{f}^*(-\mathbf{k};\omega_n);
\end{align}
\begin{align}
&2g_0(\mathbf{k};\omega_n)\mathbf{g}(\mathbf{k};\omega_n)  = i\mathbf{f}(\mathbf{k};\omega_n)\times \mathbf{f}^*(-\mathbf{k};\omega_n) \notag \\
&+f_0(\mathbf{k};\omega_n)\mathbf{f}^*(-\mathbf{k};\omega_n)-f_0^*(-\mathbf{k};\omega_n)\mathbf{f}(\mathbf{k};\omega_n).
\end{align}
In the normal state (the $0^\mathrm{th}$ order terms), one must have $g_0^2(\mathbf{k};\omega_n)+\mathbf{g}^2(\mathbf{k};\omega_n)=1$ and $2g_0(\mathbf{k};\omega_n)\mathbf{g}(\mathbf{k};\omega_n)=0$, such that $g_0(\mathbf{k};\omega_n)=\mathrm{sgn}(\omega_n)$ and $\mathbf{g}(\mathbf{k};\omega_n)=0$. These normal state $0^\mathrm{th}$ order terms together with the linearized $f_0(\mathbf{k};\omega_n)$ and $\mathbf{f}(\mathbf{k};\omega_n)$ is all we need to study the superconducting instability conditions.

\section{Limiting of triplets by SOC \label{app:limitingSOC}}

In the purely triplet case with SOC, the linearized Eilenberger equation \eqref{eil2} reads
\begin{align}
\omega_n\mathbf{f}(\mathbf{k};\omega_n) =  \mathrm{sgn}(\omega_n)\mathbf{d}(\mathbf{k})
+\boldsymbol{\gamma}(\mathbf{k})\times\mathbf{f}(\mathbf{k};\omega_n).
\end{align}
For Ising SOC, $\boldsymbol{\gamma}(\mathbf{k})=\Delta_\mathrm{so}\hat{\gamma}(\mathbf{k})\hat{\boldsymbol{z}}$ we have component-wise
\begin{align}
\omega_n f_x(\mathbf{k};\omega_n) & =  \mathrm{sgn}(\omega_n)\,d_x(\mathbf{k})
-\Delta_\mathrm{so}\hat{\gamma}(\mathbf{k})f_y(\mathbf{k};\omega_n); \\
\omega_n f_y(\mathbf{k};\omega_n) &=  \mathrm{sgn}(\omega_n)\,d_y(\mathbf{k})
+\Delta_\mathrm{so}\hat{\gamma}(\mathbf{k})f_x(\mathbf{k};\omega_n); \\
\omega_n f_z(\mathbf{k};\omega_n) &=  \mathrm{sgn}(\omega_n)\,d_z(\mathbf{k}).
\end{align}
The $z$-component pairing correlation and order parameter $\{f_z,d_z\}$ remain unaffected by SOC. We can solve for $\{f_x,f_y\}$, which gives
\begin{align}
f_x(\mathbf{k};\omega_n) & =\frac{\mathrm{sgn}(\omega_n)}{\omega_n^2+\Delta_\mathrm{so}^2\hat{\gamma}^2(\mathbf{k})}\left[\omega_n d_x(\mathbf{k})-\Delta_\mathrm{so}\hat{\gamma}(\mathbf{k})d_y(\mathbf{k}) \right ]; \label{eq:soc1}\\
f_y(\mathbf{k};\omega_n) & =\frac{\mathrm{sgn}(\omega_n)}{\omega_n^2+\Delta_\mathrm{so}^2\hat{\gamma}^2(\mathbf{k})}\left[\Delta_\mathrm{so}\hat{\gamma}(\mathbf{k})d_x(\mathbf{k}) +\omega_n d_y(\mathbf{k})\right ].
\label{eq:soc2}
\end{align}
Note that the terms with $\Delta_\mathrm{so}$ are odd in frequency $\omega_n$. However, they do not contribute to the self-consistency conditions, because they vanish in the averages, see Eq. \eqref{eq:di}. 
For each component $d_i(\mathbf{k})$ ($i=x,y$), we have the self-consistency condition as in Eq. \eqref{eq:selcCtriplet}

\small
\begin{align}
d_i(\mathbf{k})\ln\frac{T}{T_\mathrm{ct}}+\frac{\pi}{\beta}\sum_{n=-\infty}^\infty
\biggr(\frac{d_i(\mathbf{k})}{|\omega_n|}  -\hat{\gamma}(\mathbf{k})\left\langle\hat{\gamma}(\mathbf{k}')f_i(\mathbf{k}';\omega_n) \right\rangle
\biggr)
=0,
\label{eq:selfconsapp}
\end{align}
\normalsize
and we must evaluate the average
\begin{align}
\hat{\gamma}(\mathbf{k})\left\langle\hat{\gamma}(\mathbf{k}')f_i(\mathbf{k}';\omega_n) \right\rangle =\frac{|\omega_n|}{\omega_n^2+\Delta_\mathrm{so}^2}d_i(\mathbf{k}),
\label{eq:di}
\end{align}
where we approximated $\hat{\gamma}^2(\mathbf{k})\approx 1$ and the odd-frequency term vanished.
Therefore, the components decouple in the self-consistency. Performing the Matsubara sum in Eq. \eqref{eq:selfconsapp} leads to the pair-breaking equation \eqref{eq:fulde} with $T_\mathrm{c}=T_\mathrm{ct}$ and $\alpha = i\Delta_\mathrm{so}$.

\twocolumngrid

\section{Expressing \texorpdfstring{$\lambda_\mathrm{s(t)}$}{} in favour of \texorpdfstring{$T_\mathrm{cs(t)}$}{}\label{app:cpling}}

With the self-consistency condition given by Eq. \eqref{eq:selfconsistent} and the pairing interaction specified in Eq. \eqref{eq:interaction}, we obtain one self-consistency condition for each order parameter component. Below we discuss the singlet and triplet components separately. 

\subsection{Singlet part}

For the singlet order parameter, assuming a constant $s$-wave, Eq. \eqref{eq:selfconsistent} gives 
\begin{align}
\psi_0=\frac{\lambda_\mathrm{s}}{2\beta}\sum_{n=-\infty}^\infty \int_{-\infty}^\infty \mathrm{d}\xi_{\mathbf{k}}\left[F_{\uparrow\downarrow}(\mathbf{k};\omega_n)-F_{\downarrow\uparrow}(\mathbf{k};\omega_n) \right ],
\label{eq:ssing}
\end{align}
where the dimensionless coupling constant $\lambda_\mathrm{s}=2N_0 v_\mathrm{s}/V$. 
To establish the unperturbed critical temperature $T_\mathrm{cs}$
at $\mathbf{B}=\boldsymbol{\gamma}(\mathbf{k})=0$, we can solve the Gor'kov equation \eqref{eq:leftG} to find
\begin{align}
F_{\uparrow\downarrow(\downarrow\uparrow)}(\mathbf{k};\omega_n)=\mp \frac{\psi_0}{\omega_n^2+\xi_\mathbf{k}^2+|\psi_0|^2}
\label{eq:Fgork}.
\end{align}
Using Eq. \eqref{eq:Fgork} in Eq. \eqref{eq:ssing}, and performing the Matsubara sum, we obtain the gap equation
\begin{align}
1=-\lambda_\mathrm{s}\int_{-\infty}^\infty \mathrm{d}\xi_{\mathbf{k}}\,\frac{\tanh\left( \frac{\beta}{2}\sqrt{\xi_\mathbf{k}^2+|\psi_0|^2}\right )}{2\sqrt{\xi_\mathbf{k}^2+|\psi_0|^2}}.
\label{eq:gapeq}
\end{align}
If the interaction is attractive $v_\mathrm{c}<0\Rightarrow \lambda_\mathrm{s}<0$, then Eq. \eqref{eq:gapeq} admits a solution. At the singlet critical transition temperature $T_\mathrm{cs}$, $\psi_0=0$, such that
\begin{align}
1=-\lambda_\mathrm{s}\int_{-\epsilon_\mathrm{c}}^{\epsilon_\mathrm{c}} \mathrm{d}\xi_{\mathbf{k}}\,\frac{\tanh\left(\frac{|\xi_\mathbf{k}|}{2T_\mathrm{cs}}\right )}{2|\xi_\mathbf{k}|},
\label{eq:rel}
\end{align}
where we introduced the characteristic cutoff of the pairing interaction $\epsilon_\mathrm{c}$. Eq. \eqref{eq:rel} relates $\lambda_\mathrm{s}$ to $T_\mathrm{cs}$. We can further manipulate the gap equation in the following way \cite{Kita2015}
\begin{align}
 -\frac{1}{\lambda_\mathrm{s}}  & =\int_{-\epsilon_c}^{\epsilon_c}\mathrm{d}\xi_\mathbf{k}\,\frac{1}{2|\xi_\mathbf{k}|}\tanh\frac{|\xi_\mathbf{k}|}{2 T_\mathrm{cs}} \notag \\
 & =  \int_{-\epsilon_\mathrm{c}}^{\epsilon_\mathrm{c}}\mathrm{d}\xi_\mathbf{k}\left[\frac{1}{2|\xi_\mathbf{k}|}\tanh\frac{|\xi_\mathbf{k}|}{2 T_c}-\frac{1}{2|\xi_\mathbf{k}|}\tanh\frac{|\xi_\mathbf{k}|}{2 T} \right ] \notag \\
 & \quad+\frac{1}{\beta}\int_{-\epsilon_\mathrm{c}}^{\epsilon_\mathrm{c}}\mathrm{d}\xi_\mathbf{k} \sum_{n=-n_\mathrm{c}-1}^{n_\mathrm{c}}\frac{1}{\omega_n^2+\xi_\mathbf{k} ^2}  \notag \\
& \approx  \ln\frac{T}{T_\mathrm{cs}} +\frac{1}{\beta}\sum_{n=-n_\mathrm{c}-1}^{n_\mathrm{c}}\mathrm{d}\xi_\mathbf{k}\,\frac{1}{\omega_n^2+\xi_\mathbf{k}} \notag \\
& = \ln\frac{T}{T_\mathrm{cs}} +\frac{1}{\beta}\sum_{n=-n_\mathrm{c}-1}^{n_\mathrm{c}}\frac{\pi}{|\omega_n|}. \label{eq:relating}
\end{align}
To do the analytical integration, we performed an integration by parts and extended $\epsilon_\mathrm{c}\rightarrow\infty$.
This demonstrates Eq. \eqref{eq:manipulated}.

\begin{figure*} 
\centering
\includegraphics[width=0.8\textwidth]{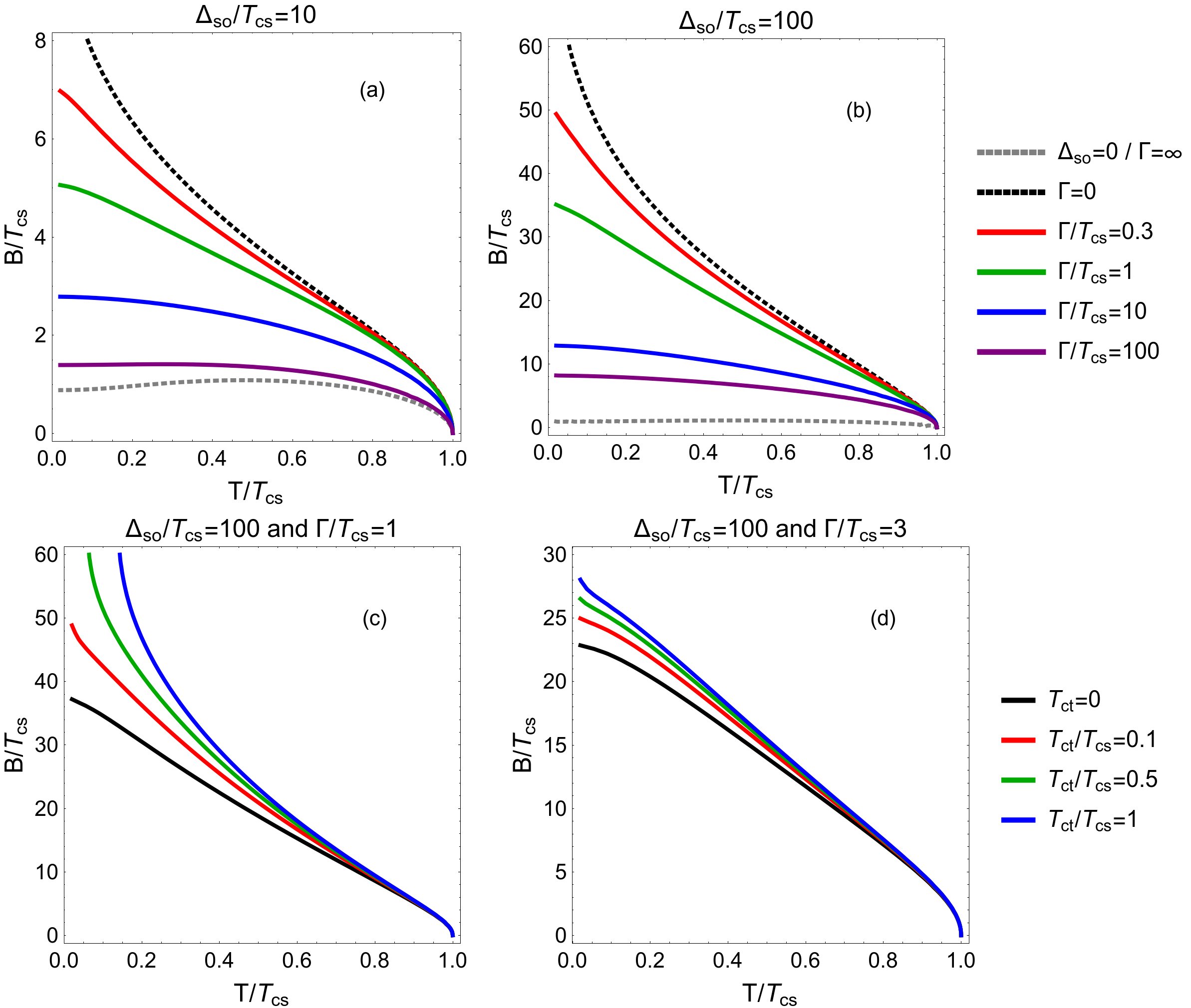}
\caption{\label{fig:4} 
The effect of the disorder on the singlet transition lines $B_\mathrm{c}(T)$ with $\hat{\gamma}(\mathbf{k})=\sqrt{2}\cos(3\varphi_\mathbf{k})$. (a) Singlets only with $\Delta_\mathrm{so}/T_\mathrm{cs}=10$. (b) Singlets only with $\Delta_\mathrm{so}/T_\mathrm{cs}=100$. (c) The effect of an increasing triplet channel on disordered curves with $\Gamma/T_\mathrm{cs}=1$ and in (d) $\Gamma/T_\mathrm{cs}=3$. 
}
\end{figure*}

\subsection{Triplet}

For the triplets, since $d_z(\mathbf{k})$ decouples from $\{d_x(\mathbf{k}),d_y(\mathbf{k})\}$, we only write the pairing channel  for the in-plane $\mathbf{d}$-vector components. In the example of $D_{3h}$, this corresponds to the channel of the irreducible representation $E''$. In this case, Eq. \eqref{eq:interaction} is
\begin{align}
V^{\sigma_1\sigma_2}_{\sigma_1'\sigma_2'}(\mathbf{k},\mathbf{k}') & =v_\mathrm{t}\left[\hat{\gamma}(\mathbf{k})\sigma_x \,i\sigma_y \right ]_{\sigma_1\sigma_2}
\left[\hat{\gamma}(\mathbf{k}')\sigma_x \,i\sigma_y \right ]^*_{\sigma_1'\sigma_2'} \notag \\
& +v_\mathrm{t}\left[\hat{\gamma}(\mathbf{k})\sigma_y \,i\sigma_y \right ]_{\sigma_1\sigma_2}
\left[\hat{\gamma}(\mathbf{k}')\sigma_y \,i\sigma_y \right ]^*_{\sigma_1'\sigma_2'}.
\end{align}
With this interaction, we rewrite Eq. \eqref{eq:selfconsistent} as
\begin{widetext}
\begin{align}
 \left[-d_x(\mathbf{k})\sigma_z+id_y(\mathbf{k})\sigma_0 \right ]_{\sigma_1\sigma_2}= 
 \frac{1}{\beta V}\sum_{n,\mathbf{k}'}\sum_{\sigma_1'\sigma_2'}v_\mathrm{t}\hat{\gamma}(\mathbf{k})\hat{\gamma}(\mathbf{k}')
\left(\left[\sigma_z \right ]_{\sigma_1\sigma_2}\left[\sigma_z \right ]_{\sigma_1'\sigma_2'}+\left[\sigma_0 \right ]_{\sigma_1\sigma_2}\left[\sigma_0 \right ]_{\sigma_1'\sigma_2'} \right )F_{\sigma_1'\sigma_2'}(\mathbf{k}';\omega_n).
\end{align}
\end{widetext}
For the $d_y(\mathbf{k})$ component, we have (and similarly for $d_x(\mathbf{k})$) 

\begin{align} 
id_y(\mathbf{k})
=\frac{1}{\beta V}\sum_{n,\mathbf{k}'}v_\mathrm{t}\hat{\gamma}(\mathbf{k})\hat{\gamma}(\mathbf{k}')
\left(F_{\uparrow\uparrow}(\mathbf{k}';\omega_n)+F_{\downarrow\downarrow}(\mathbf{k}';\omega_n) \right ).
\end{align}

For the unperturbed case, the Gor'kov equation \eqref{eq:leftG} yields 
\begin{align}
F_{\uparrow\uparrow(\downarrow\downarrow)}(\mathbf{k};\omega_n)= \frac{\pm d_x(\mathbf{k})-id_y(\mathbf{k})}{\omega_n^2+\xi_\mathbf{k}^2+|\mathbf{d}(\mathbf{k})|^2}.
\end{align}
Defining the triplet coupling constant $\lambda_\mathrm{t}=2N_0 v_\mathrm{t}/V$ and using Eq. \eqref{eq:dos}, we have (for $d_x=0$)
\small
\begin{align}
d_y(\mathbf{k}) = -\hat{\gamma}(\mathbf{k})\frac{\lambda_\mathrm{t}}{\beta}\sum_{n=-\infty}^\infty\int\mathrm{d}\xi_{\mathbf{k}'}\int\frac{\mathrm{d}\varphi_{\mathbf{k}'}}{2\pi}\frac{\hat{\gamma}(\mathbf{k}')d_y(\mathbf{k}')}{\omega_n^2+\xi_{\mathbf{k}'}^2+|d_y(\mathbf{k}')|^2}.
\end{align}
\normalsize 
Instead of performing the angular integral exactly, we write the order parameter in terms of its basis function as $d_y(\mathbf{k})=\tilde{\eta}_y \hat{\gamma}(\mathbf{k})$and approximate $\hat{\gamma}^2(\mathbf{k})\rightarrow 1$. This yields 
\begin{align}
1 & = -\frac{\lambda_\mathrm{t}}{\beta}\sum_{n=-\infty}^\infty\int\mathrm{d}\xi_{\mathbf{k}'}\,\frac{1}{\omega_n^2+\xi_{\mathbf{k}'}^2+|\tilde{\eta}_y|^2} \notag \\
& =-\lambda_\mathrm{t}\int_{-\infty}^\infty \mathrm{d}\xi_{\mathbf{k}}\,\frac{\tanh\left( \frac{\beta}{2}\sqrt{\xi_\mathbf{k}^2+|\tilde{\eta}_y|^2}\right )}{2\sqrt{\xi_\mathbf{k}^2+|\tilde{\eta}_y|^2}}.
\end{align}
In analogy to the singlet case \eqref{eq:relating}, we can write 
\begin{align}
 -\frac{1}{\lambda_\mathrm{t}}  = \ln\frac{T}{T_\mathrm{ct}} +\frac{1}{\beta}\sum_{n=-n_\mathrm{c}-1}^{n_\mathrm{c}}\frac{\pi}{|\omega_n|}.
\end{align}
This result does not rely on the approximation $\hat{\gamma}^2(\mathbf{k})\rightarrow 1$.

\section{ \label{app:relaxingDx0} Solution and decoupling of the \texorpdfstring{$d_x(\mathbf{k})$}{} component}

Here we show that the $d_x(\mathbf{k})$ component is independent of the other components. 
In the case of the $d_z(\mathbf{k})$ component, which is also independent, the $f_z(\mathbf{k};\omega_n)$ pairing correlations from which the order parameter $d_z(\mathbf{k})$ is formed is decoupled from $\{f_0,f_x,f_y\}$. The issue is more subtle for the $d_x(\mathbf{k})$ component, which is discussed below.

\subsection{The clean case}

Solving the Eilenberger matrix \eqref{eq:cleanEil} for $\{f_0,f_x,f_y\}$, we obtain 
\begin{widetext}
\begin{align}
& f_0(\mathbf{k};\omega_n) =\overbrace{\frac{1}{|\omega_n|}\frac{iB\omega_n d_x(\mathbf{k})}{\omega_n^2+B^2+\gamma^2(\mathbf{k})}}^\mathrm{odd-frequency} + \overbrace{\frac{1}{|\omega_n|}\frac{\left(\omega_n^2+\gamma^2(\mathbf{k}) \right )\psi(\mathbf{k})-iB\gamma(\mathbf{k})d_y(\mathbf{k})}{\omega_n^2+B^2+\gamma^2(\mathbf{k})}}^\mathrm{even-frequency}; \label{d1} \\
& f_x(\mathbf{k};\omega_n) =\overbrace{\frac{|\omega_n|d_x(\mathbf{k})}{\omega_n^2+B^2+\gamma^2(\mathbf{k})}}^\mathrm{even-frequency}+\overbrace{ \mathrm{sgn}(\omega_n)\frac{iB\psi(\mathbf{k}) -\gamma(\mathbf{k})d_y(\mathbf{k})}{\omega_n^2+B^2+\gamma^2(\mathbf{k})}}^\mathrm{odd-frequency};  \label{d2} \\
& f_y(\mathbf{k};\omega_n) =  \overbrace{\frac{1}{|\omega_n|}\frac{\omega_n\gamma(\mathbf{k})d_x(\mathbf{k})}{\omega_n^2+B^2+\gamma^2(\mathbf{k})}}^\mathrm{odd-frequency}+\overbrace{\frac{1}{|\omega_n|}\frac{iB\gamma(\mathbf{k})\psi(\mathbf{k})+\left(\omega_n^2+B^2 \right )d_y(\mathbf{k})}{\omega_n^2+B^2+\gamma^2(\mathbf{k})}}^\mathrm{even-frequency}.   \label{d3}
\end{align}
\end{widetext}
Here, we explicitly separate the even- from the odd-frequency parts of the pairing correlations. When feeding Eqs. (\ref{d1}-\ref{d3}) to the self-consistency conditions \eqref{eq:selfconsapp}, only the even-frequency terms contribute, since the odd-frequency terms vanish in the averages. 
Using $\hat{\gamma}(\mathbf{k})=\mathrm{sgn}[\gamma(\mathbf{k})]$, the relevant averages for the self consistency equations are
\begin{align}
& \langle f_0(\mathbf{k};\omega_n)\rangle = \frac{1}{|\omega_n|}\frac{\left(\omega_n^2+\Delta^2_\mathrm{so} \right )\psi_0-iB\Delta_\mathrm{so}\tilde{\eta}_y}{\omega_n^2+B^2+\Delta^2_\mathrm{so}}; \\
& \hat{\gamma}(\mathbf{k})\langle \hat{\gamma}(\mathbf{k}')f_x(\mathbf{k}';\omega_n)\rangle = \frac{ |\omega_ n| d_x(\mathbf{k})}{\omega_n^2+B^2+\Delta^2_\mathrm{so}}; \\
& \hat{\gamma}(\mathbf{k})\langle \hat{\gamma}(\mathbf{k}')f_y(\mathbf{k}';\omega_n)\rangle = \frac{1}{|\omega_n|}\frac{iB\gamma(\mathbf{k})\psi_0+\left(\omega_n^2+B^2\right )d_y(\mathbf{k})}{\omega_n^2+B^2+\Delta^2_\mathrm{so}}.
\end{align}
This yields a sub-system for  $\{\psi_0,d_y(\mathbf{k})\}$ with pair-breaking equation \eqref{eq:pairbreakingClean}, and an independent behavior for $d_x(\mathbf{k})$ with the pair-breaking equation given by Eq. \eqref{eq:fulde} with $\alpha = i\sqrt{B^2+\Delta_\mathrm{so}^2}$ and $T_\mathrm{c}=T_\mathrm{ct}$.
It is immaterial to keep $d_x(\mathbf{k})$ explicitly for the analysis of the more interesting $\{\psi_0,d_y(\mathbf{k})\}$ sub-system. 

\begin{figure*} 
\centering
\includegraphics[width=0.6\textwidth]{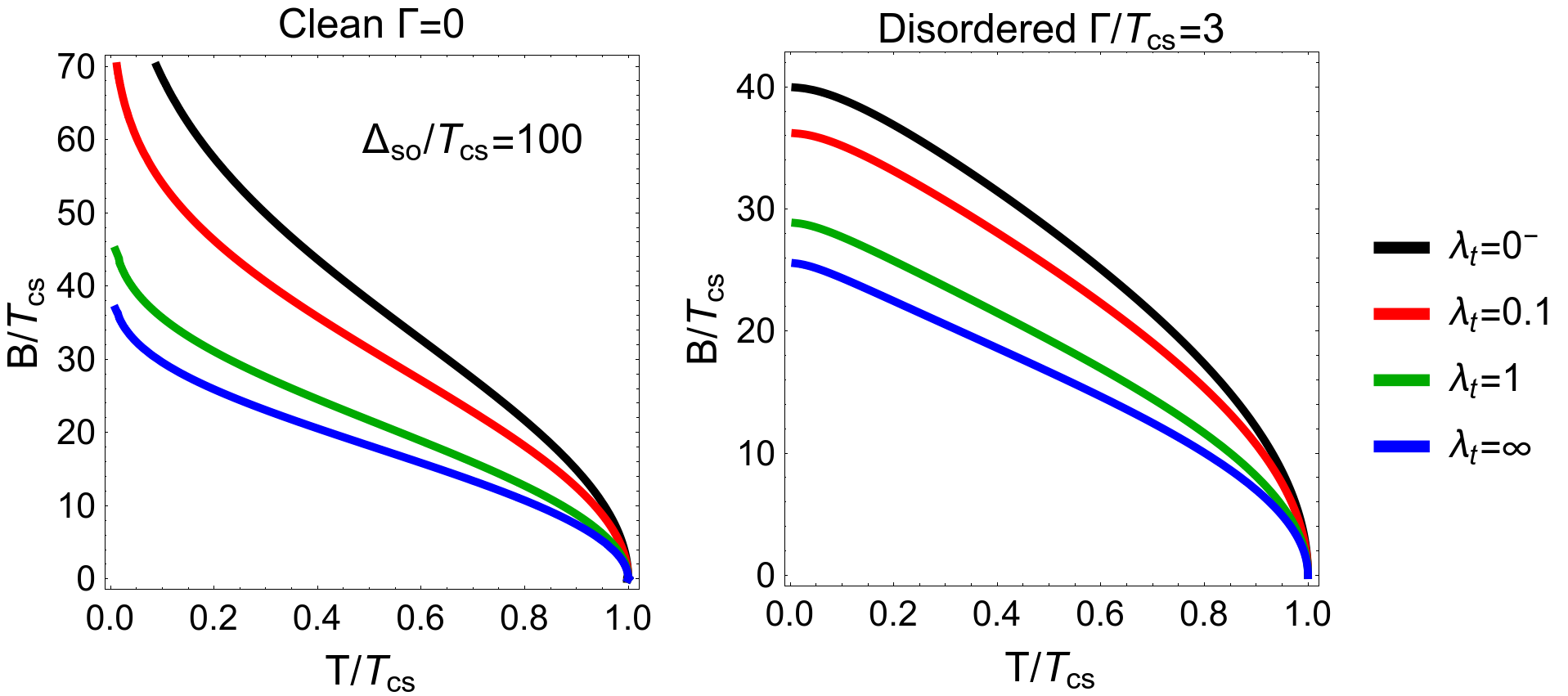}
\caption{\label{fig:repulsion} 
The effect of repulsion in the triplet channel on the transition $B_\mathrm{c}(T)$ with $\Delta_\mathrm{so}/T_\mathrm{cs}=100$. The same color legend for different coupling constants apply to both plots. We used a cutoff energy of $\epsilon_\mathrm{c}/T_\mathrm{cs}=500$. The blue lines show the case with maximum repulsion (saturation). The larger the scattering rate $\Gamma$, the smaller the effect of the triplet channel.
}
\end{figure*}

\subsection{The disordered case}

The decoupling of the $d_x(\mathbf{k})$ components also occurs in the presence of impurities. To check this explicitly, the solution of the Eilenberger matrix \eqref{eq:EilMatDis} in the case $d_x(\mathbf{k})\neq 0$ is

\begin{align}
f_x(\mathbf{k};\omega_n)  & = 
\frac{i\mathrm{sgn}(\omega_n)B}{\tilde{\omega}_n^2+B^2+\gamma^2(\mathbf{k}) }\tilde{\psi}(\omega_n) \notag \\
&+\frac{|\tilde{\omega}_n|}{\tilde{\omega}_n^2+B^2+\gamma^2(\mathbf{k}) }\tilde{d_x}(\mathbf{k};\omega_n) \notag \\
& -\frac{\mathrm{sgn}(\omega_n)\gamma(\mathbf{k})}{\tilde{\omega}_n^2+B^2+\gamma^2(\mathbf{k})}\tilde{d_y}(\mathbf{k};\omega_n),
\end{align}

with
\begin{align}
\hat{\gamma}(\mathbf{k})\langle \hat{\gamma}(\mathbf{k})f_x(\mathbf{k}';\omega_n)\rangle =\frac{|\tilde{\omega_n}|}{|\tilde{\omega_n}|^2+B^2+\Delta^2}d_x(\mathbf{k}).
\end{align}
This yields a pair-breaking strength $\alpha = \Gamma+i\sqrt{B^2+\Delta^2}$. The $d_x(\mathbf{k})$ component is the most sensitive order parameter.

\section{Repulsion in the triplet channel \label{app:repulsion}}

In Ref. \cite{Mockli2019}, we showed that in the clean case, repulsion in the triplet channel suppresses the critical field. For completeness, here we include the effect of impurities, see Fig. \ref{fig:repulsion}.
The pair-breaking equation is given by Eq. \eqref{eq:pairbreaking_disordered} with the replacement 
\begin{align}
\ln\left(\frac{T}{T_\mathrm{ct}} \right )& \rightarrow -\frac{1}{\lambda_\mathrm{t}}-\pi T\sum_{n=-n_\mathrm{c}-1}^{n_\mathrm{c}} \frac{1}{|\omega_n|} \notag \\ & =-\frac{1}{\lambda_\mathrm{t}}-\ln\left(\frac{2e^\gamma}{\pi}\frac{\epsilon_\mathrm{c}}{T} \right ),
\end{align}
where the dimensionless triplet coupling constant is positive (repulsive) $\lambda_\mathrm{t}>0$.

\twocolumngrid

\bibliographystyle{apsrev4-2}
\bibliography{biblio}

\begin{thebibliography}{68}%
\makeatletter
\providecommand \@ifxundefined [1]{%
 \@ifx{#1\undefined}
}%
\providecommand \@ifnum [1]{%
 \ifnum #1\expandafter \@firstoftwo
 \else \expandafter \@secondoftwo
 \fi
}%
\providecommand \@ifx [1]{%
 \ifx #1\expandafter \@firstoftwo
 \else \expandafter \@secondoftwo
 \fi
}%
\providecommand \natexlab [1]{#1}%
\providecommand \enquote  [1]{``#1''}%
\providecommand \bibnamefont  [1]{#1}%
\providecommand \bibfnamefont [1]{#1}%
\providecommand \citenamefont [1]{#1}%
\providecommand \href@noop [0]{\@secondoftwo}%
\providecommand \href [0]{\begingroup \@sanitize@url \@href}%
\providecommand \@href[1]{\@@startlink{#1}\@@href}%
\providecommand \@@href[1]{\endgroup#1\@@endlink}%
\providecommand \@sanitize@url [0]{\catcode `\\12\catcode `\$12\catcode
  `\&12\catcode `\#12\catcode `\^12\catcode `\_12\catcode `\%12\relax}%
\providecommand \@@startlink[1]{}%
\providecommand \@@endlink[0]{}%
\providecommand \url  [0]{\begingroup\@sanitize@url \@url }%
\providecommand \@url [1]{\endgroup\@href {#1}{\urlprefix }}%
\providecommand \urlprefix  [0]{URL }%
\providecommand \Eprint [0]{\href }%
\providecommand \doibase [0]{https://doi.org/}%
\providecommand \selectlanguage [0]{\@gobble}%
\providecommand \bibinfo  [0]{\@secondoftwo}%
\providecommand \bibfield  [0]{\@secondoftwo}%
\providecommand \translation [1]{[#1]}%
\providecommand \BibitemOpen [0]{}%
\providecommand \bibitemStop [0]{}%
\providecommand \bibitemNoStop [0]{.\EOS\space}%
\providecommand \EOS [0]{\spacefactor3000\relax}%
\providecommand \BibitemShut  [1]{\csname bibitem#1\endcsname}%
\let\auto@bib@innerbib\@empty
\bibitem [{\citenamefont {Geim}\ and\ \citenamefont
  {Grigorieva}(2013)}]{Geim2013}%
  \BibitemOpen
  \bibfield  {author} {\bibinfo {author} {\bibfnamefont {A.~K.}\ \bibnamefont
  {Geim}}\ and\ \bibinfo {author} {\bibfnamefont {I.~V.}\ \bibnamefont
  {Grigorieva}},\ }\href {https://doi.org/10.1038/nature12385} {\bibfield
  {journal} {\bibinfo  {journal} {Nature}\ }\textbf {\bibinfo {volume} {499}},\
  \bibinfo {pages} {419} (\bibinfo {year} {2013})}\BibitemShut {NoStop}%
\bibitem [{\citenamefont {Ugeda}\ \emph {et~al.}(2016)\citenamefont {Ugeda},
  \citenamefont {Bradley}, \citenamefont {Zhang}, \citenamefont {Onishi},
  \citenamefont {Chen}, \citenamefont {Ruan}, \citenamefont
  {Ojeda-Aristizabal}, \citenamefont {Ryu}, \citenamefont {Edmonds},
  \citenamefont {Tsai}, \citenamefont {Riss}, \citenamefont {Mo}, \citenamefont
  {Lee}, \citenamefont {Zettl}, \citenamefont {Hussain}, \citenamefont {Shen},\
  and\ \citenamefont {Crommie}}]{Ugeda2015}%
  \BibitemOpen
  \bibfield  {author} {\bibinfo {author} {\bibfnamefont {M.~M.}\ \bibnamefont
  {Ugeda}}, \bibinfo {author} {\bibfnamefont {A.~J.}\ \bibnamefont {Bradley}},
  \bibinfo {author} {\bibfnamefont {Y.}~\bibnamefont {Zhang}}, \bibinfo
  {author} {\bibfnamefont {S.}~\bibnamefont {Onishi}}, \bibinfo {author}
  {\bibfnamefont {Y.}~\bibnamefont {Chen}}, \bibinfo {author} {\bibfnamefont
  {W.}~\bibnamefont {Ruan}}, \bibinfo {author} {\bibfnamefont {C.}~\bibnamefont
  {Ojeda-Aristizabal}}, \bibinfo {author} {\bibfnamefont {H.}~\bibnamefont
  {Ryu}}, \bibinfo {author} {\bibfnamefont {M.~T.}\ \bibnamefont {Edmonds}},
  \bibinfo {author} {\bibfnamefont {H.-Z.}\ \bibnamefont {Tsai}}, \bibinfo
  {author} {\bibfnamefont {A.}~\bibnamefont {Riss}}, \bibinfo {author}
  {\bibfnamefont {S.-K.}\ \bibnamefont {Mo}}, \bibinfo {author} {\bibfnamefont
  {D.}~\bibnamefont {Lee}}, \bibinfo {author} {\bibfnamefont {A.}~\bibnamefont
  {Zettl}}, \bibinfo {author} {\bibfnamefont {Z.}~\bibnamefont {Hussain}},
  \bibinfo {author} {\bibfnamefont {Z.-X.}\ \bibnamefont {Shen}},\ and\
  \bibinfo {author} {\bibfnamefont {M.~F.}\ \bibnamefont {Crommie}},\ }\href
  {https://doi.org/10.1038/nphys3527} {\bibfield  {journal} {\bibinfo
  {journal} {Nature Physics}\ }\textbf {\bibinfo {volume} {12}},\ \bibinfo
  {pages} {92} (\bibinfo {year} {2016})}\BibitemShut {NoStop}%
\bibitem [{\citenamefont {Xi}\ \emph {et~al.}(2016)\citenamefont {Xi},
  \citenamefont {Wang}, \citenamefont {Zhao}, \citenamefont {Park},
  \citenamefont {Law}, \citenamefont {Berger}, \citenamefont {Forr{\'{o}}},
  \citenamefont {Shan},\ and\ \citenamefont {Mak}}]{Xi2015}%
  \BibitemOpen
  \bibfield  {author} {\bibinfo {author} {\bibfnamefont {X.}~\bibnamefont
  {Xi}}, \bibinfo {author} {\bibfnamefont {Z.}~\bibnamefont {Wang}}, \bibinfo
  {author} {\bibfnamefont {W.}~\bibnamefont {Zhao}}, \bibinfo {author}
  {\bibfnamefont {J.-H.}\ \bibnamefont {Park}}, \bibinfo {author}
  {\bibfnamefont {K.~T.}\ \bibnamefont {Law}}, \bibinfo {author} {\bibfnamefont
  {H.}~\bibnamefont {Berger}}, \bibinfo {author} {\bibfnamefont
  {L.}~\bibnamefont {Forr{\'{o}}}}, \bibinfo {author} {\bibfnamefont
  {J.}~\bibnamefont {Shan}},\ and\ \bibinfo {author} {\bibfnamefont {K.~F.}\
  \bibnamefont {Mak}},\ }\href {https://doi.org/10.1038/nphys3538} {\bibfield
  {journal} {\bibinfo  {journal} {Nature Physics}\ }\textbf {\bibinfo {volume}
  {12}},\ \bibinfo {pages} {139} (\bibinfo {year} {2016})}\BibitemShut
  {NoStop}%
\bibitem [{\citenamefont {Saito}\ \emph {et~al.}(2016)\citenamefont {Saito},
  \citenamefont {Nakamura}, \citenamefont {Bahramy}, \citenamefont {Kohama},
  \citenamefont {Ye}, \citenamefont {Kasahara}, \citenamefont {Nakagawa},
  \citenamefont {Onga}, \citenamefont {Tokunaga}, \citenamefont {Nojima},
  \citenamefont {Yanase},\ and\ \citenamefont {Iwasa}}]{Saito2016}%
  \BibitemOpen
  \bibfield  {author} {\bibinfo {author} {\bibfnamefont {Y.}~\bibnamefont
  {Saito}}, \bibinfo {author} {\bibfnamefont {Y.}~\bibnamefont {Nakamura}},
  \bibinfo {author} {\bibfnamefont {M.~S.}\ \bibnamefont {Bahramy}}, \bibinfo
  {author} {\bibfnamefont {Y.}~\bibnamefont {Kohama}}, \bibinfo {author}
  {\bibfnamefont {J.}~\bibnamefont {Ye}}, \bibinfo {author} {\bibfnamefont
  {Y.}~\bibnamefont {Kasahara}}, \bibinfo {author} {\bibfnamefont
  {Y.}~\bibnamefont {Nakagawa}}, \bibinfo {author} {\bibfnamefont
  {M.}~\bibnamefont {Onga}}, \bibinfo {author} {\bibfnamefont {M.}~\bibnamefont
  {Tokunaga}}, \bibinfo {author} {\bibfnamefont {T.}~\bibnamefont {Nojima}},
  \bibinfo {author} {\bibfnamefont {Y.}~\bibnamefont {Yanase}},\ and\ \bibinfo
  {author} {\bibfnamefont {Y.}~\bibnamefont {Iwasa}},\ }\href
  {https://doi.org/10.1038/nphys3580} {\bibfield  {journal} {\bibinfo
  {journal} {Nature Physics}\ }\textbf {\bibinfo {volume} {12}},\ \bibinfo
  {pages} {144} (\bibinfo {year} {2016})}\BibitemShut {NoStop}%
\bibitem [{\citenamefont {Dvir}\ \emph {et~al.}(2018)\citenamefont {Dvir},
  \citenamefont {Massee}, \citenamefont {Attias}, \citenamefont {Khodas},
  \citenamefont {Aprili}, \citenamefont {Quay},\ and\ \citenamefont
  {Steinberg}}]{Dvir2017}%
  \BibitemOpen
  \bibfield  {author} {\bibinfo {author} {\bibfnamefont {T.}~\bibnamefont
  {Dvir}}, \bibinfo {author} {\bibfnamefont {F.}~\bibnamefont {Massee}},
  \bibinfo {author} {\bibfnamefont {L.}~\bibnamefont {Attias}}, \bibinfo
  {author} {\bibfnamefont {M.}~\bibnamefont {Khodas}}, \bibinfo {author}
  {\bibfnamefont {M.}~\bibnamefont {Aprili}}, \bibinfo {author} {\bibfnamefont
  {C.~H.~L.}\ \bibnamefont {Quay}},\ and\ \bibinfo {author} {\bibfnamefont
  {H.}~\bibnamefont {Steinberg}},\ }\href
  {https://doi.org/10.1038/s41467-018-03000-w} {\bibfield  {journal} {\bibinfo
  {journal} {Nature Communications}\ }\textbf {\bibinfo {volume} {9}},\
  \bibinfo {pages} {598} (\bibinfo {year} {2018})}\BibitemShut {NoStop}%
\bibitem [{\citenamefont {Liu}\ \emph {et~al.}(2018)\citenamefont {Liu},
  \citenamefont {Wang}, \citenamefont {Zhang}, \citenamefont {Liu},
  \citenamefont {Liu}, \citenamefont {Zhou}, \citenamefont {Wang},
  \citenamefont {Wang}, \citenamefont {Liu}, \citenamefont {Xi}, \citenamefont
  {Tian}, \citenamefont {Liu}, \citenamefont {Feng}, \citenamefont {Xie},\ and\
  \citenamefont {Wang}}]{Liu2018}%
  \BibitemOpen
  \bibfield  {author} {\bibinfo {author} {\bibfnamefont {Y.}~\bibnamefont
  {Liu}}, \bibinfo {author} {\bibfnamefont {Z.}~\bibnamefont {Wang}}, \bibinfo
  {author} {\bibfnamefont {X.}~\bibnamefont {Zhang}}, \bibinfo {author}
  {\bibfnamefont {C.}~\bibnamefont {Liu}}, \bibinfo {author} {\bibfnamefont
  {Y.}~\bibnamefont {Liu}}, \bibinfo {author} {\bibfnamefont {Z.}~\bibnamefont
  {Zhou}}, \bibinfo {author} {\bibfnamefont {J.}~\bibnamefont {Wang}}, \bibinfo
  {author} {\bibfnamefont {Q.}~\bibnamefont {Wang}}, \bibinfo {author}
  {\bibfnamefont {Y.}~\bibnamefont {Liu}}, \bibinfo {author} {\bibfnamefont
  {C.}~\bibnamefont {Xi}}, \bibinfo {author} {\bibfnamefont {M.}~\bibnamefont
  {Tian}}, \bibinfo {author} {\bibfnamefont {H.}~\bibnamefont {Liu}}, \bibinfo
  {author} {\bibfnamefont {J.}~\bibnamefont {Feng}}, \bibinfo {author}
  {\bibfnamefont {X.~C.}\ \bibnamefont {Xie}},\ and\ \bibinfo {author}
  {\bibfnamefont {J.}~\bibnamefont {Wang}},\ }\href
  {https://doi.org/10.1103/PhysRevX.8.021002} {\bibfield  {journal} {\bibinfo
  {journal} {Physical Review X}\ }\textbf {\bibinfo {volume} {8}},\ \bibinfo
  {pages} {021002} (\bibinfo {year} {2018})}\BibitemShut {NoStop}%
\bibitem [{\citenamefont {Sohn}\ \emph {et~al.}(2018)\citenamefont {Sohn},
  \citenamefont {Xi}, \citenamefont {He}, \citenamefont {Jiang}, \citenamefont
  {Wang}, \citenamefont {Kang}, \citenamefont {Park}, \citenamefont {Berger},
  \citenamefont {Forr{\'{o}}}, \citenamefont {Law}, \citenamefont {Shan},\ and\
  \citenamefont {Mak}}]{Sohn2018}%
  \BibitemOpen
  \bibfield  {author} {\bibinfo {author} {\bibfnamefont {E.}~\bibnamefont
  {Sohn}}, \bibinfo {author} {\bibfnamefont {X.}~\bibnamefont {Xi}}, \bibinfo
  {author} {\bibfnamefont {W.-Y.}\ \bibnamefont {He}}, \bibinfo {author}
  {\bibfnamefont {S.}~\bibnamefont {Jiang}}, \bibinfo {author} {\bibfnamefont
  {Z.}~\bibnamefont {Wang}}, \bibinfo {author} {\bibfnamefont {K.}~\bibnamefont
  {Kang}}, \bibinfo {author} {\bibfnamefont {J.-H.}\ \bibnamefont {Park}},
  \bibinfo {author} {\bibfnamefont {H.}~\bibnamefont {Berger}}, \bibinfo
  {author} {\bibfnamefont {L.}~\bibnamefont {Forr{\'{o}}}}, \bibinfo {author}
  {\bibfnamefont {K.~T.}\ \bibnamefont {Law}}, \bibinfo {author} {\bibfnamefont
  {J.}~\bibnamefont {Shan}},\ and\ \bibinfo {author} {\bibfnamefont {K.~F.}\
  \bibnamefont {Mak}},\ }\href {https://doi.org/10.1038/s41563-018-0061-1}
  {\bibfield  {journal} {\bibinfo  {journal} {Nature Materials}\ }\textbf
  {\bibinfo {volume} {17}},\ \bibinfo {pages} {504} (\bibinfo {year}
  {2018})}\BibitemShut {NoStop}%
\bibitem [{\citenamefont {Nakata}\ \emph {et~al.}(2018)\citenamefont {Nakata},
  \citenamefont {Sugawara}, \citenamefont {Ichinokura}, \citenamefont {Okada},
  \citenamefont {Hitosugi}, \citenamefont {Koretsune}, \citenamefont {Ueno},
  \citenamefont {Hasegawa}, \citenamefont {Takahashi},\ and\ \citenamefont
  {Sato}}]{Nakata2018}%
  \BibitemOpen
  \bibfield  {author} {\bibinfo {author} {\bibfnamefont {Y.}~\bibnamefont
  {Nakata}}, \bibinfo {author} {\bibfnamefont {K.}~\bibnamefont {Sugawara}},
  \bibinfo {author} {\bibfnamefont {S.}~\bibnamefont {Ichinokura}}, \bibinfo
  {author} {\bibfnamefont {Y.}~\bibnamefont {Okada}}, \bibinfo {author}
  {\bibfnamefont {T.}~\bibnamefont {Hitosugi}}, \bibinfo {author}
  {\bibfnamefont {T.}~\bibnamefont {Koretsune}}, \bibinfo {author}
  {\bibfnamefont {K.}~\bibnamefont {Ueno}}, \bibinfo {author} {\bibfnamefont
  {S.}~\bibnamefont {Hasegawa}}, \bibinfo {author} {\bibfnamefont
  {T.}~\bibnamefont {Takahashi}},\ and\ \bibinfo {author} {\bibfnamefont
  {T.}~\bibnamefont {Sato}},\ }\href
  {https://doi.org/10.1038/s41699-018-0057-3} {\bibfield  {journal} {\bibinfo
  {journal} {npj 2D Materials and Applications}\ }\textbf {\bibinfo {volume}
  {2}},\ \bibinfo {pages} {12} (\bibinfo {year} {2018})}\BibitemShut {NoStop}%
\bibitem [{\citenamefont {de~la Barrera}\ \emph {et~al.}(2018)\citenamefont
  {de~la Barrera}, \citenamefont {Sinko}, \citenamefont {Gopalan},
  \citenamefont {Sivadas}, \citenamefont {Seyler}, \citenamefont {Watanabe},
  \citenamefont {Taniguchi}, \citenamefont {Tsen}, \citenamefont {Xu},
  \citenamefont {Xiao},\ and\ \citenamefont {Hunt}}]{DelaBarrera2018}%
  \BibitemOpen
  \bibfield  {author} {\bibinfo {author} {\bibfnamefont {S.~C.}\ \bibnamefont
  {de~la Barrera}}, \bibinfo {author} {\bibfnamefont {M.~R.}\ \bibnamefont
  {Sinko}}, \bibinfo {author} {\bibfnamefont {D.~P.}\ \bibnamefont {Gopalan}},
  \bibinfo {author} {\bibfnamefont {N.}~\bibnamefont {Sivadas}}, \bibinfo
  {author} {\bibfnamefont {K.~L.}\ \bibnamefont {Seyler}}, \bibinfo {author}
  {\bibfnamefont {K.}~\bibnamefont {Watanabe}}, \bibinfo {author}
  {\bibfnamefont {T.}~\bibnamefont {Taniguchi}}, \bibinfo {author}
  {\bibfnamefont {A.~W.}\ \bibnamefont {Tsen}}, \bibinfo {author}
  {\bibfnamefont {X.}~\bibnamefont {Xu}}, \bibinfo {author} {\bibfnamefont
  {D.}~\bibnamefont {Xiao}},\ and\ \bibinfo {author} {\bibfnamefont {B.~M.}\
  \bibnamefont {Hunt}},\ }\href {https://doi.org/10.1038/s41467-018-03888-4}
  {\bibfield  {journal} {\bibinfo  {journal} {Nature Communications}\ }\textbf
  {\bibinfo {volume} {9}},\ \bibinfo {pages} {1427} (\bibinfo {year}
  {2018})}\BibitemShut {NoStop}%
\bibitem [{\citenamefont {Fulde}\ and\ \citenamefont
  {Ferrell}(1964)}]{Fulde1964}%
  \BibitemOpen
  \bibfield  {author} {\bibinfo {author} {\bibfnamefont {P.}~\bibnamefont
  {Fulde}}\ and\ \bibinfo {author} {\bibfnamefont {R.~A.}\ \bibnamefont
  {Ferrell}},\ }\href {https://doi.org/10.1103/PhysRev.135.A550} {\bibfield
  {journal} {\bibinfo  {journal} {Physical Review}\ }\textbf {\bibinfo {volume}
  {135}},\ \bibinfo {pages} {A550} (\bibinfo {year} {1964})}\BibitemShut
  {NoStop}%
\bibitem [{\citenamefont {Shimozawa}\ \emph {et~al.}(2016)\citenamefont
  {Shimozawa}, \citenamefont {Goh}, \citenamefont {Shibauchi},\ and\
  \citenamefont {Matsuda}}]{Shimozawa2016}%
  \BibitemOpen
  \bibfield  {author} {\bibinfo {author} {\bibfnamefont {M.}~\bibnamefont
  {Shimozawa}}, \bibinfo {author} {\bibfnamefont {S.~K.}\ \bibnamefont {Goh}},
  \bibinfo {author} {\bibfnamefont {T.}~\bibnamefont {Shibauchi}},\ and\
  \bibinfo {author} {\bibfnamefont {Y.}~\bibnamefont {Matsuda}},\ }\href@noop
  {} {\bibfield  {journal} {\bibinfo  {journal} {Reports on Progress in
  Physics}\ }\textbf {\bibinfo {volume} {79}} (\bibinfo {year}
  {2016})}\BibitemShut {NoStop}%
\bibitem [{\citenamefont {Smidman}\ \emph {et~al.}(2017)\citenamefont
  {Smidman}, \citenamefont {Salamon}, \citenamefont {Yuan},\ and\ \citenamefont
  {Agterberg}}]{Smidman2017}%
  \BibitemOpen
  \bibfield  {author} {\bibinfo {author} {\bibfnamefont {M.}~\bibnamefont
  {Smidman}}, \bibinfo {author} {\bibfnamefont {M.~B.}\ \bibnamefont
  {Salamon}}, \bibinfo {author} {\bibfnamefont {H.~Q.}\ \bibnamefont {Yuan}},\
  and\ \bibinfo {author} {\bibfnamefont {D.~F.}\ \bibnamefont {Agterberg}},\
  }\href {https://doi.org/10.1088/1361-6633/80/3/036501} {\bibfield  {journal}
  {\bibinfo  {journal} {Reports on Progress in Physics}\ }\textbf {\bibinfo
  {volume} {80}},\ \bibinfo {pages} {036501} (\bibinfo {year}
  {2017})}\BibitemShut {NoStop}%
\bibitem [{\citenamefont {Gor'kov}\ and\ \citenamefont
  {Rashba}(2001)}]{Gorkov2001}%
  \BibitemOpen
  \bibfield  {author} {\bibinfo {author} {\bibfnamefont {L.~P.}\ \bibnamefont
  {Gor'kov}}\ and\ \bibinfo {author} {\bibfnamefont {E.~I.}\ \bibnamefont
  {Rashba}},\ }\href {https://doi.org/10.1103/PhysRevLett.87.037004} {\bibfield
   {journal} {\bibinfo  {journal} {Physical Review Letters}\ }\textbf {\bibinfo
  {volume} {87}},\ \bibinfo {pages} {037004} (\bibinfo {year}
  {2001})}\BibitemShut {NoStop}%
\bibitem [{\citenamefont {Frigeri}\ \emph
  {et~al.}(2004{\natexlab{a}})\citenamefont {Frigeri}, \citenamefont
  {Agterberg}, \citenamefont {Koga},\ and\ \citenamefont
  {Sigrist}}]{Frigeri2004d}%
  \BibitemOpen
  \bibfield  {author} {\bibinfo {author} {\bibfnamefont {P.~A.}\ \bibnamefont
  {Frigeri}}, \bibinfo {author} {\bibfnamefont {D.~F.}\ \bibnamefont
  {Agterberg}}, \bibinfo {author} {\bibfnamefont {A.}~\bibnamefont {Koga}},\
  and\ \bibinfo {author} {\bibfnamefont {M.}~\bibnamefont {Sigrist}},\ }\href
  {https://doi.org/10.1103/PhysRevLett.92.097001} {\bibfield  {journal}
  {\bibinfo  {journal} {Physical Review Letters}\ }\textbf {\bibinfo {volume}
  {92}},\ \bibinfo {pages} {097001} (\bibinfo {year}
  {2004}{\natexlab{a}})}\BibitemShut {NoStop}%
\bibitem [{\citenamefont {Bulaevskii}\ \emph {et~al.}(1976)\citenamefont
  {Bulaevskii}, \citenamefont {Guseinov},\ and\ \citenamefont
  {Rusinov}}]{Bulaevskii976}%
  \BibitemOpen
  \bibfield  {author} {\bibinfo {author} {\bibfnamefont {L.}~\bibnamefont
  {Bulaevskii}}, \bibinfo {author} {\bibfnamefont {A.}~\bibnamefont
  {Guseinov}},\ and\ \bibinfo {author} {\bibfnamefont {A.}~\bibnamefont
  {Rusinov}},\ }\href@noop {} {\bibfield  {journal} {\bibinfo  {journal} {Zh.
  Eksp. Teor. Fiz.}\ }\textbf {\bibinfo {volume} {71}},\ \bibinfo {pages}
  {2356} (\bibinfo {year} {1976})}\BibitemShut {NoStop}%
\bibitem [{\citenamefont {Samokhin}(2008)}]{Samokhin2008}%
  \BibitemOpen
  \bibfield  {author} {\bibinfo {author} {\bibfnamefont {K.~V.}\ \bibnamefont
  {Samokhin}},\ }\href {https://doi.org/10.1103/PhysRevB.78.224520} {\bibfield
  {journal} {\bibinfo  {journal} {Physical Review B}\ }\textbf {\bibinfo
  {volume} {78}},\ \bibinfo {pages} {224520} (\bibinfo {year}
  {2008})}\BibitemShut {NoStop}%
\bibitem [{\citenamefont {Ili{\'{c}}}\ \emph {et~al.}(2017)\citenamefont
  {Ili{\'{c}}}, \citenamefont {Meyer},\ and\ \citenamefont
  {Houzet}}]{Ilic2017}%
  \BibitemOpen
  \bibfield  {author} {\bibinfo {author} {\bibfnamefont {S.}~\bibnamefont
  {Ili{\'{c}}}}, \bibinfo {author} {\bibfnamefont {J.~S.}\ \bibnamefont
  {Meyer}},\ and\ \bibinfo {author} {\bibfnamefont {M.}~\bibnamefont
  {Houzet}},\ }\href {https://doi.org/10.1103/PhysRevLett.119.117001}
  {\bibfield  {journal} {\bibinfo  {journal} {Physical Review Letters}\
  }\textbf {\bibinfo {volume} {119}},\ \bibinfo {pages} {117001} (\bibinfo
  {year} {2017})}\BibitemShut {NoStop}%
\bibitem [{\citenamefont {M{\"{o}}ckli}\ and\ \citenamefont
  {Khodas}(2018)}]{Mockli2018}%
  \BibitemOpen
  \bibfield  {author} {\bibinfo {author} {\bibfnamefont {D.}~\bibnamefont
  {M{\"{o}}ckli}}\ and\ \bibinfo {author} {\bibfnamefont {M.}~\bibnamefont
  {Khodas}},\ }\href {https://doi.org/10.1103/PhysRevB.98.144518} {\bibfield
  {journal} {\bibinfo  {journal} {Physical Review B}\ }\textbf {\bibinfo
  {volume} {98}},\ \bibinfo {pages} {144518} (\bibinfo {year}
  {2018})}\BibitemShut {NoStop}%
\bibitem [{\citenamefont {M{\"{o}}ckli}\ and\ \citenamefont
  {Khodas}(2019)}]{Mockli2019}%
  \BibitemOpen
  \bibfield  {author} {\bibinfo {author} {\bibfnamefont {D.}~\bibnamefont
  {M{\"{o}}ckli}}\ and\ \bibinfo {author} {\bibfnamefont {M.}~\bibnamefont
  {Khodas}},\ }\href {https://doi.org/10.1103/PhysRevB.99.180505} {\bibfield
  {journal} {\bibinfo  {journal} {Physical Review B}\ }\textbf {\bibinfo
  {volume} {99}},\ \bibinfo {pages} {180505(R)} (\bibinfo {year}
  {2019})}\BibitemShut {NoStop}%
\bibitem [{\citenamefont {Sosenko}\ \emph {et~al.}(2017)\citenamefont
  {Sosenko}, \citenamefont {Zhang},\ and\ \citenamefont {Aji}}]{Sosenko2017}%
  \BibitemOpen
  \bibfield  {author} {\bibinfo {author} {\bibfnamefont {E.}~\bibnamefont
  {Sosenko}}, \bibinfo {author} {\bibfnamefont {J.}~\bibnamefont {Zhang}},\
  and\ \bibinfo {author} {\bibfnamefont {V.}~\bibnamefont {Aji}},\ }\href
  {https://doi.org/10.1103/PhysRevB.95.144508} {\bibfield  {journal} {\bibinfo
  {journal} {Phys. Rev. B}\ }\textbf {\bibinfo {volume} {95}},\ \bibinfo
  {pages} {144508} (\bibinfo {year} {2017})}\BibitemShut {NoStop}%
\bibitem [{\citenamefont {Bauer}\ and\ \citenamefont
  {Sigrist}(2012)}]{Bauer2012}%
  \BibitemOpen
  \bibinfo {editor} {\bibfnamefont {E.}~\bibnamefont {Bauer}}\ and\ \bibinfo
  {editor} {\bibfnamefont {M.}~\bibnamefont {Sigrist}},\ eds.,\ \href
  {https://doi.org/10.1007/978-3-642-24624-1} {\emph {\bibinfo {title}
  {{Non-Centrosymmetric Superconductors}}}},\ \bibinfo {series} {Lecture Notes
  in Physics}, Vol.\ \bibinfo {volume} {847}\ (\bibinfo  {publisher} {Springer
  Berlin Heidelberg},\ \bibinfo {year} {2012})\BibitemShut {NoStop}%
\bibitem [{\citenamefont {Balian}\ and\ \citenamefont
  {Werthamer}(1963)}]{Balian963}%
  \BibitemOpen
  \bibfield  {author} {\bibinfo {author} {\bibfnamefont {R.}~\bibnamefont
  {Balian}}\ and\ \bibinfo {author} {\bibfnamefont {N.~R.}\ \bibnamefont
  {Werthamer}},\ }\href {https://doi.org/10.1103/PhysRev.131.1553} {\bibfield
  {journal} {\bibinfo  {journal} {Physical Review}\ }\textbf {\bibinfo {volume}
  {131}},\ \bibinfo {pages} {1553} (\bibinfo {year} {1963})}\BibitemShut
  {NoStop}%
\bibitem [{\citenamefont {Sigrist}\ and\ \citenamefont
  {Ueda}(1991)}]{Sigrist1991}%
  \BibitemOpen
  \bibfield  {author} {\bibinfo {author} {\bibfnamefont {M.}~\bibnamefont
  {Sigrist}}\ and\ \bibinfo {author} {\bibfnamefont {K.}~\bibnamefont {Ueda}},\
  }\href {https://doi.org/10.1103/RevModPhys.63.239} {\bibfield  {journal}
  {\bibinfo  {journal} {Reviews of Modern Physics}\ }\textbf {\bibinfo {volume}
  {63}},\ \bibinfo {pages} {239} (\bibinfo {year} {1991})}\BibitemShut
  {NoStop}%
\bibitem [{\citenamefont {Yip}(2014)}]{Yip2014a}%
  \BibitemOpen
  \bibfield  {author} {\bibinfo {author} {\bibfnamefont {S.}~\bibnamefont
  {Yip}},\ }\href {https://doi.org/10.1146/annurev-conmatphys-031113-133912}
  {\bibfield  {journal} {\bibinfo  {journal} {Annual Review of Condensed Matter
  Physics}\ }\textbf {\bibinfo {volume} {5}},\ \bibinfo {pages} {15} (\bibinfo
  {year} {2014})}\BibitemShut {NoStop}%
\bibitem [{\citenamefont {Ramires}\ and\ \citenamefont
  {Sigrist}(2016)}]{Ramires2016}%
  \BibitemOpen
  \bibfield  {author} {\bibinfo {author} {\bibfnamefont {A.}~\bibnamefont
  {Ramires}}\ and\ \bibinfo {author} {\bibfnamefont {M.}~\bibnamefont
  {Sigrist}},\ }\href {https://doi.org/10.1103/PhysRevB.94.104501} {\bibfield
  {journal} {\bibinfo  {journal} {Physical Review B}\ }\textbf {\bibinfo
  {volume} {94}},\ \bibinfo {pages} {104501} (\bibinfo {year}
  {2016})}\BibitemShut {NoStop}%
\bibitem [{\citenamefont {Ramires}\ \emph {et~al.}(2018)\citenamefont
  {Ramires}, \citenamefont {Agterberg},\ and\ \citenamefont
  {Sigrist}}]{aline_fitness2}%
  \BibitemOpen
  \bibfield  {author} {\bibinfo {author} {\bibfnamefont {A.}~\bibnamefont
  {Ramires}}, \bibinfo {author} {\bibfnamefont {D.~F.}\ \bibnamefont
  {Agterberg}},\ and\ \bibinfo {author} {\bibfnamefont {M.}~\bibnamefont
  {Sigrist}},\ }\href {https://doi.org/10.1103/PhysRevB.98.024501} {\bibfield
  {journal} {\bibinfo  {journal} {Phys. Rev. B}\ }\textbf {\bibinfo {volume}
  {98}},\ \bibinfo {pages} {024501} (\bibinfo {year} {2018})}\BibitemShut
  {NoStop}%
\bibitem [{\citenamefont {Anderson}(1959)}]{Anderson1959}%
  \BibitemOpen
  \bibfield  {author} {\bibinfo {author} {\bibfnamefont {P.}~\bibnamefont
  {Anderson}},\ }\href {https://doi.org/10.1016/0022-3697(59)90036-8}
  {\bibfield  {journal} {\bibinfo  {journal} {Journal of Physics and Chemistry
  of Solids}\ }\textbf {\bibinfo {volume} {11}},\ \bibinfo {pages} {26}
  (\bibinfo {year} {1959})}\BibitemShut {NoStop}%
\bibitem [{\citenamefont {Mackenzie}\ \emph {et~al.}(1998)\citenamefont
  {Mackenzie}, \citenamefont {Haselwimmer}, \citenamefont {Tyler},
  \citenamefont {Lonzarich}, \citenamefont {Mori}, \citenamefont {Nishizaki},\
  and\ \citenamefont {Maeno}}]{Mackenzie1998}%
  \BibitemOpen
  \bibfield  {author} {\bibinfo {author} {\bibfnamefont {A.~P.}\ \bibnamefont
  {Mackenzie}}, \bibinfo {author} {\bibfnamefont {R.~K.~W.}\ \bibnamefont
  {Haselwimmer}}, \bibinfo {author} {\bibfnamefont {A.~W.}\ \bibnamefont
  {Tyler}}, \bibinfo {author} {\bibfnamefont {G.~G.}\ \bibnamefont
  {Lonzarich}}, \bibinfo {author} {\bibfnamefont {Y.}~\bibnamefont {Mori}},
  \bibinfo {author} {\bibfnamefont {S.}~\bibnamefont {Nishizaki}},\ and\
  \bibinfo {author} {\bibfnamefont {Y.}~\bibnamefont {Maeno}},\ }\href
  {https://doi.org/10.1103/PhysRevLett.80.161} {\bibfield  {journal} {\bibinfo
  {journal} {Physical Review Letters}\ }\textbf {\bibinfo {volume} {80}},\
  \bibinfo {pages} {161} (\bibinfo {year} {1998})}\BibitemShut {NoStop}%
\bibitem [{\citenamefont {Kopnin}(2001)}]{kopnin2001}%
  \BibitemOpen
  \bibfield  {author} {\bibinfo {author} {\bibfnamefont {N.}~\bibnamefont
  {Kopnin}},\ }\href@noop {} {\emph {\bibinfo {title} {Theory of Nonequilibrium
  Superconductivity}}},\ International Series of Monographs on Physics\
  (\bibinfo  {publisher} {Clarendon Press},\ \bibinfo {year}
  {2001})\BibitemShut {NoStop}%
\bibitem [{\citenamefont {Bruus}\ and\ \citenamefont
  {Flensberg}(2004)}]{bruus2004many}%
  \BibitemOpen
  \bibfield  {author} {\bibinfo {author} {\bibfnamefont {H.}~\bibnamefont
  {Bruus}}\ and\ \bibinfo {author} {\bibnamefont {Flensberg}},\ }\href@noop {}
  {\emph {\bibinfo {title} {Many-Body Quantum Theory in Condensed Matter
  Physics: An Introduction}}},\ Oxford Graduate Texts\ (\bibinfo  {publisher}
  {OUP Oxford},\ \bibinfo {year} {2004})\BibitemShut {NoStop}%
\bibitem [{\citenamefont {Kita}(2015)}]{Kita2015}%
  \BibitemOpen
  \bibfield  {author} {\bibinfo {author} {\bibfnamefont {T.}~\bibnamefont
  {Kita}},\ }\bibinfo {title} {{Gor'kov, Eilenberger, and Ginzburg-Landau
  Equations}},\ in\ \href {https://doi.org/10.1007/978-4-431-55405-9_14} {\emph
  {\bibinfo {booktitle} {Statistical Mechanics of Superconductivity}}}\
  (\bibinfo  {publisher} {Springer Japan},\ \bibinfo {address} {Tokyo},\
  \bibinfo {year} {2015})\ pp.\ \bibinfo {pages} {201--227}\BibitemShut
  {NoStop}%
\bibitem [{\citenamefont {Eilenberger}(1968)}]{Eilenberger1968}%
  \BibitemOpen
  \bibfield  {author} {\bibinfo {author} {\bibfnamefont {G.}~\bibnamefont
  {Eilenberger}},\ }\href {https://doi.org/10.1007/BF01379803} {\bibfield
  {journal} {\bibinfo  {journal} {Zeitschrift f{\"{u}}r Physik A Hadrons and
  nuclei}\ }\textbf {\bibinfo {volume} {214}},\ \bibinfo {pages} {195}
  (\bibinfo {year} {1968})}\BibitemShut {NoStop}%
\bibitem [{\citenamefont {Larkin}\ and\ \citenamefont
  {Ovchinnikov}(1969)}]{Larkin}%
  \BibitemOpen
  \bibfield  {author} {\bibinfo {author} {\bibfnamefont {A.}~\bibnamefont
  {Larkin}}\ and\ \bibinfo {author} {\bibfnamefont {Y.~N.}\ \bibnamefont
  {Ovchinnikov}},\ }\href
  {http://www.jetp.ac.ru/cgi-bin/e/index/e/28/6/p1200?a=list} {\bibfield
  {journal} {\bibinfo  {journal} {Journal of Experimental and Theoretical
  Physics}\ }\textbf {\bibinfo {volume} {28}} (\bibinfo {year}
  {1969})}\BibitemShut {NoStop}%
\bibitem [{\citenamefont {Tinkham}(2004)}]{tinkham}%
  \BibitemOpen
  \bibfield  {author} {\bibinfo {author} {\bibfnamefont {M.}~\bibnamefont
  {Tinkham}},\ }\href@noop {} {\emph {\bibinfo {title} {Introduction to
  Superconductivity: Second Edition}}},\ Dover Books on Physics\ (\bibinfo
  {publisher} {Dover Publications},\ \bibinfo {year} {2004})\BibitemShut
  {NoStop}%
\bibitem [{\citenamefont {Clogston}(1962)}]{Clogston1962}%
  \BibitemOpen
  \bibfield  {author} {\bibinfo {author} {\bibfnamefont {A.~M.}\ \bibnamefont
  {Clogston}},\ }\href {https://doi.org/10.1103/PhysRevLett.9.266} {\bibfield
  {journal} {\bibinfo  {journal} {Physical Review Letters}\ }\textbf {\bibinfo
  {volume} {9}},\ \bibinfo {pages} {266} (\bibinfo {year} {1962})}\BibitemShut
  {NoStop}%
\bibitem [{\citenamefont {Saint-James}\ \emph {et~al.}(1969)\citenamefont
  {Saint-James}, \citenamefont {Sarma},\ and\ \citenamefont
  {Thomas}}]{saint1969type}%
  \BibitemOpen
  \bibfield  {author} {\bibinfo {author} {\bibfnamefont {D.}~\bibnamefont
  {Saint-James}}, \bibinfo {author} {\bibfnamefont {G.}~\bibnamefont {Sarma}},\
  and\ \bibinfo {author} {\bibfnamefont {E.}~\bibnamefont {Thomas}},\ }\href
  {https://books.google.com.br/books?id=qlhjuQAACAAJ} {\emph {\bibinfo {title}
  {Type II Superconductivity}}},\ Commonwealth and International Library.
  Liberal Studies Divi\ (\bibinfo  {publisher} {Elsevier Science \&
  Technology},\ \bibinfo {year} {1969})\BibitemShut {NoStop}%
\bibitem [{\citenamefont {Frigeri}\ \emph
  {et~al.}(2004{\natexlab{b}})\citenamefont {Frigeri}, \citenamefont
  {Agterberg},\ and\ \citenamefont {Sigrist}}]{Frigeri2004}%
  \BibitemOpen
  \bibfield  {author} {\bibinfo {author} {\bibfnamefont {P.~A.}\ \bibnamefont
  {Frigeri}}, \bibinfo {author} {\bibfnamefont {D.~F.}\ \bibnamefont
  {Agterberg}},\ and\ \bibinfo {author} {\bibfnamefont {M.}~\bibnamefont
  {Sigrist}},\ }\href {https://doi.org/10.1088/1367-2630/6/1/115} {\bibfield
  {journal} {\bibinfo  {journal} {New Journal of Physics}\ }\textbf {\bibinfo
  {volume} {6}},\ \bibinfo {pages} {115} (\bibinfo {year}
  {2004}{\natexlab{b}})}\BibitemShut {NoStop}%
\bibitem [{\citenamefont {Frigeri}\ \emph {et~al.}(2006)\citenamefont
  {Frigeri}, \citenamefont {Agterberg}, \citenamefont {Milat},\ and\
  \citenamefont {Sigrist}}]{Frigeri2006}%
  \BibitemOpen
  \bibfield  {author} {\bibinfo {author} {\bibfnamefont {P.~A.}\ \bibnamefont
  {Frigeri}}, \bibinfo {author} {\bibfnamefont {D.~F.}\ \bibnamefont
  {Agterberg}}, \bibinfo {author} {\bibfnamefont {I.}~\bibnamefont {Milat}},\
  and\ \bibinfo {author} {\bibfnamefont {M.}~\bibnamefont {Sigrist}},\ }\href
  {https://doi.org/10.1140/epjb/e2007-00019-5} {\bibfield  {journal} {\bibinfo
  {journal} {The European Physical Journal B}\ }\textbf {\bibinfo {volume}
  {54}},\ \bibinfo {pages} {435} (\bibinfo {year} {2006})}\BibitemShut
  {NoStop}%
\bibitem [{\citenamefont {Xiao}\ \emph {et~al.}(2012)\citenamefont {Xiao},
  \citenamefont {Liu}, \citenamefont {Feng}, \citenamefont {Xu},\ and\
  \citenamefont {Yao}}]{Xiao2012}%
  \BibitemOpen
  \bibfield  {author} {\bibinfo {author} {\bibfnamefont {D.}~\bibnamefont
  {Xiao}}, \bibinfo {author} {\bibfnamefont {G.-B.}\ \bibnamefont {Liu}},
  \bibinfo {author} {\bibfnamefont {W.}~\bibnamefont {Feng}}, \bibinfo {author}
  {\bibfnamefont {X.}~\bibnamefont {Xu}},\ and\ \bibinfo {author}
  {\bibfnamefont {W.}~\bibnamefont {Yao}},\ }\href
  {https://doi.org/10.1103/PhysRevLett.108.196802} {\bibfield  {journal}
  {\bibinfo  {journal} {Physical Review Letters}\ }\textbf {\bibinfo {volume}
  {108}},\ \bibinfo {pages} {196802} (\bibinfo {year} {2012})}\BibitemShut
  {NoStop}%
\bibitem [{\citenamefont {Radtke}\ \emph {et~al.}(1993)\citenamefont {Radtke},
  \citenamefont {Levin}, \citenamefont {Sch{\"{u}}ttler},\ and\ \citenamefont
  {Norman}}]{Radtke1993}%
  \BibitemOpen
  \bibfield  {author} {\bibinfo {author} {\bibfnamefont {R.~J.}\ \bibnamefont
  {Radtke}}, \bibinfo {author} {\bibfnamefont {K.}~\bibnamefont {Levin}},
  \bibinfo {author} {\bibfnamefont {H.-B.}\ \bibnamefont {Sch{\"{u}}ttler}},\
  and\ \bibinfo {author} {\bibfnamefont {M.~R.}\ \bibnamefont {Norman}},\
  }\href {https://doi.org/10.1103/PhysRevB.48.653} {\bibfield  {journal}
  {\bibinfo  {journal} {Physical Review B}\ }\textbf {\bibinfo {volume} {48}},\
  \bibinfo {pages} {653} (\bibinfo {year} {1993})}\BibitemShut {NoStop}%
\bibitem [{Note1()}]{Note1}%
  \BibitemOpen
  \bibinfo {note} {See Supplemental Material at [URL will be inserted by
  publisher] for an animated version \label {foot}}\BibitemShut {NoStop}%
\bibitem [{\citenamefont {Samokhin}(2004)}]{Samokhin2004}%
  \BibitemOpen
  \bibfield  {author} {\bibinfo {author} {\bibfnamefont {K.~V.}\ \bibnamefont
  {Samokhin}},\ }\href {https://doi.org/10.1103/PhysRevB.70.104521} {\bibfield
  {journal} {\bibinfo  {journal} {Physical Review B}\ }\textbf {\bibinfo
  {volume} {70}},\ \bibinfo {pages} {104521} (\bibinfo {year}
  {2004})}\BibitemShut {NoStop}%
\bibitem [{\citenamefont {Samokhin}(2009)}]{Samokhin2009}%
  \BibitemOpen
  \bibfield  {author} {\bibinfo {author} {\bibfnamefont {K.}~\bibnamefont
  {Samokhin}},\ }\href {https://doi.org/10.1016/j.aop.2009.08.008} {\bibfield
  {journal} {\bibinfo  {journal} {Annals of Physics}\ }\textbf {\bibinfo
  {volume} {324}},\ \bibinfo {pages} {2385} (\bibinfo {year}
  {2009})}\BibitemShut {NoStop}%
\bibitem [{\citenamefont {Samokhin}(2015)}]{Samokhin2015}%
  \BibitemOpen
  \bibfield  {author} {\bibinfo {author} {\bibfnamefont {K.}~\bibnamefont
  {Samokhin}},\ }\href {https://doi.org/10.1016/j.aop.2015.04.024} {\bibfield
  {journal} {\bibinfo  {journal} {Annals of Physics}\ }\textbf {\bibinfo
  {volume} {359}},\ \bibinfo {pages} {385} (\bibinfo {year}
  {2015})}\BibitemShut {NoStop}%
\bibitem [{Note2()}]{Note2}%
  \BibitemOpen
  \bibinfo {note} {The illustration is a corrected version of a similar figure
  in Ref. \cite {Mockli2018}}\BibitemShut {NoStop}%
\bibitem [{\citenamefont {Fischer}\ \emph {et~al.}(2018)\citenamefont
  {Fischer}, \citenamefont {Sigrist},\ and\ \citenamefont
  {Agterberg}}]{Fischer2018}%
  \BibitemOpen
  \bibfield  {author} {\bibinfo {author} {\bibfnamefont {M.~H.}\ \bibnamefont
  {Fischer}}, \bibinfo {author} {\bibfnamefont {M.}~\bibnamefont {Sigrist}},\
  and\ \bibinfo {author} {\bibfnamefont {D.~F.}\ \bibnamefont {Agterberg}},\
  }\href {https://doi.org/10.1103/PhysRevLett.121.157003} {\bibfield  {journal}
  {\bibinfo  {journal} {Physical Review Letters}\ }\textbf {\bibinfo {volume}
  {121}},\ \bibinfo {pages} {157003} (\bibinfo {year} {2018})}\BibitemShut
  {NoStop}%
\bibitem [{\citenamefont {Lu}\ \emph {et~al.}(2015)\citenamefont {Lu},
  \citenamefont {Zheliuk}, \citenamefont {Leermakers}, \citenamefont {Yuan},
  \citenamefont {Zeitler}, \citenamefont {Law},\ and\ \citenamefont
  {Ye}}]{Lu2015}%
  \BibitemOpen
  \bibfield  {author} {\bibinfo {author} {\bibfnamefont {J.~M.}\ \bibnamefont
  {Lu}}, \bibinfo {author} {\bibfnamefont {O.}~\bibnamefont {Zheliuk}},
  \bibinfo {author} {\bibfnamefont {I.}~\bibnamefont {Leermakers}}, \bibinfo
  {author} {\bibfnamefont {N.~F.~Q.}\ \bibnamefont {Yuan}}, \bibinfo {author}
  {\bibfnamefont {U.}~\bibnamefont {Zeitler}}, \bibinfo {author} {\bibfnamefont
  {K.~T.}\ \bibnamefont {Law}},\ and\ \bibinfo {author} {\bibfnamefont {J.~T.}\
  \bibnamefont {Ye}},\ }\href {https://doi.org/10.1126/science.aab2277}
  {\bibfield  {journal} {\bibinfo  {journal} {Science}\ }\textbf {\bibinfo
  {volume} {350}},\ \bibinfo {pages} {1353} (\bibinfo {year}
  {2015})}\BibitemShut {NoStop}%
\bibitem [{\citenamefont {Bawden}\ \emph {et~al.}(2016)\citenamefont {Bawden},
  \citenamefont {Cooil}, \citenamefont {Mazzola}, \citenamefont {Riley},
  \citenamefont {Collins-McIntyre}, \citenamefont {Sunko}, \citenamefont
  {Hunvik}, \citenamefont {Leandersson}, \citenamefont {Polley}, \citenamefont
  {Balasubramanian}, \citenamefont {Kim}, \citenamefont {Hoesch}, \citenamefont
  {Wells}, \citenamefont {Balakrishnan}, \citenamefont {Bahramy},\ and\
  \citenamefont {King}}]{Bawden2016}%
  \BibitemOpen
  \bibfield  {author} {\bibinfo {author} {\bibfnamefont {L.}~\bibnamefont
  {Bawden}}, \bibinfo {author} {\bibfnamefont {S.~P.}\ \bibnamefont {Cooil}},
  \bibinfo {author} {\bibfnamefont {F.}~\bibnamefont {Mazzola}}, \bibinfo
  {author} {\bibfnamefont {J.~M.}\ \bibnamefont {Riley}}, \bibinfo {author}
  {\bibfnamefont {L.~J.}\ \bibnamefont {Collins-McIntyre}}, \bibinfo {author}
  {\bibfnamefont {V.}~\bibnamefont {Sunko}}, \bibinfo {author} {\bibfnamefont
  {K.~W.~B.}\ \bibnamefont {Hunvik}}, \bibinfo {author} {\bibfnamefont
  {M.}~\bibnamefont {Leandersson}}, \bibinfo {author} {\bibfnamefont {C.~M.}\
  \bibnamefont {Polley}}, \bibinfo {author} {\bibfnamefont {T.}~\bibnamefont
  {Balasubramanian}}, \bibinfo {author} {\bibfnamefont {T.~K.}\ \bibnamefont
  {Kim}}, \bibinfo {author} {\bibfnamefont {M.}~\bibnamefont {Hoesch}},
  \bibinfo {author} {\bibfnamefont {J.~W.}\ \bibnamefont {Wells}}, \bibinfo
  {author} {\bibfnamefont {G.}~\bibnamefont {Balakrishnan}}, \bibinfo {author}
  {\bibfnamefont {M.~S.}\ \bibnamefont {Bahramy}},\ and\ \bibinfo {author}
  {\bibfnamefont {P.~D.~C.}\ \bibnamefont {King}},\ }\href
  {https://doi.org/10.1038/ncomms11711} {\bibfield  {journal} {\bibinfo
  {journal} {Nature Communications}\ }\textbf {\bibinfo {volume} {7}},\
  \bibinfo {pages} {11711} (\bibinfo {year} {2016})}\BibitemShut {NoStop}%
\bibitem [{\citenamefont {Mazin}\ \emph {et~al.}(2002)\citenamefont {Mazin},
  \citenamefont {Andersen}, \citenamefont {Jepsen}, \citenamefont {Dolgov},
  \citenamefont {Kortus}, \citenamefont {Golubov}, \citenamefont {Kuz'menko},\
  and\ \citenamefont {van~der Marel}}]{Mazin2002}%
  \BibitemOpen
  \bibfield  {author} {\bibinfo {author} {\bibfnamefont {I.~I.}\ \bibnamefont
  {Mazin}}, \bibinfo {author} {\bibfnamefont {O.~K.}\ \bibnamefont {Andersen}},
  \bibinfo {author} {\bibfnamefont {O.}~\bibnamefont {Jepsen}}, \bibinfo
  {author} {\bibfnamefont {O.~V.}\ \bibnamefont {Dolgov}}, \bibinfo {author}
  {\bibfnamefont {J.}~\bibnamefont {Kortus}}, \bibinfo {author} {\bibfnamefont
  {A.~A.}\ \bibnamefont {Golubov}}, \bibinfo {author} {\bibfnamefont {A.~B.}\
  \bibnamefont {Kuz'menko}},\ and\ \bibinfo {author} {\bibfnamefont
  {D.}~\bibnamefont {van~der Marel}},\ }\href
  {https://doi.org/10.1103/PhysRevLett.89.107002} {\bibfield  {journal}
  {\bibinfo  {journal} {Phys. Rev. Lett.}\ }\textbf {\bibinfo {volume} {89}},\
  \bibinfo {pages} {107002} (\bibinfo {year} {2002})}\BibitemShut {NoStop}%
\bibitem [{\citenamefont {Golubov}\ and\ \citenamefont
  {Mazin}(1997)}]{Golubov1997}%
  \BibitemOpen
  \bibfield  {author} {\bibinfo {author} {\bibfnamefont {A.~A.}\ \bibnamefont
  {Golubov}}\ and\ \bibinfo {author} {\bibfnamefont {I.~I.}\ \bibnamefont
  {Mazin}},\ }\href {https://doi.org/10.1103/PhysRevB.55.15146} {\bibfield
  {journal} {\bibinfo  {journal} {Phys. Rev. B}\ }\textbf {\bibinfo {volume}
  {55}},\ \bibinfo {pages} {15146} (\bibinfo {year} {1997})}\BibitemShut
  {NoStop}%
\bibitem [{\citenamefont {Cho}\ \emph {et~al.}(2018)\citenamefont {Cho},
  \citenamefont {Ko{\'n}czykowski}, \citenamefont {Teknowijoyo}, \citenamefont
  {Tanatar}, \citenamefont {Guss}, \citenamefont {Gartin}, \citenamefont
  {Wilde}, \citenamefont {Kreyssig}, \citenamefont {McQueeney}, \citenamefont
  {Goldman}, \citenamefont {Mishra}, \citenamefont {Hirschfeld},\ and\
  \citenamefont {Prozorov}}]{Cho2018}%
  \BibitemOpen
  \bibfield  {author} {\bibinfo {author} {\bibfnamefont {K.}~\bibnamefont
  {Cho}}, \bibinfo {author} {\bibfnamefont {M.}~\bibnamefont
  {Ko{\'n}czykowski}}, \bibinfo {author} {\bibfnamefont {S.}~\bibnamefont
  {Teknowijoyo}}, \bibinfo {author} {\bibfnamefont {M.~A.}\ \bibnamefont
  {Tanatar}}, \bibinfo {author} {\bibfnamefont {J.}~\bibnamefont {Guss}},
  \bibinfo {author} {\bibfnamefont {P.~B.}\ \bibnamefont {Gartin}}, \bibinfo
  {author} {\bibfnamefont {J.~M.}\ \bibnamefont {Wilde}}, \bibinfo {author}
  {\bibfnamefont {A.}~\bibnamefont {Kreyssig}}, \bibinfo {author}
  {\bibfnamefont {R.~J.}\ \bibnamefont {McQueeney}}, \bibinfo {author}
  {\bibfnamefont {A.~I.}\ \bibnamefont {Goldman}}, \bibinfo {author}
  {\bibfnamefont {V.}~\bibnamefont {Mishra}}, \bibinfo {author} {\bibfnamefont
  {P.~J.}\ \bibnamefont {Hirschfeld}},\ and\ \bibinfo {author} {\bibfnamefont
  {R.}~\bibnamefont {Prozorov}},\ }\href
  {https://doi.org/10.1038/s41467-018-05153-0} {\bibfield  {journal} {\bibinfo
  {journal} {Nature Communications}\ }\textbf {\bibinfo {volume} {9}},\
  \bibinfo {pages} {2796} (\bibinfo {year} {2018})}\BibitemShut {NoStop}%
\bibitem [{\citenamefont {Eom}\ \emph {et~al.}(2006)\citenamefont {Eom},
  \citenamefont {Qin}, \citenamefont {Chou},\ and\ \citenamefont
  {Shih}}]{Eom2006}%
  \BibitemOpen
  \bibfield  {author} {\bibinfo {author} {\bibfnamefont {D.}~\bibnamefont
  {Eom}}, \bibinfo {author} {\bibfnamefont {S.}~\bibnamefont {Qin}}, \bibinfo
  {author} {\bibfnamefont {M.-Y.}\ \bibnamefont {Chou}},\ and\ \bibinfo
  {author} {\bibfnamefont {C.~K.}\ \bibnamefont {Shih}},\ }\href
  {https://doi.org/10.1103/PhysRevLett.96.027005} {\bibfield  {journal}
  {\bibinfo  {journal} {Physical Review Letters}\ }\textbf {\bibinfo {volume}
  {96}},\ \bibinfo {pages} {027005} (\bibinfo {year} {2006})}\BibitemShut
  {NoStop}%
\bibitem [{\citenamefont {Qin}\ \emph {et~al.}(2009)\citenamefont {Qin},
  \citenamefont {Kim}, \citenamefont {Niu},\ and\ \citenamefont
  {Shih}}]{Qin2009}%
  \BibitemOpen
  \bibfield  {author} {\bibinfo {author} {\bibfnamefont {S.}~\bibnamefont
  {Qin}}, \bibinfo {author} {\bibfnamefont {J.}~\bibnamefont {Kim}}, \bibinfo
  {author} {\bibfnamefont {Q.}~\bibnamefont {Niu}},\ and\ \bibinfo {author}
  {\bibfnamefont {C.-K.}\ \bibnamefont {Shih}},\ }\href
  {https://doi.org/10.1126/science.1170775} {\bibfield  {journal} {\bibinfo
  {journal} {Science}\ }\textbf {\bibinfo {volume} {324}},\ \bibinfo {pages}
  {1314} (\bibinfo {year} {2009})}\BibitemShut {NoStop}%
\bibitem [{\citenamefont {Zhang}\ \emph {et~al.}(2010)\citenamefont {Zhang},
  \citenamefont {Cheng}, \citenamefont {Li}, \citenamefont {Sun}, \citenamefont
  {Wang}, \citenamefont {Zhu}, \citenamefont {He}, \citenamefont {Wang},
  \citenamefont {Ma}, \citenamefont {Chen}, \citenamefont {Wang}, \citenamefont
  {Liu}, \citenamefont {Lin}, \citenamefont {Jia},\ and\ \citenamefont
  {Xue}}]{Zhang2010}%
  \BibitemOpen
  \bibfield  {author} {\bibinfo {author} {\bibfnamefont {T.}~\bibnamefont
  {Zhang}}, \bibinfo {author} {\bibfnamefont {P.}~\bibnamefont {Cheng}},
  \bibinfo {author} {\bibfnamefont {W.-J.}\ \bibnamefont {Li}}, \bibinfo
  {author} {\bibfnamefont {Y.-J.}\ \bibnamefont {Sun}}, \bibinfo {author}
  {\bibfnamefont {G.}~\bibnamefont {Wang}}, \bibinfo {author} {\bibfnamefont
  {X.-G.}\ \bibnamefont {Zhu}}, \bibinfo {author} {\bibfnamefont
  {K.}~\bibnamefont {He}}, \bibinfo {author} {\bibfnamefont {L.}~\bibnamefont
  {Wang}}, \bibinfo {author} {\bibfnamefont {X.}~\bibnamefont {Ma}}, \bibinfo
  {author} {\bibfnamefont {X.}~\bibnamefont {Chen}}, \bibinfo {author}
  {\bibfnamefont {Y.}~\bibnamefont {Wang}}, \bibinfo {author} {\bibfnamefont
  {Y.}~\bibnamefont {Liu}}, \bibinfo {author} {\bibfnamefont {H.-Q.}\
  \bibnamefont {Lin}}, \bibinfo {author} {\bibfnamefont {J.-F.}\ \bibnamefont
  {Jia}},\ and\ \bibinfo {author} {\bibfnamefont {Q.-K.}\ \bibnamefont {Xue}},\
  }\href {https://doi.org/10.1038/nphys1499} {\bibfield  {journal} {\bibinfo
  {journal} {Nature Physics}\ }\textbf {\bibinfo {volume} {6}},\ \bibinfo
  {pages} {104} (\bibinfo {year} {2010})}\BibitemShut {NoStop}%
\bibitem [{\citenamefont {Yamada}\ \emph {et~al.}(2013)\citenamefont {Yamada},
  \citenamefont {Hirahara},\ and\ \citenamefont {Hasegawa}}]{Yamada2013}%
  \BibitemOpen
  \bibfield  {author} {\bibinfo {author} {\bibfnamefont {M.}~\bibnamefont
  {Yamada}}, \bibinfo {author} {\bibfnamefont {T.}~\bibnamefont {Hirahara}},\
  and\ \bibinfo {author} {\bibfnamefont {S.}~\bibnamefont {Hasegawa}},\ }\href
  {https://doi.org/10.1103/PhysRevLett.110.237001} {\bibfield  {journal}
  {\bibinfo  {journal} {Physical Review Letters}\ }\textbf {\bibinfo {volume}
  {110}},\ \bibinfo {pages} {237001} (\bibinfo {year} {2013})}\BibitemShut
  {NoStop}%
\bibitem [{\citenamefont {Brun}\ \emph {et~al.}(2014)\citenamefont {Brun},
  \citenamefont {Cren}, \citenamefont {Cherkez}, \citenamefont {Debontridder},
  \citenamefont {Pons}, \citenamefont {Fokin}, \citenamefont {Tringides},
  \citenamefont {Bozhko}, \citenamefont {Ioffe}, \citenamefont {Altshuler},\
  and\ \citenamefont {Roditchev}}]{Brun2014}%
  \BibitemOpen
  \bibfield  {author} {\bibinfo {author} {\bibfnamefont {C.}~\bibnamefont
  {Brun}}, \bibinfo {author} {\bibfnamefont {T.}~\bibnamefont {Cren}}, \bibinfo
  {author} {\bibfnamefont {V.}~\bibnamefont {Cherkez}}, \bibinfo {author}
  {\bibfnamefont {F.}~\bibnamefont {Debontridder}}, \bibinfo {author}
  {\bibfnamefont {S.}~\bibnamefont {Pons}}, \bibinfo {author} {\bibfnamefont
  {D.}~\bibnamefont {Fokin}}, \bibinfo {author} {\bibfnamefont {M.~C.}\
  \bibnamefont {Tringides}}, \bibinfo {author} {\bibfnamefont {S.}~\bibnamefont
  {Bozhko}}, \bibinfo {author} {\bibfnamefont {L.~B.}\ \bibnamefont {Ioffe}},
  \bibinfo {author} {\bibfnamefont {B.~L.}\ \bibnamefont {Altshuler}},\ and\
  \bibinfo {author} {\bibfnamefont {D.}~\bibnamefont {Roditchev}},\ }\href
  {https://doi.org/10.1038/nphys2937} {\bibfield  {journal} {\bibinfo
  {journal} {Nature Physics}\ }\textbf {\bibinfo {volume} {10}},\ \bibinfo
  {pages} {444} (\bibinfo {year} {2014})}\BibitemShut {NoStop}%
\bibitem [{\citenamefont {Maki}\ and\ \citenamefont
  {Tsuneto}(1964)}]{Maki1964}%
  \BibitemOpen
  \bibfield  {author} {\bibinfo {author} {\bibfnamefont {K.}~\bibnamefont
  {Maki}}\ and\ \bibinfo {author} {\bibfnamefont {T.}~\bibnamefont {Tsuneto}},\
  }\href {https://doi.org/10.1143/PTP.31.945} {\bibfield  {journal} {\bibinfo
  {journal} {Progress of Theoretical Physics}\ }\textbf {\bibinfo {volume}
  {31}},\ \bibinfo {pages} {945} (\bibinfo {year} {1964})}\BibitemShut
  {NoStop}%
\bibitem [{\citenamefont {Matsuda}\ and\ \citenamefont
  {Shimahara}(2007)}]{Matsuda2007}%
  \BibitemOpen
  \bibfield  {author} {\bibinfo {author} {\bibfnamefont {Y.}~\bibnamefont
  {Matsuda}}\ and\ \bibinfo {author} {\bibfnamefont {H.}~\bibnamefont
  {Shimahara}},\ }\href {https://doi.org/10.1143/JPSJ.76.051005} {\bibfield
  {journal} {\bibinfo  {journal} {Journal of the Physical Society of Japan}\
  }\textbf {\bibinfo {volume} {76}},\ \bibinfo {pages} {051005} (\bibinfo
  {year} {2007})}\BibitemShut {NoStop}%
\bibitem [{\citenamefont {Samokhin}(2005)}]{Samokhin2005}%
  \BibitemOpen
  \bibfield  {author} {\bibinfo {author} {\bibfnamefont {K.~V.}\ \bibnamefont
  {Samokhin}},\ }\href {https://doi.org/10.1103/PhysRevLett.94.027004}
  {\bibfield  {journal} {\bibinfo  {journal} {Physical Review Letters}\
  }\textbf {\bibinfo {volume} {94}},\ \bibinfo {pages} {027004} (\bibinfo
  {year} {2005})}\BibitemShut {NoStop}%
\bibitem [{\citenamefont {Samokhin}(2007)}]{Samokhin2007}%
  \BibitemOpen
  \bibfield  {author} {\bibinfo {author} {\bibfnamefont {K.~V.}\ \bibnamefont
  {Samokhin}},\ }\href {https://doi.org/10.1103/PhysRevB.76.094516} {\bibfield
  {journal} {\bibinfo  {journal} {Physical Review B}\ }\textbf {\bibinfo
  {volume} {76}},\ \bibinfo {pages} {094516} (\bibinfo {year}
  {2007})}\BibitemShut {NoStop}%
\bibitem [{\citenamefont {Song}\ and\ \citenamefont
  {Koshelev}(2019)}]{Song2019}%
  \BibitemOpen
  \bibfield  {author} {\bibinfo {author} {\bibfnamefont {K.~W.}\ \bibnamefont
  {Song}}\ and\ \bibinfo {author} {\bibfnamefont {A.~E.}\ \bibnamefont
  {Koshelev}},\ }\href {https://doi.org/10.1103/PhysRevX.9.021025} {\bibfield
  {journal} {\bibinfo  {journal} {Physical Review X}\ }\textbf {\bibinfo
  {volume} {9}},\ \bibinfo {pages} {021025} (\bibinfo {year}
  {2019})}\BibitemShut {NoStop}%
\bibitem [{\citenamefont {Zhou}\ \emph {et~al.}(2016)\citenamefont {Zhou},
  \citenamefont {Yuan}, \citenamefont {Jiang},\ and\ \citenamefont
  {Law}}]{Zhou2016}%
  \BibitemOpen
  \bibfield  {author} {\bibinfo {author} {\bibfnamefont {B.~T.}\ \bibnamefont
  {Zhou}}, \bibinfo {author} {\bibfnamefont {N.~F.~Q.}\ \bibnamefont {Yuan}},
  \bibinfo {author} {\bibfnamefont {H.-L.}\ \bibnamefont {Jiang}},\ and\
  \bibinfo {author} {\bibfnamefont {K.~T.}\ \bibnamefont {Law}},\ }\href
  {https://doi.org/10.1103/PhysRevB.93.180501} {\bibfield  {journal} {\bibinfo
  {journal} {Physical Review B}\ }\textbf {\bibinfo {volume} {93}},\ \bibinfo
  {pages} {180501(R)} (\bibinfo {year} {2016})}\BibitemShut {NoStop}%
\bibitem [{\citenamefont {Linder}\ and\ \citenamefont
  {Balatsky}(2019)}]{Linder2017}%
  \BibitemOpen
  \bibfield  {author} {\bibinfo {author} {\bibfnamefont {J.}~\bibnamefont
  {Linder}}\ and\ \bibinfo {author} {\bibfnamefont {A.~V.}\ \bibnamefont
  {Balatsky}},\ }\href {https://doi.org/10.1103/RevModPhys.91.045005}
  {\bibfield  {journal} {\bibinfo  {journal} {Reviews of Modern Physics}\
  }\textbf {\bibinfo {volume} {91}},\ \bibinfo {pages} {045005} (\bibinfo
  {year} {2019})}\BibitemShut {NoStop}%
\bibitem [{\citenamefont {Gentile}\ \emph {et~al.}(2011)\citenamefont
  {Gentile}, \citenamefont {Noce}, \citenamefont {Romano}, \citenamefont
  {Annunziata}, \citenamefont {Linder},\ and\ \citenamefont
  {Cuoco}}]{Gentile2011}%
  \BibitemOpen
  \bibfield  {author} {\bibinfo {author} {\bibfnamefont {P.}~\bibnamefont
  {Gentile}}, \bibinfo {author} {\bibfnamefont {C.}~\bibnamefont {Noce}},
  \bibinfo {author} {\bibfnamefont {A.}~\bibnamefont {Romano}}, \bibinfo
  {author} {\bibfnamefont {G.}~\bibnamefont {Annunziata}}, \bibinfo {author}
  {\bibfnamefont {J.}~\bibnamefont {Linder}},\ and\ \bibinfo {author}
  {\bibfnamefont {M.}~\bibnamefont {Cuoco}},\ }\href
  {http://arxiv.org/abs/1109.4885} {\bibfield  {journal} {\bibinfo  {journal}
  {Preprint}\ ,\ \bibinfo {pages} {1}} (\bibinfo {year} {2011})},\ \Eprint
  {https://arxiv.org/abs/1109.4885} {arXiv:1109.4885} \BibitemShut {NoStop}%
\bibitem [{\citenamefont {Black-Schaffer}\ and\ \citenamefont
  {Balatsky}(2013)}]{Black-Schaffer2013}%
  \BibitemOpen
  \bibfield  {author} {\bibinfo {author} {\bibfnamefont {A.~M.}\ \bibnamefont
  {Black-Schaffer}}\ and\ \bibinfo {author} {\bibfnamefont {A.~V.}\
  \bibnamefont {Balatsky}},\ }\href
  {https://doi.org/10.1103/PhysRevB.88.104514} {\bibfield  {journal} {\bibinfo
  {journal} {Physical Review B}\ }\textbf {\bibinfo {volume} {88}},\ \bibinfo
  {pages} {104514} (\bibinfo {year} {2013})}\BibitemShut {NoStop}%
\bibitem [{\citenamefont {Rahimi}\ \emph {et~al.}(2017)\citenamefont {Rahimi},
  \citenamefont {Moghaddam}, \citenamefont {Dykstra}, \citenamefont
  {Governale},\ and\ \citenamefont {Z\"ulicke}}]{Moghaddam}%
  \BibitemOpen
  \bibfield  {author} {\bibinfo {author} {\bibfnamefont {M.~A.}\ \bibnamefont
  {Rahimi}}, \bibinfo {author} {\bibfnamefont {A.~G.}\ \bibnamefont
  {Moghaddam}}, \bibinfo {author} {\bibfnamefont {C.}~\bibnamefont {Dykstra}},
  \bibinfo {author} {\bibfnamefont {M.}~\bibnamefont {Governale}},\ and\
  \bibinfo {author} {\bibfnamefont {U.}~\bibnamefont {Z\"ulicke}},\ }\href
  {https://doi.org/10.1103/PhysRevB.95.104515} {\bibfield  {journal} {\bibinfo
  {journal} {Phys. Rev. B}\ }\textbf {\bibinfo {volume} {95}},\ \bibinfo
  {pages} {104515} (\bibinfo {year} {2017})}\BibitemShut {NoStop}%
\bibitem [{\citenamefont {Cayao}\ and\ \citenamefont
  {Black-Schaffer}(2017)}]{cayao2017}%
  \BibitemOpen
  \bibfield  {author} {\bibinfo {author} {\bibfnamefont {J.}~\bibnamefont
  {Cayao}}\ and\ \bibinfo {author} {\bibfnamefont {A.~M.}\ \bibnamefont
  {Black-Schaffer}},\ }\href {https://doi.org/10.1103/PhysRevB.96.155426}
  {\bibfield  {journal} {\bibinfo  {journal} {Phys. Rev. B}\ }\textbf {\bibinfo
  {volume} {96}},\ \bibinfo {pages} {155426} (\bibinfo {year}
  {2017})}\BibitemShut {NoStop}%
\bibitem [{\citenamefont {Cayao}\ and\ \citenamefont
  {Black-Schaffer}(2018)}]{cayao2018}%
  \BibitemOpen
  \bibfield  {author} {\bibinfo {author} {\bibfnamefont {J.}~\bibnamefont
  {Cayao}}\ and\ \bibinfo {author} {\bibfnamefont {A.~M.}\ \bibnamefont
  {Black-Schaffer}},\ }\href {https://doi.org/10.1103/PhysRevB.98.075425}
  {\bibfield  {journal} {\bibinfo  {journal} {Phys. Rev. B}\ }\textbf {\bibinfo
  {volume} {98}},\ \bibinfo {pages} {075425} (\bibinfo {year}
  {2018})}\BibitemShut {NoStop}%
\end{thebibliography}%

\end{document}